\documentclass[12pt]{article}%
\usepackage[nosort]{cite}
\usepackage[usenames, dvipsnames]{xcolor}
\usepackage{graphicx}
\usepackage{multicol}
\usepackage{amsfonts}
\usepackage{amssymb}
\usepackage{amsmath}
\usepackage{whatthe}
\usepackage{afterpage}
\usepackage{braket}
\usepackage{setspace}
\usepackage{verbatim}
\usepackage{color}
\usepackage{longtable}
\usepackage{float}
\usepackage{subcaption}
\usepackage{epsfig}
\usepackage{enumerate}
\usepackage{epstopdf}
\usepackage[enableskew, vcentermath]{youngtab}
\usepackage{adjustbox}
\usepackage{multirow}
\usepackage{tikz}
\usepackage[margin=1in]{geometry}
\usepackage{titletoc}
\usepackage[percent]{overpic}
\usepackage{gensymb}
\usepackage{bbm}
\usepackage{mathtools}
\usepackage{tikz-cd}%
\newsavebox{\mysavebox}

\usetikzlibrary{decorations.markings}

\newcommand{\Z}{\mathbb{Z}}

\numberwithin{equation}{section}

\usetikzlibrary{chains}
\allowdisplaybreaks
\tikzset{node distance=2em, ch/.style={circle,draw,on chain,inner sep=2pt},chj/.style={ch,join},every path/.style={shorten >=4pt,shorten <=4pt},line width=1pt,baseline=-1ex}

\newcommand{\ff}{\mathfrak{f}}
\newcommand{\fg}{\mathfrak{g}}
\newcommand{\fh}{\mathfrak{h}}
\newcommand{\fso}{\mathfrak{so}}
\newcommand{\fsp}{\mathfrak{sp}}
\newcommand{\fsu}{\mathfrak{su}}
\newcommand{\fe}{\mathfrak{e}}

\makeatletter \@addtoreset{equation}{section} \makeatother

\colorlet{darkblue}{blue!70!black}
\colorlet{darkgreen}{green!70!black}

\usepackage[colorlinks=true,urlcolor=darkblue,linktocpage=true,linkcolor=darkblue,citecolor=darkblue]{hyperref}
\hypersetup{
colorlinks=true,
linkcolor=blue,
citecolor=magenta,
}

\begin{document}

%% Report number
\vspace*{-2cm}
\begin{flushright}
{\tt UPR-1332-T}\\
{\tt LMU-ASC 16/24}\\
{\tt CERN-TH-2024-145}
\end{flushright}

\date{October 2024}

\title{Frozen Generalized Symmetries}

\institution{PENN}{\centerline{${}^{1}$Department of Physics and Astronomy, University of Pennsylvania, Philadelphia, PA 19104, USA}}

\institution{PENNmath}{\centerline{${}^{2}$Department of Mathematics, University of Pennsylvania, Philadelphia, PA 19104, USA}}

\institution{MARIBOR}{\centerline{${}^{3}$Center for Applied Mathematics and Theoretical Physics, University of Maribor, Maribor, Slovenia}}

\institution{LMU}{\centerline{${}^{4}$Arnold Sommerfeld Center for Theoretical Physics, LMU, Munich, 80333, Germany}}

\institution{BOLOGNA}{\centerline{${}^{5}$Dipartimento di Fisica e Astronomia, Universita di Bologna, via Irnerio 46, Bologna, Italy}}

\institution{INFN}{\centerline{${}^{6}$INFN, Sezione di Bologna, viale Berti Pichat 6/2, Bologna, Italy}}

\institution{CERN}{\centerline{${}^{7}$Theoretical Physics Department, CERN, 1211 Geneva 23, Switzerland}}

\institution{IPMU}{\centerline{${}^{8}$Kavli IPMU, University of Tokyo, Kashiwa, Chiba 277-8583, Japan}}

\authors{
Mirjam Cveti\v{c}\worksat{\PENN,\PENNmath,\MARIBOR},
Markus Dierigl\worksat{\LMU,\CERN},
Ling Lin\worksat{\BOLOGNA,\INFN},\\[3mm]
Ethan Torres\worksat{\CERN}, and
Hao Y. Zhang\worksat{\IPMU}}

\abstract{\noindent M-theory frozen singularities are (locally) $D$- or $E$-type orbifold singularities with a background fractional $C_3$-monodromy surrounding them. In this paper, we revisit such backgrounds and address several puzzling features of their physics. We first give a top-down derivation of how the $D$- or $E$-type 7D $\mathcal{N}=1$ gauge theory directly ``freezes" to a lower rank gauge theory due to the $C_3$-background. This relies on a Hanany--Witten effect of fractional M5 branes and the presence of a gauge anomaly of fractional D$p$ probes in the circle reduction. Additionally, we compute defect groups and 8D symmetry topological field theories (SymTFTs) of the 7D frozen theories in several duality frames. We apply our results to understanding the evenness condition of strings ending on O$7^+$-planes, and calculating the global forms of supergravity gauge groups of M-theory compactified on $T^4/\Gamma$ with frozen singularities. In an Appendix, we also revisit IIA $ADE$ singularities with a $C_1$-monodromy along a 1-cycle in the boundary lens space and show that this freezes the gauge degrees-of-freedom via confinement.}

\maketitle

\setcounter{tocdepth}{2}

\tableofcontents

\newpage

\section{Introduction}

Frozen singularities present an interesting corner of consistent string theory backgrounds that remains relatively unexplored compared to their unfrozen cousins. In their M-theory description, frozen singularities are realized as fractional $G_4$ fluxes stuck at geometric singularities, which can be detected via their non-trivial $C_3$ holonomy at the asymptotic boundary \cite{Witten:1997bs,deBoer:2001wca, Atiyah:2001qf}. Dimensional reduction on these geometrical singularities typically engineer non-Abelian supersymmetric gauge theories, but the presence of the fractional fluxes {\it freezes} some of the original gauge degrees of freedom \cite{Witten:1997bs,deBoer:2001wca, Atiyah:2001qf, Tachikawa:2015wka, Bhardwaj:2018jgp, Fraiman:2021hma, ParraDeFreitas:2022wnz, Cecotti:2023mlc, Morrison:2023hqx, Donagi:2023sbk, Oehlmann:2024cyn}.
While this freezing mechanism is more apparent from other string duality frames, a detailed explanation directly in M-theory remained mysterious.
Furthermore, it has become apparent in recent years that studying the local dynamics of gauge theories only captures part of the theory, with more refined data encoded in the generalized (categorical) symmetries of the system \cite{Aharony:2013hda, Gaiotto:2014kfa}, see also \cite{Cordova:2022ruw, Schafer-Nameki:2023jdn, Bhardwaj:2023kri, Luo:2023ive, Shao:2023gho} for reviews.
The goal of this work is to address both these aspects, by explicitly deriving the freezing of $ADE$ singularities in M-theory as a consequence of the background holonomy/flux
\begin{equation}\label{eq:frozenbkrdintro}
 r \equiv \int_{S^3/\Gamma} C_3=\int_{\mathbb{C}^2/\Gamma} G_4= \frac{n}{d} \enspace \mathrm{mod} \enspace 1 \,, \quad \mathrm{gcd}(n, d) = 1 \,,
\end{equation}
and additionally understand how this affects the 1-/4-form symmetries of the associated 7D super-Yang--Mills (SYM) theory.

To extract the global symmetries of a geometrically engineered gauge sector in string and M-theory one heavily utilizes the properties of the geometrical background, see , e.g., \cite{DelZotto:2015isa, Garcia-Etxebarria:2019cnb, Morrison:2020ool, Albertini:2020mdx, Bah:2020uev, DelZotto:2020esg,  Apruzzi:2020zot, Bhardwaj:2020phs, Cvetic:2020kuw, DelZotto:2020sop,Bhardwaj:2021pfz, Apruzzi:2021vcu, Cvetic:2021sxm, Braun:2021sex, Bhardwaj:2021wif, Tian:2021cif, Bhardwaj:2021mzl, DelZotto:2022fnw, Cvetic:2022imb,DelZotto:2022joo,Hubner:2022kxr, Bashmakov:2022jtl , vanBeest:2022fss,Bashmakov:2022uek, Acharya:2023bth, Dierigl:2023jdp, Cvetic:2023plv,Bashmakov:2023kwo, Apruzzi:2023uma, Closset:2023pmc}. In particular, the branes of the underlying theory wrapped on various cycles of the geometry (which can be more general than singular homology cycles, e.g., K-theory cycles \cite{Heckman:2022xgu,Zhang:2024oas,Torres:2024sbl}) produce the topological symmetry operators as well as the charged objects \cite{Apruzzi:2021nmk, Apruzzi:2022rei, GarciaEtxebarria:2022vzq, Heckman:2022muc}, whose properties are encoded in the symmetry topological field theory (SymTFT) \cite{Freed:2012bs, Apruzzi:2021nmk, Freed:2022qnc}.
The study of this more general framework has found much attention within string compactifications as well, see \cite{Witten:1998wy,Belov:2006xj,Gukov:2020btk,Apruzzi:2022dlm,Bergman:2022otk,Lawrie:2023tdz, Cvetic:2023pgm, Baume:2023kkf, Yu:2023nyn, Gould:2023wgl,Heckman:2024oot,Braeger:2024jcj,Cvetic:2024dzu,Apruzzi:2024htg,GarciaEtxebarria:2024fuk, Heckman:2024zdo} in addition to the references above.

The SymTFT of a $D$-dimensional system is a topological theory in $(D+1)$-dimensions, where the extra dimension is taken to be an interval. One end of the interval contains the gapless modes, e.g., the degrees of freedom of a $\mathfrak{g}$ gauge theory, whose generalized global symmetries are encoded on the other end of the interval, the topological boundary, via the implementation of gapped boundary conditions. These boundary conditions fix which of the topological operators of the SymTFT can end on the boundary and which have to be parallel to it. These, in turn, encode the ending charged operators, that stretch along the extra dimension, and the symmetry operators, localized on a point in the extra direction, respectively.
In geometrically engineered gauge theories, i.e., a dimensional reduction of string or M-theory on a local neighborhood $X$ of the singularity, the SymTFT is dimensional reduction on the asymptotic boundary $\partial X$ \cite{Apruzzi:2021nmk}. For this paper, we are interested in $(D+1)=8$ since we obtain the SymTFT by reducing M-theory on $S^3/\Gamma$.
\begin{figure}[ht]
    \centering
    \includegraphics[width=1\textwidth, trim = {0cm 2cm 0cm 3cm}]{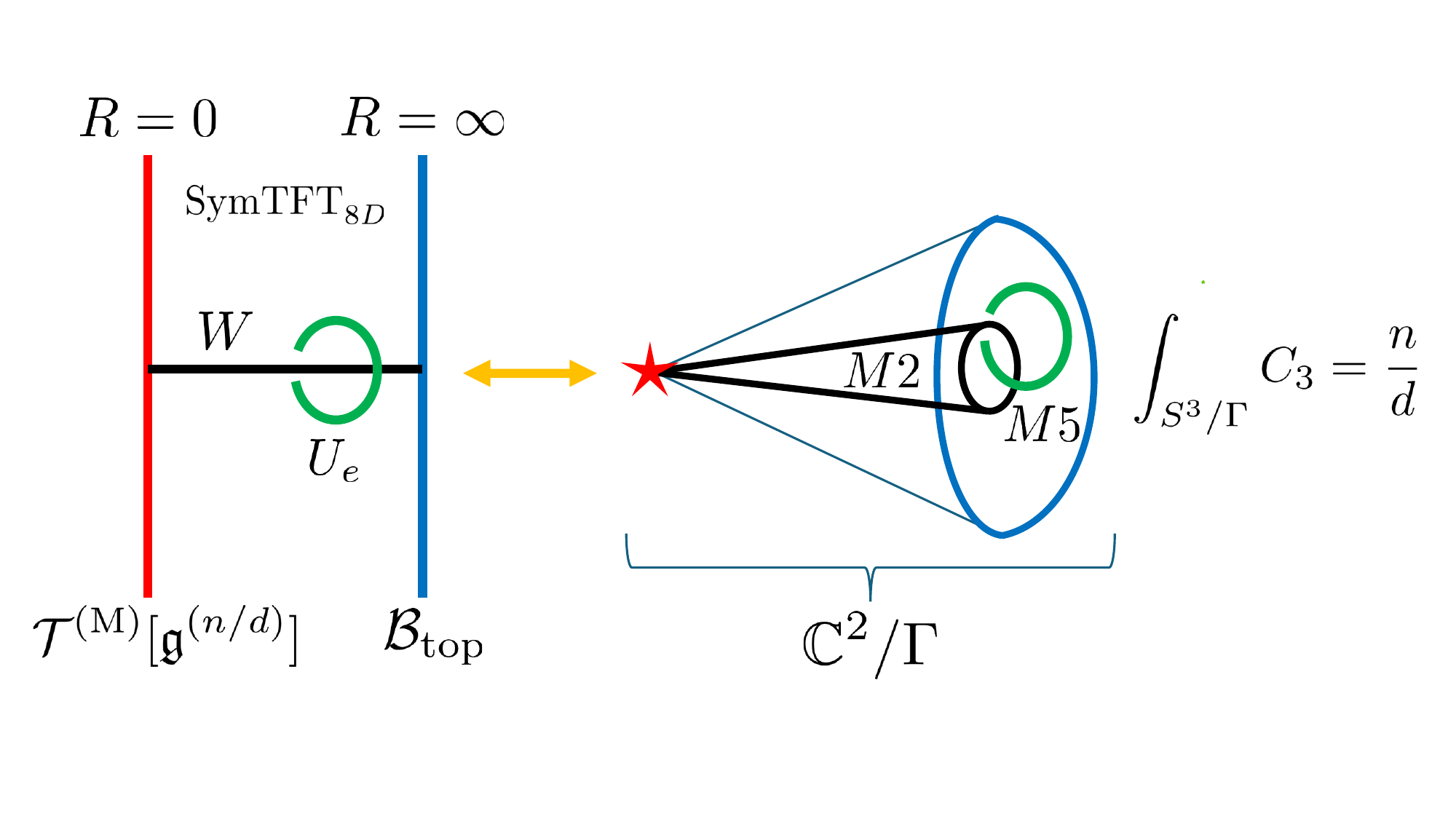}
    \caption{ On the left, we illustrate an 8D SymTFT construction which has a non-topological boundary $\mathcal{T}^{(\mathrm{M})}[\mathfrak{g}^{(n/d)}]$ and a topological boundary condition $\mathcal{B}_{\mathrm{top.}}$. The boundary condition is such that there exist and operators $W$ and $U_e$ which respectively produce a Wilson line and $\mathbb{Z}^{(1)}_2$ symmetry operator in the 7D gauge theory. On the right, we show the M-theory frozen singularity geometry where $W$ is engineered from an M2 brane on a relative 2-cycle and $U_e$ from an M5 brane on a boundary 1-cycle.}
    \label{fig:symtfte7}
\end{figure}
The interval direction is described by the radial coordinate $R$ in $X = \mathbb{C}^2 /\Gamma$ with respect to the singularity. The gapless boundary at $R=0$ describes the $\mathfrak{g} \equiv \mathfrak{g}_\Gamma$ degrees of freedom associated with the singularity, and the topological boundary conditions are implemented at the asymptotic boundary at $R = \infty$, see Figure \ref{fig:symtfte7}.

For 7D SYM theories one expects a class of invertible generalized symmetries encoded in the spectrum of electric Wilson line operators and their magnetically dual four-dimensional 't Hooft operators. In the M-theory realization, these originate from M2 and M5 branes wrapping relative 2-cycles in $X$, i.e., they stretch from the asymptotic boundary all the way to the singularity. These lead to 1-form and 4-form symmetries given by the center $Z(G)$ of the simply-connected group $G$ associated to the gauge dynamics. The topological symmetry operators on the other hand are described by M5 and M2 branes wrapping boundary 1-cycles. The SymTFT additionally captures the 't Hooft anomaly between them related to the mutual non-locality between Wilson and 't Hooft operators, which can be derived from reduction of the kinetic term of the 11D M-theory action \cite{GarciaEtxebarria:2024fuk}.

An application of this general recipe to frozen singularities raises an immediate problem. Since frozen singularities share the geometrical backgrounds with their unfrozen versions (i.e., $r=0$), all conclusions about the generalized symmetries extracted purely from geometry should be identical. However, when $r \neq 0$ the gauge sector changes (see Table \ref{tab:tachikawaintro}), and one would expect the generalized symmetries to be modified as well.
\begin{table}[ht]
\begin{equation*}
\begin{array}{c@{\,}l||c|cc|ccc|ccccc}
&\fg_\Gamma &   \fso(2k+8)  &   \fe_6 & \fe_6 & \fe_7 &\fe_7 &\fe_7&  \fe_8& \fe_8& \fe_8& \fe_8& \fe_8 \\[1pt]
\hline
&\vphantom{\Big|} r = \frac{n}{d}&  \frac{1}{2} &    \frac{1}{2} & \frac{1}{3}, \frac{2}{3} & \frac{1}{2} & \frac{1}{3},\frac{2}{3} & \frac{1}{4},\frac{3}{4} &   \frac{1}{2} & \frac{1}{3},\frac{2}{3} & \frac{1}{4},\frac{3}{4} & \frac{1}{5},\frac{2}{5},\frac{3}{5},\frac{4}{5} & \frac{1}{6},\frac{5}{6}\\
&\fh_{\Gamma, d} &  \fsp(k)  & \fsu(3) & \varnothing &  \fso(7) & \fsu(2) & \varnothing &  \ff_4 & \fg_2 & \fsu(2) & \varnothing & \varnothing\end{array}
\end{equation*}
\caption{Frozen half-BPS M-theory singularities of the form $\mathbb{C}^2/\Gamma$ \cite{deBoer:2001wca, Tachikawa:2015wka}. The top-line gives the Lie algebra $\mathfrak{g}_{\Gamma}$ of the 7D SYM with no frozen flux, and $\mathfrak{h}_{\Gamma, d}$ is the frozen gauge algebra.}
\label{tab:tachikawaintro}
\end{table}

To illustrate the mismatch between our expectations and geometry, it is easiest to consider a fully frozen singularity, i.e., a singularity whose frozen version hosts no continuous gauge degrees of freedom. If we denote an (un)frozen singularity by $\mathfrak{g}^{(n/d)}$ to indicate the $ADE$-type with flux $(n/d)$, then $\mathfrak{e}^{(1/3)}_6$ is an instance of such a fully frozen singularity.
In such cases, we then do not expect any non-trivial 1-form and 4-form symmetries associated to the gauge theory as described above. Nevertheless the asymptotic geometry allows the definition of the associated topological operators, suggesting 1-form and 4-form symmetries identical to the unfrozen gauge theory. In the following we resolve this mismatch using two complementary techniques.

One approach uses the definition of a local \emph{freezing map} that specifies which M2 and M5 brane states are allowed after the inclusion of the fractional flux.\footnote{An analogous map has been previously constructed for 8D theories realized on 7-branes with O7$^\pm$-planes in type IIB \cite{Cvetic:2022uuu}.
Furthermore, ``global'' freezing maps for theories with dynamical gravity, i.e., compact internal spaces, which were derived in dual heterotic compactifications, have appeared in \cite{Font:2021uyw, Cvetic:2021sjm,Fraiman:2021soq,Fraiman:2021hma,Fraiman:2022aik}.
Note that the construction of these ``global'' maps are inherently tied to the presence of dynamical gravity in these models, and they do not inform the most general local freezings that are possible in M-theory on non-compact internal spaces.
}
We show how these freezing rules come about from a top-down perspective using the relation between M2 branes within the 7D theory and gauge instantons, in cases the frozen gauge algebra is non-trivial.
However, it can equally be applied in the case of fully frozen gauge dynamics.
Extending the freezing map to the charged extended operators allows the identification of the higher-form symmetries of the frozen singularities, which can be elegantly captured in terms of the charge lattices of the unfrozen singularities.
We find that while the magnetic 4-form symmetries are identical to the unfrozen theory, the 1-form symmetry sector is modified. For the fully frozen theories the 1-form symmetries are broken completely, while for other models they can be partially broken ($D$-type singularities). Throughout this derivation the only duality we use is a circle reduction to IIA so one can view our top-down derivation of why the $C_3$ holonomy at infinity freezes the $ADE$ singularity in IIA without appealing to the long chain of dualities currently invoked to argue to the freezing in the literature, e.g. \cite{deBoer:2001wca, Atiyah:2001qf, Tachikawa:Slides}.

Naively, this result leads to a puzzle for the SymTFT:
since the singularity and the boundary topology are not modified, the 8D bulk theory should not change, especially since the flux responsible for the freezing is localized at the singularity.
To accommodate the reduction of the 1-form symmetry in the 7D theory, we therefore expect a modification of the physical boundary.
While it is clear that this modification must trace back to the non-trivial flux in the M-theory realization, it is not immediately clear how to extract this using the methods we use to derive the local freezing rules.

To clarify how the flux changes the physical boundary of the SymTFT, we utilize the duality between frozen M-theory singularities and twisted circle compactifications of F-theory, see \cite{deBoer:2001wca,Tachikawa:2015wka}. This complementary derivation precisely recovers the 1- and 4-form symmetries of the frozen singularities and identifies the charged states in terms of strings and 5-brane states that behave appropriately under an automorphism of the central elliptic fiber. Again, we find that the 4-form symmetries are identical to the unfrozen setting while the 1-form symmetries will in general be modified. This translates into the fact that while the M2 brane configurations are restricted in the presence of fractional fluxes the M5 brane states are not, which corroborates our construction of the freezing maps directly in the M-theory setting.

The main advantage of this second approach is that we now have a purely geometric background $Z \cong (Y \times S^1)/\mathbb{Z}_d$ on which M-theory compactifies to the 6D theory that is the (untwisted) circle compactification of the frozen M-theory model.
For this 6D theory, we can apply the usual machinery to derive the SymTFT and the boundary conditions.
In this approach we can clearly trace the modification of the 1-form symmetry to a topological term on the physical boundary at $R=0$ that arises from \emph{compact} torsional 1-cycles in $Z$.
The physical implications of this sector, which apply also to fully frozen cases, are consistent with the interpretation of compact torsion cycles in previous works \cite{Cvetic:2023pgm,Gould:2023wgl}.

These techniques provide evidence for the existence of a topological counterterm on the physical boundary, which we also relate to modifications in correlation functions of the extended operators in the case of the $\mathfrak{e}^{(1/2)}_6$ frozen singularity and to modified boundary conditions for the 8D SymTFT operators on the $R=0$ boundary. The case of $E_8$ singularity with flux $r = \tfrac{1}{4}$, which we denote by $\mathfrak{e}^{(1/4)}_8$ in our nomenclature, has perhaps the most mysterious symmetry properties.
The geometrical structure, both in the frozen M-theory and the twisted circle compactification description,
does not allow for the construction of symmetry operators, suggesting a trivial 1- and 4-form symmetry sector.
Yet, the frozen gauge algebra is $\mathfrak{su}(2)$ and one expects Wilson and 't Hooft operators accounting for $Z(SU(2)) = \mathbb{Z}_2$ 1- and 4-form symmetries.
Motivated by the other examples where we find from top-down a topological term that modifies the 1-form symmetry, we suggest a similar solution to this mismatch from the bottom-up.
Namely, we propose the existence of a counterterm on the gapless physical boundary which breaks both 1- and 4-form symmetries, explaining why they cannot be recovered from geometry.
Moreover, the counterterm arises as a boundary term of an 8D TFT sector which trivializes the SymTFT of the 7D frozen gauge theory, thus resolving the discrepancy.

To summarize, in our analysis of frozen M-theory singularities we obtain the following results:
\begin{itemize}
 \item{A top-down derivation of how the frozen flux \eqref{eq:frozenbkrdintro} causes the $\mathfrak{g} \equiv \mathfrak{g}_\Gamma$ gauge theory degrees of freedom localized on $\mathbb{C}^2/\Gamma$ to freeze to a lower rank algebra $\mathfrak{h}_{\Gamma, d}$. }
    \item{Freezing maps that are used to extract the generalized symmetries of frozen singularities.}
    \item{A geometric derivation of the same generalized symmetries using twisted circle compactifications in the F-theory dual.}
    \item{A top-down derivation of SymTFT descriptions of the invertible 1- and 4-form symmetry sector of the frozen gauge theories.}
    \item{A bottom-up solution to the frozen $\mathfrak{su}(2)$ theory originating from an $E_8$ singularity via a local counterterm on the gapless boundary that trivializes the SymTFT, which can have wider field theory applications.}
\end{itemize}
All of these results demonstrate that frozen singularities not only modify the local gauge dynamics captured in terms of the gauge algebra, but also the generalized symmetries of the system in a non-intuitive fashion.

The manuscript is organized as follows: In Section \ref{sec:recap} we recall known facts about frozen singularities in M-theory and how their gauge algebra can be extracted from the precise value of the fractional localized flux. We define and motivate the local freezing map, producing the correct gauge degrees of freedom, in Section \ref{sec:freezemap}. This freezing map is then applied to extract the generalized symmetries of the frozen 7D SYM theory in Section \ref{sec:defectgrps}. The results are confirmed geometrically in the F-theory dual description. Section \ref{sec:symtft} reformulates the realization of these symmetries at the level of the 8D SymTFT, and discusses a solution to the mysterious situation of the $\mathfrak{e}^{(1/4)}_8$ frozen singularity which engineers an $\mathfrak{su}(2)$ gauge algebra without 1- and 4-form symmetries. These general results can be used in order to understand predicted properties of O7$^+$ branes in type IIB and the construction of gravitational 7D theories in the presence of frozen fluxes, which we summarize in Section \ref{sec:applications}. We conclude in Section \ref{sec:conc} and point towards some interesting questions for future work. In Appendix \ref{app:appA} we give an argument for how the gauge algebras of IIA $ADE$ singularities with a 2-form RR flux, i.e., $\int_{\gamma_1}C_1 \neq 0$ for a boundary 1-cycle $\gamma_1\in H_1(S^3/\Gamma)$, freeze via a confinement mechanism. This argument is independent of (and is consistent with) the argument of \cite{deBoer:2001wca} which relies on a dual version of the Freed--Witten anamoly. In Appendix \ref{app:fullyfrozen}, we give a geometrical derivation of the presence of a counterterm for fully frozen singularities. Finally, Appendix \ref{app:resolving} gives homology computations of the twisted F-theory geometry under resolution of the singularities.

\section{Review of Frozen Singularities}
\label{sec:recap}

In this section we recall the construction and known properties of frozen singularities in M-theory. We will mainly recall the facts from \cite{deBoer:2001wca, Atiyah:2001qf, Tachikawa:2015wka}, but see also \cite{Witten:1997bs, Bhardwaj:2018jgp, Fraiman:2021hma, ParraDeFreitas:2022wnz, Cecotti:2023mlc, Morrison:2023hqx, Donagi:2023sbk, Oehlmann:2024cyn} for other works on frozen singularities, possibly in other duality frames.

\subsection{Unfrozen Singularities}
\label{subsec:unfrozen}

The starting point of a frozen singularity in M-theory is an $ADE$ singularity, \cite{deBoer:2001wca, Atiyah:2001qf}, described by the quotient $\mathbb{C}^2 / \Gamma$. The discrete groups $\Gamma$ are subgroups of $SU(2)$ and allow for an $ADE$ classification. The $A$-series corresponds to $\Gamma = \mathbb{Z}_n$, the $D$-series to the binary dihedral group, and the exceptional groups are given by the binary tetrahedral group for $E_6$, the binary octahedral group for $E_7$, and the binary icosahedral group for $E_8$, respectively. In the following we will specify which group we refer to by the notation $\Gamma_G$.

Each of the unfrozen $ADE$ singularities can be resolved into a smooth asymptotic locally Euclidean (ALE) space, which we will denote by $\widetilde{X}_{\Gamma}$. This involves the blow-up of a number $\text{rank}(\mathfrak{g}_{\Gamma})$ curves $E_i$, which topologically are 2-spheres $S^2 \simeq \mathbb{P}^1$, where $\mathfrak{g}_{\Gamma}$ is the non-Abelian Lie algebra associated to $\Gamma$. These curves intersect according to the associated Dynkin diagram, producing the (negative of) the Cartan matrix $C_{ij}$ of $\mathfrak{g}_{\Gamma}$:
\begin{equation}
E_i \cdot E_j = - C_{ij} \,.
\label{eq:Cartanintersection}
\end{equation}
The asymptotic geometry of such a singularity is described by
\begin{equation}
\partial (\mathbb{C}^2 / \Gamma) = S^3 / \Gamma \,,
\end{equation}
a generalized lens space whose homology (we consider integer valued (co)homology in this work unless stated otherwise) is
\begin{equation}
    H_*(S^3/\Gamma)=\left\{ \mathbb{Z}, \mathrm{Ab}(\Gamma), 0, \mathbb{Z}\right\} \,,
\end{equation}
where $\mathrm{Ab}(\Gamma):=\Gamma/[\Gamma, \Gamma]$ denotes the Abelianization of $\Gamma$.

Putting M-theory on such a singular background produces an $\mathcal{N} = 1$ super-Yang--Mills (SYM) gauge theory with gauge algebra given by $\mathfrak{g}_{\Gamma}$. Going to the resolution $\widetilde{X}_{\Gamma}$ corresponds to an adjoint Higgs mechanism breaking the gauge theory to its maximal torus, the Cartan subalgebra. This identifies the Cartan generators to be associated to the reduction of the M-theory 3-form $C_3$ with respect to the harmonic forms dual to the blow-up curves $E_i$, which we denote by $\omega_i$. This can be written as the decomposition
\begin{equation}
C_3 = \sqrt{-1} \sum_{i = 1}^{\text{rank}(\mathfrak{g}_{\Gamma})} A_i \wedge \omega_i + \dots \,,
\end{equation}
where we omit the other components such as the resulting 7D 3-form field. We will use a quantization condition for $C_3$, such that $\oint d C_3 \in \mathbb{Z}$ is an integer, at least in the absence of a shifted quantization condition \cite{Witten:1996md}. The topological term in M-theory given by\footnote{Throughout this paper, we use conventions where path-integral integrand is $\exp(iS)$.}
\begin{equation}\label{eq:11DCSterm}
  S^{\mathrm{top.}}_{11\text{D}} = \frac{2\pi}{6}\int C_3\wedge G_4 \wedge G_4 \,,
\end{equation}
produces a 7D coupling, on the Coulomb branch of the gauge theory, given by
\begin{equation}
S_{\text{7D}}^{\text{top.}} \supset \frac{2 \pi}{2} C_{i j} \int C_3 \wedge F_i \wedge F_j \,,
\label{eq:Cartaninstanton}
\end{equation}
where we used the duality of $\omega_i$ to $E_i$ with intersection matrix given by \eqref{eq:Cartanintersection}. In the singular limit, i.e., in the limit of vanishing volumes for the curves $E_i$ one obtains more massless states originating from wrapped M2 branes on $E_i$. These provide the W-bosons necessary for the non-Abelian gauge enhancement to gauge algebra $\mathfrak{g}_{\Gamma}$ on $\mathbb{C}^2 / \Gamma$. In this limit the term \eqref{eq:Cartaninstanton} enhances to
\begin{equation}
S_{\text{7D}}^{\text{top.}} \supset 2 \pi \int C_3 \wedge \frac{1}{4} \text{Tr}_{G_{\Gamma}} F^2 \,,
\end{equation}
where the trace $\text{Tr}_{G_{\gamma}} = \tfrac{1}{h^{\vee}_{G}} \text{Tr}_{\text{adjoint}}$ is normalized in such a way that a unit $G_{\Gamma}$ instanton on $\mathbb{R}^4$ satisfies
\begin{equation}
\frac{1}{4} \int_{\mathbb{R}^4} \text{Tr}_{G_{\Gamma}} F^2 = 1 \,.
\end{equation}
Here, $G_{\Gamma}$ denotes the simply-connected Lie group associated to the Lie algebra $\mathfrak{g}_{\Gamma}$. Recall that a unit instanton configuration in $\mathbb{R}^4$ implies that the gauge field along the asymptotic boundary $\partial \mathbb{R}^4=S^3$ is $A=g^{-1}dg$ where $g:S^3\rightarrow G_{\Gamma}$ has a homotopy class $1\in \pi_3(G_{\Gamma})$. For more details on deriving topological terms on $ADE$ singularities see Appendix B of \cite{Ohmori:2014kda}.

\subsection{Frozen Singularities}

The frozen singularities have the identical spacetime geometry, given by $\mathbb{C}^2/\Gamma$, with the only difference that one includes a non-trivial holonomy of $C_3$ on the asymptotic boundary $S^3 / \Gamma$
\begin{equation}\label{eq:frozenbkrd}
 r \equiv \int_{S^3/\Gamma} C_3= \frac{n}{d} \enspace \mathrm{mod} \enspace 1 \,, \quad \mathrm{gcd}(n, d) = 1 \,.
\end{equation}
Extending this to the interior, we can equivalently write $r= \int_{\mathbb{C}^2/\Gamma} G_4$ and will often refer to the value of $r$ as the frozen flux of the singularity. More generally, one can consider a non-compact K3 manifold $X$, such that $\int_{\partial X}C_3=n/d$. For example, $X$ can be a type $I_0^*$ Kodaira surface with $\int_{\partial X}C_3=1/2$.

The inclusion of this flux has drastic consequences. In particular it produces a potential energy which obstructs the full resolution of the singularity to $\widetilde{X}_{\Gamma}$, i.e., some of the moduli fields are frozen at their singular value. Considering the identification of the moduli space with the Coulomb branch, this indicates a modification of the gauge theory sector localized on the frozen singularities. This is indeed the case as argued for in \cite{deBoer:2001wca, Atiyah:2001qf}, and we provide the frozen gauge algebras in Table \ref{tab:tachikawa}.
\begin{table}[H]
\begin{equation*}
\begin{array}{c@{\,}l||c|cc|ccc|ccccc}
&\fg_\Gamma &   \fso(2k+8)  &   \fe_6 & \fe_6 & \fe_7 &\fe_7 &\fe_7&  \fe_8& \fe_8& \fe_8& \fe_8& \fe_8 \\[1pt]
\hline
&\vphantom{\Big|} r = \frac{n}{d}&  \frac{1}{2} &    \frac{1}{2} & \frac{1}{3}, \frac{2}{3} & \frac{1}{2} & \frac{1}{3},\frac{2}{3} & \frac{1}{4},\frac{3}{4} &   \frac{1}{2} & \frac{1}{3},\frac{2}{3} & \frac{1}{4},\frac{3}{4} & \frac{1}{5},\frac{2}{5},\frac{3}{5},\frac{4}{5} & \frac{1}{6},\frac{5}{6}\\
&\fh_{\Gamma, d} &  \fsp(k)  & \fsu(3) & \varnothing &  \fso(7) & \fsu(2) & \varnothing &  \ff_4 & \fg_2 & \fsu(2) & \varnothing & \varnothing\end{array}
\end{equation*}
\caption{Frozen half-BPS M-theory singularities of the form $\mathbb{C}^2/\Gamma$ as they appeared in \cite{Tachikawa:2015wka} (note, however, that in our conventions $\mathfrak{sp}(k)$ denotes the rank $k$ $C$-series).}
\label{tab:tachikawa}
\end{table}
We denote the frozen gauge algebra, which can be empty, by $\mathfrak{h}_{\Gamma,d}$. The associated 7D $\mathcal{N} = 1$ SYM sectors localized on the frozen singularity are denoted by $\mathcal{T}^{\text{(M)}} [D_k^{(1/2)}]$ for the $D$-type cases and $\mathcal{T}^{\text{(M)}} [\mathfrak{e}_k^{(n/d)}]$ for the $E$-type cases.\footnote{We are using different fonts for aesthetic reasons only.}

We see from Table \ref{tab:tachikawa} that the allowed values of $r$ depend on the type of singularity determined by $\Gamma$, \cite{deBoer:2001wca, Atiyah:2001qf}. A necessary and sufficient condition for a fraction $r$ to be a valid monodromy for a given $\mathfrak{g}_{\Gamma}$ according to \cite{deBoer:2001wca} is for $d$ to be a co-mark (also known as a dual Coxeter label) of the Lie algebra $\mathfrak{g}_{\Gamma}$. Since $r$ is defined modulo integers, $d$ must be bigger than one, which means that $A$-type Lie algebras cannot appear as frozen backgrounds as $A_n$ has co-marks given by $1$ for all of its nodes. The possible frozen singularity backgrounds come from enumerating all possible $n$ and $d$ subject to this condition (see Figure \ref{fig:affine_dynkin_DE} for the co-marks of nodes for $D$- and $E$-type Lie algebras).

To illustrate this rank reduction let us recall the example of the frozen $D$-type singularity as discussed in \cite{Atiyah:2001qf}. There it was noted that if one embeds the M-theory $D_{4+k}$ frozen singularity into an Atiyah--Hitchin manifold, one can reduce on a transverse circle to go to a type IIA description. In particular, one finds a type IIA configuration with $k$ D6 branes and an O$6^+$ plane, as opposed to an O$6^-$ plane, due to the $B_2$ holonomy induced by the $C_3$ field in M-theory. Indeed, such a system has a perturbative gauge algebra given by $\mathfrak{sp}(k)$. The same argument, however, does not apply for the $E$-type singularities for which one needs to go to dual descriptions.

One way of doing so, described in \cite{deBoer:2001wca}, is to use duality between the heterotic string on $T^3$ and M-theory on a compact K3 manifold. In case the K3 manifold contains frozen $ADE$ singularities the dual heterotic theory contains non-trivial gauge configurations of the heterotic gauge fields on $T^3$, so-called triples, which induce the required rank reduction. However, it is not clear how to extend this to non-compact backgrounds. Perhaps one hint is that $d$ always divides the order of $|\Gamma|$ but this does not account for the intricacies of the Lie algebras in Table \ref{tab:tachikawa}. Another alternative to extract the frozen gauge algebras is discussed in \cite{Tachikawa:2015wka}, see also \cite{Tachikawa:Slides}. For that one needs to reduce the 7D gauge theory on $T^3$ to four dimensions and T-dualize along all three circle directions. The frozen flux is then captured by a triple for $\mathfrak{g}_{\Gamma}$ gauge fields on $T^3$, whose Chern--Simons invariant precisely reproduces the allowed values of $r$ in Table \ref{tab:tachikawa}. The vacuum expectation values for the gauge fields on $T^3$ break the gauge algebra to $\mathfrak{h}_{\Gamma,d}^{\vee}$, the commutant of the triple, which translates to its Langlands dual $\mathfrak{h}_{\Gamma,d}$ after the three T-dualities, producing the entries in Table \ref{tab:tachikawa}.

We see that the extraction of the frozen gauge dynamics is rather indirect and makes heavy use of several dualities. In the next section we provide a more direct path to obtaining these using a local freezing map and motivate it from a top-down perspective.

\section{M-theory on Frozen ADE Singularities}\label{sec:freezemap}

In this section we present our so-called freezing map which takes as input a Lie algebra $\mathfrak{g}_{\Gamma}$ corresponding to the $ADE$ singularity $\mathbb{C}^2/\Gamma$, an integer $d$ which appears in the denominator of the frozen flux $r\equiv \int_{S^3/\Gamma}C_3=n/d$, and outputs the Lie algebra $\mathfrak{h}_{\Gamma,d}$ of the theory associated to the frozen singularity. While the specific type of frozen theory also depends on the numerator $n$, this does not affect the frozen gauge algebra $\mathfrak{h}_{\Gamma,d}$ as long as $n$ and $d$ are co-prime. Global versions of such freezing maps have appeared in the literature as far back as \cite{deBoer:2001wca,Atiyah:2001qf} by relating M-theory frozen singularities (embedded inside of a compact K3 manifold) to heterotic/CHL strings on $T^3$, possibly without vector structure \cite{Witten:1997bs}\footnote{See also \cite{Fraiman:2021soq,Fraiman:2021hma, ParraDeFreitas:2022wnz, Fraiman:2022aik} for more recent analyses of freezing maps for heterotic strings on tori.}. Additionally, a local picture of the freezing $\mathfrak{g}_{\Gamma}\rightarrow \mathfrak{h}_{\Gamma,d}$ can be argued by relating the frozen backgrounds compactified on $T^3$ to a Higgsing of $\mathfrak{g}$ after a T-duality transformation on $T^3$ (see \cite{Tachikawa:Slides} for more details), as mentioned above. Each of these arguments are, in some way, indirect for the purpose of understanding why the $C_3$-monodromy modifies the gauge algebra. In particular, given a background $C_3$ monodromy why should M2 brane states wrapped on exceptional cycles be restricted?

Our freezing map, while following from a fairly simple ansatz, recovers each of the frozen algebras in Table \ref{tab:tachikawa} and gives an explicit embedding of the root/weight lattices of $\mathfrak{h}_{\Gamma,d}$ into the respective lattices of $\mathfrak{g}_\Gamma$. We then give an argument for how this rule comes about from a top-down perspective in Section \ref{ssec:topdown}.

\subsection{A Local Freezing Map}\label{ssec:locfreezing}

To describe this freezing rule, first recall that a resolved $ADE$ singularity $\widetilde{X}_{\Gamma}$ contains a collection of curves $E_i$ intersecting according to the negative Cartan matrix, see \eqref{eq:Cartanintersection} above. Denoting the root lattice of $\mathfrak{g}_{\Gamma}$ by $\Lambda^{\mathfrak{g}}_{\text{root}}$ we can choose a basis of roots, $\{ \alpha_i \}$, such that
\begin{equation}\label{eq:galphapairing}
2 \frac{(\alpha_i,\alpha_j)}{(\alpha_i, \alpha_i)} = (\alpha_i,\alpha_j)=+C_{ij} \,.
\end{equation}
Here, $(-,-)$ is a norm on $\Lambda^{\mathfrak{g}}_{\text{root}}$, and we have used the convention that the norm-squared of all of the simple roots of $\mathfrak{g}_\Gamma$ are $2$. That they can be given the same norm is possible because $\mathfrak{g}_\Gamma$ is simply-laced. This means that we can identify the homology lattice $H_2(\widetilde{X}_{\Gamma},\mathbb{Z})$ with $\Lambda^{\mathfrak{g}}_{\text{root}}$, the lattice pairing with the intersection product, $(v,w) = - v \cdot w$, and we will often write $E_i$ and $\alpha_i$ interchangeably.

\begin{figure}
    \centering
    \includegraphics[width=0.8\textwidth]{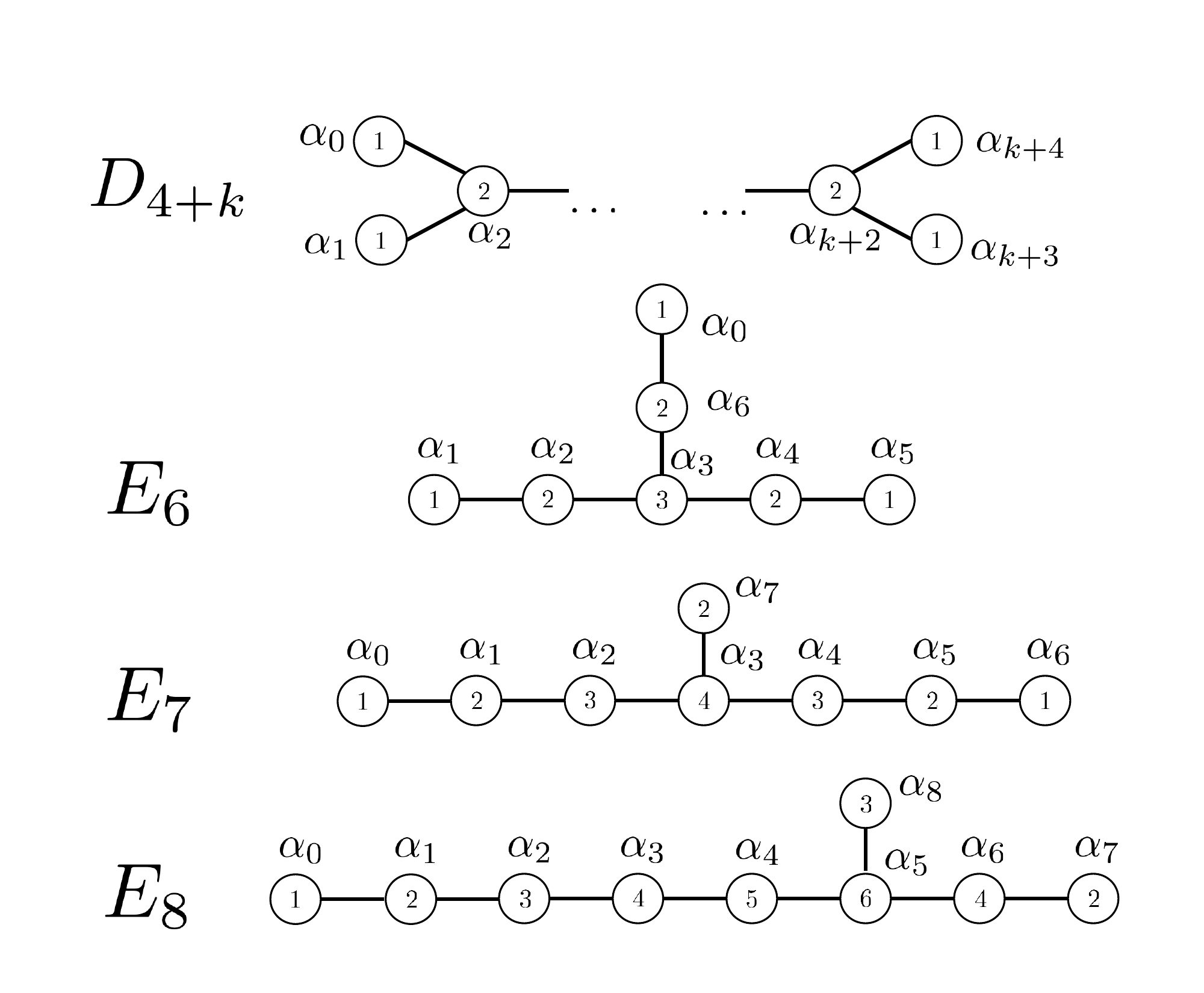}
    \caption{$D$- and $E$-type affine Dynkin diagrams. The dual Coxeter labels, $d_i$, are written in the center of the corresponding node. We also label the simple roots $\alpha_i,\ i = 0, 1, \dots, \mathrm{rank}(\mathfrak{g})$. We note for completeness that $A$-type affine Dynkin diagram nodes all have a dual Coxeter label $d_i = 1$.}
    \label{fig:affine_dynkin_DE}
\end{figure}

In Figure \ref{fig:affine_dynkin_DE}, we denote the affine Dynkin diagrams for the $D$- and $E$-type simple Lie algebras. The nodes are labeled by certain integers called \textit{dual Coxeter labels} (also known as co-marks) which we denote by $d_i$, $i=0,...,\mathrm{rank}(\mathfrak{g}_\Gamma)$. These integers appear in several algebraic identities (for a helpful review see \cite{Fuchs:1997jv}) and play a central role in our freezing map. Recall first that for a non-affine Lie algebra $\mathfrak{g}_\Gamma$, we can identify the affine node of the affine Dynkin diagram with the highest root vector $\theta(\mathfrak{g}_{\Gamma})\in \Lambda^{\mathfrak{g}}_{\text{root}}$. Such a vector is the highest weight vector of the adjoint representation of $\mathfrak{g}_{\Gamma}$ and it has an expansion in the $\alpha$-basis as
\begin{equation}\label{eq:highestrootexp}
\theta(\mathfrak{g}_{\Gamma})=\sum_{i=1}^{\text{rank}(\mathfrak{g}_{\Gamma})} d_i \, \alpha_i \,.
\end{equation}
Strictly speaking, \eqref{eq:highestrootexp} defines the Coxeter labels but these are identical to the dual Coxeter labels because $\mathfrak{g}_\Gamma$ is simply-laced. If we define $\alpha_0\equiv - \theta(\mathfrak{g}_{\Gamma})$ then we have an identity
\begin{equation}\label{eq:ellipticfibhomologyrel}
  \sum_{i=0}^{\text{rank}(\mathfrak{g}_{\Gamma})} d_i \, \alpha_i=0 \,,
\end{equation}
which has a natural geometric interpretation when $\mathbb{C}^2/\Gamma$ is included in an elliptic fibration over $\mathbb{C}$. Namely, \eqref{eq:ellipticfibhomologyrel} relates the homology class of the generic elliptic torus to a linear combination of the exceptional cycles. Additionally, the dual Coxeter labels of $\mathfrak{g}_\Gamma$ are related to its dual Coxeter number, $h^\vee_{\mathfrak{g}_{\Gamma}}$, by
\begin{equation}\label{eq:dualcox}
  h^\vee_{\mathfrak{g}_{\Gamma}}=\sum_{i=0}^{\text{rank}(\mathfrak{g}_{\Gamma})} d_i \,.
\end{equation}
It has been previously noted (see for instance Section 4.6.6 of \cite{deBoer:2001wca} and Section 6.4 of \cite{Atiyah:2001qf}) that when turning on a frozen flux $r=\frac{n}{d} \neq 0$, the algebra $\mathfrak{h}_{\Gamma, d}$ can intuitively be obtained by dividing the dual Coxeter labels of $\mathfrak{g}_{\Gamma}$ by $d$ and throwing away the nodes in the affine Dynkin diagram whose dual Coxeter labels are not divisible by $d$. This intuition was motivated by the fact that M-theory on a compact singular K3 with frozen localized fluxes is dual to heterotic string theory on $T^3$ with a background $\mathfrak{e}_8$, $\mathfrak{e}_7$, $\mathfrak{e}_6$, or $\mathfrak{so}(8+2k)$ connection whose Chern--Simons invariant evaluates to the fraction $r$. The algebras $\mathfrak{h}_{\Gamma,d}$ arise as the commutator subgroups of such connections and the dependence on the dual Coxeter labels follows from the properties of such flat connections on $T^3$ \cite{Borel:1999bx}.

Determining the frozen algebras as a commutator subgroup is a bit roundabout from the M-theory point-of-view however, and is not even technically possible for non-compact frozen singularities since there is no duality to heterotic string theory without a compact embedding. In the M-theory frame, the W-bosons of the $\mathfrak{h}_{\Gamma,d}$ vector multiplet will still arise from M2 branes wrapping some of the exceptional cycles of $\widetilde{X}_{\Gamma}$. This is clear from the fact that on a generic point of the Coulomb branch of the 7D $\mathfrak{h}_{\Gamma,d}$ gauge theory, the resolution $\widetilde{X}_{\Gamma}\rightarrow X$ contains $\mathrm{rank}(\mathfrak{h}_{\Gamma, d})$ independent exceptional cycles (one for each low-energy $U(1)$ gauge factor) and $\widetilde{X}_{\Gamma}$ contains the minimal frozen singularity for a given value of $r$. From Table \ref{tab:tachikawa} the minimal singularity is $D_4$ for $r=1/2$, $E_6$ for $r=\pm 1/3$, and $E_7$ for $r=\pm 1/4$. Values of $r=\pm 1/5$ and $\pm 1/6$ always fix the gauge algebra to be trivial. Since M2 branes wrapping exceptional cycles fill out the root lattices of $\mathfrak{g}$ and $\mathfrak{h}_{\Gamma,d}$ and the unfrozen case includes all possible hyper-K\"ahler resolutions on its Coulomb branch we can conclude that that
\begin{equation}\label{eq:sublattice}
    \Lambda^{\mathfrak{h}}_{\text{root}}\subset \Lambda^{\mathfrak{g}}_{\text{root}} \,.
\end{equation}

We can then formalize the intuition of ``throwing away Dynkin nodes whose dual Coxeter labels are not divisible by $d$" by the following ansatz:
\begin{equation}\label{eq:freezingrule}
    \boxed{\textnormal{Freezing Rule:  }\beta \in \Lambda^{\mathfrak{h}}_{\text{root}}  \quad \Longleftrightarrow \quad \beta \cdot \alpha_i = 0 \; \;  \text{ whenever } d \not |\  d_i \quad (i=0,..., \text{rank}(\mathfrak{g})) .}
\end{equation}
Together with requirement that the norm of long roots $\beta$ is $(\beta, \beta) = 2d$, something which we prove in Section \ref{ssec:topdown}, this ansatz recovers all of the root lattices for the frozen algebras with just the input of $\mathfrak{g}_\Gamma$ and the integer $d$. We list the root systems for all of the non-trivial frozen algebras obtained this way in Table \ref{tab:frozen_algebras_roots}. The trivial frozen algebras lattices are also recovered in the sense that the minimal solution to \eqref{eq:freezingrule} is the sum $\beta_0=\sum_{i} d_i\alpha_i$ which equals $0$ by \eqref{eq:ellipticfibhomologyrel}.

One can also perform a similar freezing on the weight lattice of $\mathfrak{g}_{\Gamma}$, $\Lambda^{\mathfrak{g}}_{\mathrm{wt}}\supset \Lambda^{\mathfrak{g}}_{\text{root}}$, to obtain the weight lattice of $\mathfrak{h}_{\Gamma,d}$. Physically, $\Lambda^{\mathfrak{g}}_{\text{wt}}$ corresponds to M2 branes wrapping relative homology 2-cycles $H_2(\widetilde{X}_\Gamma, \partial \widetilde{X}_\Gamma)$ which include non-compact 2-cycles with non-trivial image in $H_1(\partial \widetilde{X}_\Gamma)$ in addition to the compact exceptional 2-cycles. We will have more to say about such non-compact cycles in Section \ref{sec:defectgrps} where they will play a central role in calculating the defect groups/higher-form symmetries of the frozen theories $\mathcal{T}^{(\mathrm{M})}[\mathfrak{g}^{(n/d)}]$. One can also generalize the freezing rule \eqref{eq:freezingrule} to solve for the coroot and coweight lattices of $\mathfrak{h}_{\Gamma,d}$ for which we will have more to say in Section \ref{ssec:topdown}. Physically, such lattices capture the charges of M5 branes wrapped on exceptional 2-cycles and relative 2-cycles, respectively, and are also key in calculating the 7D defect groups.

\begin{figure}
    \centering
    \includegraphics[width=0.8\textwidth]{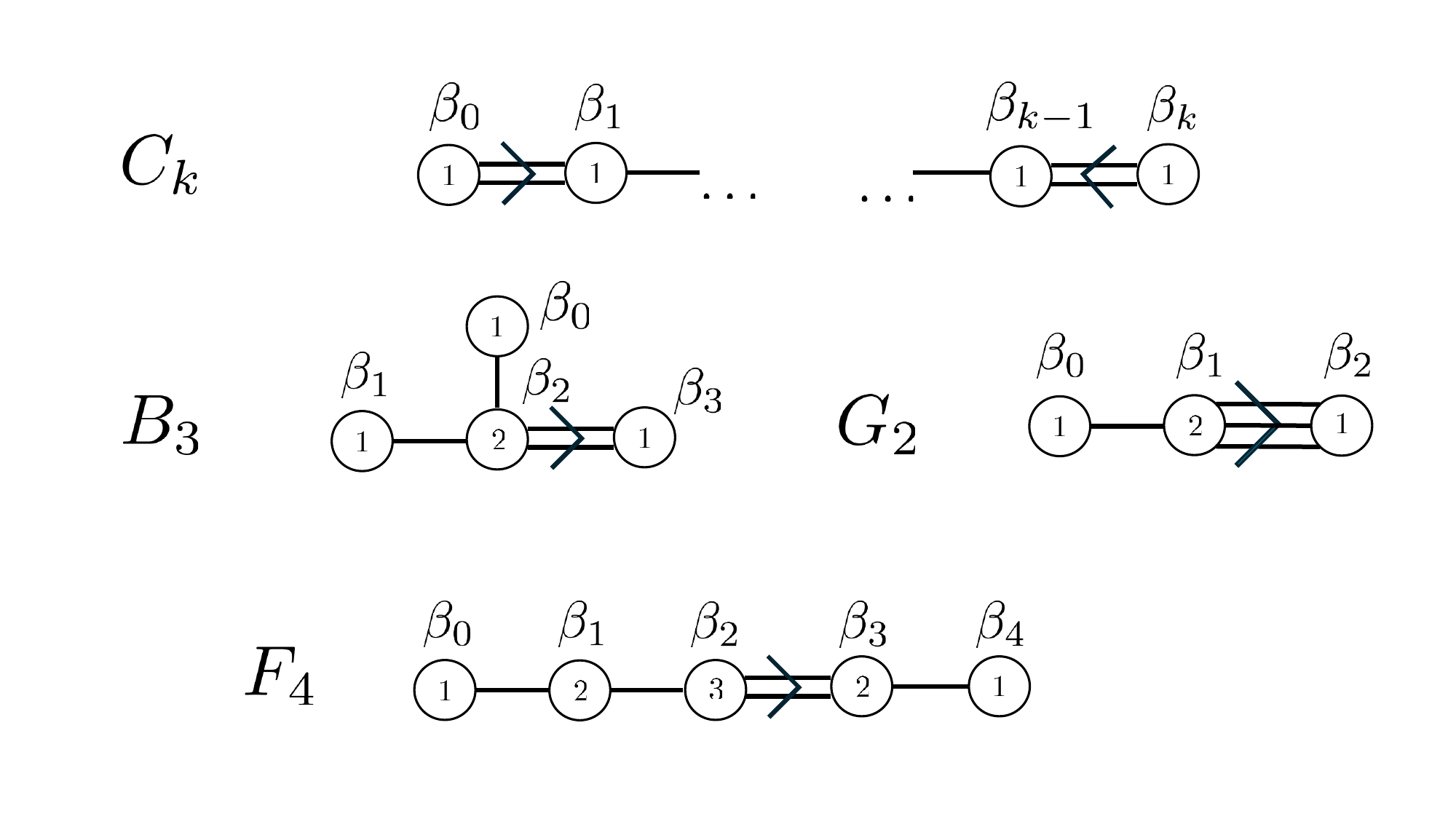}
    \caption{Non-Simply laced affine Dynkin diagrams which arise from M-theory frozen singularities. The dual Coxeter labels, $d_i$, are written in the center of its corresponding node. We also label the simple roots $\beta_i,\ i = 0, 1, \dots, \mathrm{rank}(\mathfrak{h})$.}
    \label{fig:affine_dynkin_NSL}
\end{figure}

\begin{table}[t!]
\centering
\begin{tabular}{|c|c|c|}
\hline
 Frozen Singularity & Frozen Algebra & Frozen Root System \\
 \hline\hline
$D^{(1/2)}_{5}$ & $\fsp(1)$ & $\beta_0 = \alpha_0 + \alpha_1 + 2\alpha_2$ \\ & &
$\beta_{1} = 2\alpha_{3} + \alpha_{4} + \alpha_5$ \\
\hline
$D^{(1/2)}_{4+k}$\; ($k>1$) & $\fsp(k)$ & $\beta_0 = \alpha_0 + \alpha_1 + 2\alpha_2$ \\ & &  $\beta_{i} = \alpha_{i+2}\ ( 1\leq  i \leq k-1)$ \\ & &  $\beta_{k} = 2\alpha_{k+2} + \alpha_{k+3} + \alpha_{k+4}$ \\
\hline
$\mathfrak{e}^{(1/2)}_6$ & $\mathfrak{su}(3)$ & $\beta_0=\alpha_0+\alpha_3+2\alpha_6$ \\ & &
$\beta_1=\alpha_1+2\alpha_2+\alpha_3$ \\ & & $\beta_2=\alpha_3+2\alpha_4+\alpha_5$ \\
\hline
$\mathfrak{e}^{(1/2)}_7$ & $\mathfrak{so}(7)$ & $\beta_0=\alpha_0+2\alpha_1+\alpha_2$\\ & &
$\beta_1=\alpha_4+2\alpha_5+\alpha_6$ \\ & & $\beta_2=\alpha_2+2\alpha_3+\alpha_4$ \\ & &
$\beta_3=\alpha_7$ \\
\hline
$\mathfrak{e}^{(1/3)}_7$ & $\mathfrak{su}(2)$ & $\beta_0=\alpha_0 + 2\alpha_1 + 3\alpha_2 + 2\alpha_3 + \alpha_7$ \\ & &
$\beta_1=\alpha_7 + 2\alpha_3 + 3\alpha_4 + 2\alpha_5 + \alpha_6$ \\
\hline
$\mathfrak{e}^{(1/2)}_8$ & $\mathfrak{f}_4$ & $\beta_0=\alpha_0+2\alpha_1+\alpha_2$ \\ & &
$\beta_1=\alpha_2+2\alpha_3+\alpha_4$ \\ & &
$\beta_2=\alpha_4+2\alpha_5+\alpha_8$  \\ & &
$\beta_3=\alpha_6$ \\ & &
$\beta_4=\alpha_7$ \\
\hline
$\mathfrak{e}^{(1/3)}_8$ & $\mathfrak{g}_2$ & $\beta_0=\alpha_0+2\alpha_1 + 3\alpha_2 + 2\alpha_3 + \alpha_4$ \\ & &
$\beta_1=\alpha_3 + 2 \alpha_4 + 3 \alpha_5 + 2\alpha_6 + \alpha_7$ \\ & &
$\beta_2=\alpha_8$ \\
\hline
$\mathfrak{e}^{(1/4)}_8$ & $\mathfrak{su}(2)$ & $\beta_0=\alpha_0+2\alpha_1+3\alpha_2+4\alpha_3+2\alpha_4$ \\ & &
$\beta_1=2\times \left( \alpha_4+2\alpha_5+2\alpha_6+\alpha_7+\alpha_8\right)$ \\
\hline
\end{tabular}
\caption{Here we list the vectors in the affine root system for the non-trivial frozen algebras which we label by $\beta_i$. For labeling conventions, see Figure \ref{fig:affine_dynkin_DE} and \ref{fig:affine_dynkin_NSL}. In each case, the non-affine vectors ($i\neq 0$) span the frozen root lattice $\Lambda^{\mathfrak{h}}_{\text{root}}$ which is a solution to the freezing rule \eqref{eq:freezingrule}. These are the maximal lattices which satisfy such a constraint with the exception of the $\mathfrak{e}^{(1/4)}_8$ frozen theory. The overall factor of $2$ in $\beta_1$ for this case is required for later considerations of instanton fractionalization in Section \ref{ssec:topdown}.}
\label{tab:frozen_algebras_roots}
\end{table}

\subsection{Top-Down Derivation}\label{ssec:topdown}

We now give a top-down argument for how to obtain the frozen algebras which relies directly on the physical effects resulting from frozen flux $r=\int_{\mathbb{C}^2/\Gamma}G_4$ in the M-theory or type IIA frame after compactifying on a circle. Outlining the discussion, we will first argue that the Coulomb branch of an M2 brane probing a frozen singularity is a moduli space of $d$ instantons in $\mathbb{R}^4$ with gauge algebra $\mathfrak{h}_{\Gamma,d}$, whatever $\mathfrak{h}_{\Gamma,d}$ may be. This statement gives a constraint on the index of the sublattice $\Lambda^{\mathfrak{h}}_{\text{root}}\subset \Lambda^{\mathfrak{g}}_{\text{root}}$ which fixes $\mathfrak{h}_{\Gamma,d}$ in certain cases. After reducing on a circle to IIA, we find that the frozen flux causes some of the $U(1)$ gauge factors in the quiver gauge theories of D2 and D0 probes to be anomalous. Such anomalous $U(1)$ factors do not survive at low-energies which allows us to write down a general formula for $\mathrm{rank}(\mathfrak{g}_\Gamma)-\mathrm{rank}(\mathfrak{h}_{\Gamma, d})$ and to solve to for $\mathfrak{h}_{\Gamma, d}$ in general.

\paragraph{Instanton Fractionalization}

Remember the reduction of M-theory on $ADE$ singularities discussed in Section \ref{subsec:unfrozen}, which produced the term
\begin{equation}
S^{\text{top.}}_{\text{7D}} \supset 2 \pi \int C_3 \wedge \frac{1}{4} \text{Tr}_{G_{\Gamma}} F^2 \,,
\label{eq:7Dtopaction}
\end{equation}
in the singular limit. It tells us that integer instantons in the gauge theory sector have an integer M2 brane charge, since they couple to the 3-form field $C_3$.

Turning now to frozen singularities, we naively cannot use the same argument to derive \eqref{eq:7Dtopaction} because we suspect that the rules for blowing up the $ADE$ singularity are restricted by a potential energy in the presence of the $G_4$ flux. What we can do, however, is to understand what kind of object a single M2 brane probe is from the point-of-view of the frozen gauge theory. When $\mathfrak{h}_{\Gamma, d}\neq \varnothing$, we can perform a partial blowup of $\mathbb{C}^2/\Gamma$ and obtain a topological term proportional to \eqref{eq:Cartaninstanton}. Blowing down the singularity and using $\mathfrak{h}_{\Gamma, d}$ gauge invariance, we obtain
\begin{align}\label{eq:7Dtopactionfrozen}
  S^{\mathrm{top.}}_{\text{7D}}\supset 2\pi K_{\Gamma, d}\int_{M_7} C_3 \wedge \frac{1}{4}\mathrm{Tr}_{H_\Gamma} F^2.
\end{align}
where the constant $K_{\Gamma, d}$ is the M2 brane charge of a unit charge instanton of the $\mathfrak{h}_{\Gamma, d}$ gauge theory\footnote{We know that $K_{\Gamma, d}\neq 0$ because if $\mathfrak{h}_{\Gamma, d}\neq \varnothing$, one can partially resolve the singularity and obtain a $C_3\wedge F\wedge F$ term again from reducing the $C_3\wedge G_4\wedge G_4$ on the resolved 2-cycles.}. As mentioned previously, we know that the root lattice of a frozen gauge theory is a sublattice of the corresponding unfrozen theory \eqref{eq:sublattice}. We can express $K_{\Gamma, d}$ in terms of this data as
\begin{equation}\label{eq:Knormratio}
    K_{\Gamma, d}=\frac{\left(\theta(\mathfrak{g}_{\Gamma}), \theta(\mathfrak{g}_{\Gamma}) \right)}{\left(\theta(\mathfrak{h}_{\Gamma,d }), \theta(\mathfrak{h}_{\Gamma, d}) \right)} \,,
\end{equation}
where $\theta(\mathfrak{g}_{\Gamma}),\theta(\mathfrak{h}_{\Gamma,d }) \in \Lambda^{\mathfrak{g}}_{\text{root}}$ are respectively the highest root vectors of $\mathfrak{g}_{\Gamma}$, and $\mathfrak{h}_{\Gamma, d}$ and $(-,-)$ is the norm on $\Lambda^{\mathfrak{g}}_{\text{root}}$ mentioned in \eqref{eq:galphapairing}. To understand how equation \eqref{eq:Knormratio} follows from \eqref{eq:7Dtopaction} and \eqref{eq:7Dtopactionfrozen}, we see that roughly $K_{\Gamma, d}\mathrm{Tr}_{H_\Gamma} F^2=\mathrm{Tr}_{G_{\Gamma}}F^2$. Such a relation is sensible if one restricts to gauge configurations in the maximal torus of the frozen gauge algebra. Equation \eqref{eq:Knormratio} then follows from the fact that the Killing form of a Lie algebra is fixed by the normalization of $(\; ,\; )$ (see for instance Chapter 18 of \cite{Fuchs:1997jv}). The frozen roots $\beta_i$ obtained from the freezing rule in Table \ref{tab:frozen_algebras_roots} suggests that
\begin{equation}\label{eq:Kequalsoneoverd}
    K_{\Gamma, d}=\frac{1}{d} \,,
\end{equation}
as one can check using the fact that $\theta(\mathfrak{h}_{\Gamma,d })$ is a long root and computing the norm squared of any one of the long frozen roots.

Our goal now is to show why $K_{\Gamma, d}=1/d$ from a stringy perspective. Physically, this means that M2 branes can fractionate into $d$ fractional M2 branes on frozen singularities as one can always consider a unit instanton configuration of the frozen gauge theory. In other words, the 3D $\mathcal{N}=4$ Coulomb branch of the probe M2 brane is the moduli space of $d$ $H_{\Gamma, d}$-instantons in $\mathbb{R}^4$. Such a fractionalization is apparent in various dual frames. As noted in \cite{Atiyah:2001qf}, embedding a $D^{(1/2)}_{4+k}$ singularity into an Atiyah--Hitchin manifold produces $k$ D6 branes probing an O$6^+$ plane and the fractionalization follows from the fact that the enhancement due to the orientifold plane, $\mathfrak{u}(k)\hookrightarrow \mathfrak{sp}(k)$, specifies a group embedding of Dynkin index two. By definition \cite{Meyers:1979pd}, we have a relation of the traces $2\mathrm{Tr}_{U(k)}=\mathrm{Tr}_{Sp(k)}$ which implies that a D2 can fractionate in two on the O$6^+$ plane. Our goal here is to not appeal to dualities.

\begin{figure}
    \centering
    \includegraphics[width=0.8\textwidth]{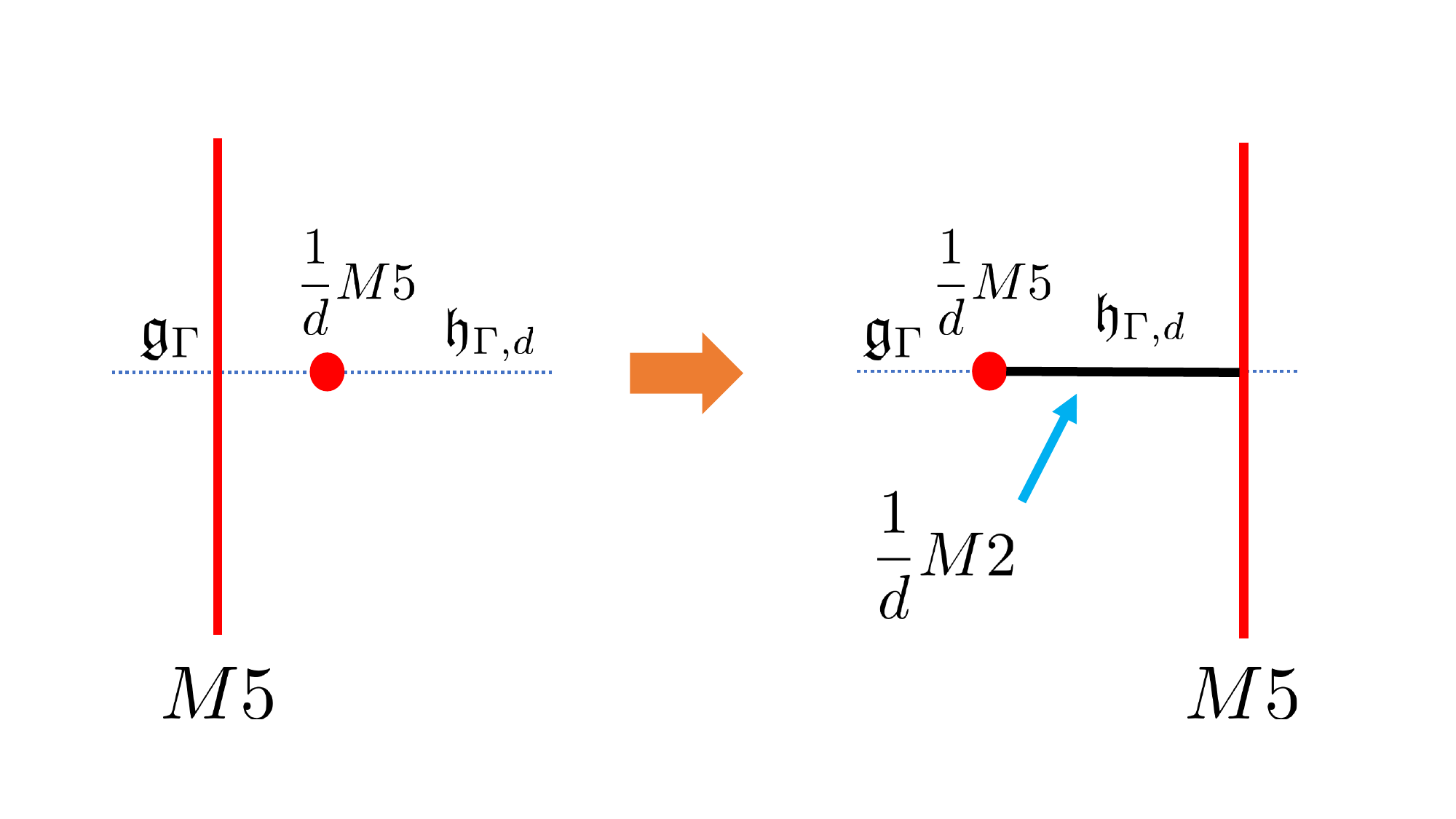}
    \caption{A Hanany--Witten transition between an M5 brane filling the space $\mathbb{C}^2/\Gamma$ and a $\frac{1}{d}$M5 filling a codimension-one subspace of the $ADE$ singularity which creates a $\frac{1}{d}$M2 brane (solid black line). The $\mathbb{C}^2/\Gamma$ singularity is illustrated by the blue dotted line and to the left of the $\frac{1}{d}$M5 the gauge algebra is $\mathfrak{g}_\Gamma$ and is $\mathfrak{h}_{\Gamma, d}$ to the right of it.}
    \label{fig:HWfrozen4}
\end{figure}

One can first consider the scenario depicted in Figure \ref{fig:HWfrozen4}. On both sides of the Figure we have a $\frac{1}{d}$M5 brane acting as a interface separating a non-frozen $D$- or $E$-type singularity with gauge algebra $\mathfrak{g}_\Gamma$ from a frozen singularity with gauge algebra $\mathfrak{h}_{\Gamma, d}$; for details on the worldvolume theories of such fractional M5 branes see \cite{DelZotto:2014hpa}. If we place an M5 brane with worldvolume $\Sigma_2\times \mathbb{C}^2/\Gamma$ on the unfrozen singularity and drag it across the $\frac{1}{d}$M5 brane domain wall to the frozen singularity, then a $\frac{1}{d}$M2 brane is created due to the Hanany--Witten (HW) effect\footnote{Recall that in M-theory, the HW effect is just a consequence of the 11D bulk equation of motion $dG_7=G_4\wedge G_4$.} \cite{Hanany:1996ie}. The table below indicates where the branes in this setup are located (the 7D theory spacetime directions are $0,1,...,6$):
\begin{equation}%
\begin{array}
[c]{cccccccccccc}
  & 0 & 1 & 2 & 3 & 4 & 5 & 6 & 7 & 8 & 9 & 10\\
\frac{1}{d}\text{M}5  & \times & \times & \times & \times & \times & \times &  &  &  & \\
\text{M}5  & \times & \times &  &  &  &  &  & \times & \times & \times & \times \\
\frac{1}{d}\text{M}2  & \times & \times &  &  &  &  & \times &  &  &
\end{array}
\end{equation}

Importantly, notice that it is not possible to place an M5 brane filling $\mathbb{C}^2/\Gamma$ to the right of the $\frac{1}{2}$M5 domain wall and create a $\frac{1}{d}$M2 to the left of the domain wall because of the M5 worldvolume coupling $\int_{\text{M}5}C_3\wedge db_2$ where $b_2$ couples to an M2 brane ending on the M5. From
\begin{equation}\label{eq:M5coupling}
  \int_{\text{M}5}C_3\wedge db_2=\int_{\text{M}5} G_4\wedge b_2=\frac{1}{d}\int_{\Sigma_2} b_2 \,,
\end{equation}
we see that the M5 brane with worldvolume $\Sigma_2\times \mathbb{C}^2/\Gamma$ is forced to have a $\frac{1}{d}$M2-brane ending on it. This addresses a conjecture of \cite{deBoer:2001wca} (see Section 4.6.6) that only a multiple of $d$ M5 branes can fill all four real directions of a frozen $ADE$ singularity with flux $r=n/d$. We see that this condition can be relaxed provided we include the fractional $r\times \text{M}2$ ending on the M5.

Now consider a frozen singularity by itself without any domain wall to an unfrozen singularity. If we have an M5 and an anti-M5 brane on the frozen singularity both with worldvolumes $\Sigma_2\times \mathbb{C}^2/\Gamma$, albeit separated along a transverse direction, one then has a $\frac{1}{d}$M2 brane stretched between them. We can then send these M5 endpoints to infinity to see that we can simply have a probe $\frac{1}{d}$M2 brane localized on the frozen singularity. Putting $d$ of these fractional branes together just produces an M2 brane and of course reversing the process shows that it can fractionate. This shows that \eqref{eq:Kequalsoneoverd} holds. More generally, for frozen singularities of type $\mathfrak{g}^{(n/d)}$, for $n \neq 1$, the previous argument shows that an M5 with worldvolume $\Sigma_2\times \mathbb{C}^2/\Gamma$ has a $\frac{n}{d}$M2 brane attached to it. Since $n$ and $d$ are coprime, one can bring a sufficient number of these fractional branes together to create a $\frac{1}{d}$M2 plus some integer amount of M2 branes which we can separate away.

\paragraph{Consequences of Instanton Fractionalization}
Given the instanton fractionalization statement, we are left with the following mathematical question: given a simply-laced Lie algebra $\mathfrak{g}_{\Gamma}$, what are the sublattices of $\Lambda^{\mathfrak{g}}_{\text{root}}$ which are the root lattice for a Lie algebra $\mathfrak{h}_{\Gamma, d}$ such that the highest weight vector of $\mathfrak{h}_{\Gamma, d}$ satisfies $\left(\theta(\mathfrak{h}_{\Gamma,d}), \theta(\mathfrak{h}_{\Gamma,d}) \right)=d\cdot \left(\theta(\mathfrak{g}_{\Gamma}), \theta(\mathfrak{g}_{\Gamma}) \right)=2d$?

Note that if $d>1$, such sublattices cannot be obtained from taking subalgebras of $\mathfrak{g}_{\Gamma}$. This is because if $\mathfrak{k}\subset \mathfrak{g}_{\Gamma}$, then
\begin{equation}
    N= \frac{\left(\theta(\mathfrak{g}_{\Gamma}), \theta(\mathfrak{g}_{\Gamma}) \right)}{\left(\theta(\mathfrak{k}), \theta(\mathfrak{k}) \right)}
\end{equation}
where $N$ is the Dynkin index of the subalgebra\footnote{One way to define the Dynkin index of a subgroup of a Lie group (which descends to the definition of Dynkin index at the Lie algebra level) is that $H\hookrightarrow G$ induces a map $H^3(BG, \mathbb{Z})\rightarrow H^3(BH, \mathbb{Z})$. Assuming $G$ and $H$ are simple groups, then we have a map $\mathbb{Z}\rightarrow \mathbb{Z}$ and the Dynkin index is the image of $1\in \mathbb{Z}$.} (see \cite{Fuchs:1997jv}). Because $N\in \mathbb{N}_+$, this means that a frozen algebra $\mathfrak{h}_{\Gamma, d}$ cannot be a subalgebra of $\mathfrak{g}_{\Gamma}$ as it would be inconsistent with the instanton fractionalization.

It turns out that we can characterize root/weight sublattices with the correct instanton fractionalization in a simple manner by their coroot/coweight lattice. Recall that a vector $v$ of the root system of any Lie algebra can be transformed into an element $v^\vee$ of the coroot system by
\begin{equation}\label{eq:coroot}
    v^\vee:=\frac{2v}{(v,v)} \, .
\end{equation}
The generators of the coroot lattice $\Lambda_{\text{coroot}}$ are obtained from applying \eqref{eq:coroot} to the simple roots in $\Lambda_{\text{root}}$. We see that the choice of norm \eqref{eq:galphapairing} is convenient in the sense that it specifies a lattice isomorphism $\Lambda^{\mathfrak{g}}_{\text{root}}\simeq \Lambda^{\mathfrak{g}}_{\text{coroot}}$ because $\mathfrak{g}_{\Gamma}$ is simply-laced.
On the other hand, acting by \eqref{eq:coroot} on the simple roots of $\mathfrak{h}$, i.e., the generators of the frozen sublattice $\Lambda^{\mathfrak{h}}_{\text{root}}$ will not specify a lattice isomorphism, simply because the norm-squared of $\theta(\mathfrak{h}_{\Gamma,d})$ is larger than $2$. It is well-known that the lattice $\Lambda^{\mathfrak{h}}_{\text{coroot}}$ can be thought of as a root lattice for an algebra $\mathfrak{h}^\vee$ which is Langlands dual to $\mathfrak{h}$. We emphasize that even if $\mathfrak{h}$ and $\mathfrak{h}^\vee$ are isomorphic to the same simple Lie algebra, their root lattices will be distinct as sublattices of $\Lambda^{\mathfrak{g}}_{\text{wt}}$.

The upshot of studying $\Lambda^{\mathfrak{h}}_{\text{cowt}}=\Lambda^{\mathfrak{h}^\vee}_{\text{wt}}$ is that it specifies a subalgebra $\mathfrak{h}^\vee\subset \mathfrak{g}$. Specifically, from
\begin{align*}
\Lambda^{\mathfrak{h}^\vee}_{\text{wt}}\simeq \mathrm{Hom}(\Lambda^{\mathfrak{h}}_{\text{root}}, \mathbb{Z}) & \\
    v^\vee \; \mapsto \; (-,v^\vee) \quad \quad  \;  &
\end{align*}
we see that the embedding $\Lambda^{\mathfrak{h}}_{\text{root}}\subset \Lambda^{\mathfrak{g}}_{\text{root}}$ can dually be presented as a projection map
\begin{equation}\label{eq:proj}
    \pi: \; \Lambda^{\mathfrak{g}}_{\text{wt}}\twoheadrightarrow \Lambda^{\mathfrak{h}^\vee}_{\text{wt}}
\end{equation}
To see the utility of \eqref{eq:proj}, recall that for a given Lie algebra $\mathfrak{f}$ with $\mathfrak{t}^*_{\mathfrak{f}}:=\mathrm{Span}_{\mathbb{R}}\Lambda^\mathfrak{f}_{\text{wt}}$ as the weight space of $\mathfrak{f}$, the Cartan subalgebra of $\mathfrak{f}$ is the dual of this, $(\mathfrak{t}^*_{\mathfrak{f}})^*=\mathfrak{t}_{\mathfrak{f}}$. By taking the real span and the dual of \eqref{eq:proj}, we have an explicit embedding of the Cartan subalgebra of $\mathfrak{h}^\vee$ into the Cartan subalgebra of $\mathfrak{g}$. This data is enough to specify the full embedding $\mathfrak{h}^\vee\subset \mathfrak{g}$ because \eqref{eq:proj} also specifies how $\mathfrak{g}$-representations decomposes into $\mathfrak{h}^\vee$-representations. That we can characterize the frozen algebra by its Langlands dual is not new as one can explicitly understand how $\mathfrak{h}^\vee$ arises as a Higgsing of $\mathfrak{g}$ after placing the 7D theory on $T^3$ and performing three T-dualities, see \cite{Tachikawa:Slides} for details.

Let us now compute the Dynkin index of the embedding $\mathfrak{h}^\vee\subset \mathfrak{g}$. This can be done straightforwardly by applying \eqref{eq:Knormratio} and \eqref{eq:coroot}:
\begin{equation}
   \frac{\left(\theta(\mathfrak{g}_{\Gamma}), \theta(\mathfrak{g}_{\Gamma}) \right)}{\left(\theta(\mathfrak{h}^\vee_{\Gamma,d }), \theta(\mathfrak{h}^\vee_{\Gamma, d}) \right)}
   =\frac{1}{N_\mathfrak{h}}\cdot \frac{\left(\theta(\mathfrak{g}_{\Gamma}), \theta(\mathfrak{g}_{\Gamma}) \right)}{\left(\theta(\mathfrak{h}_{\Gamma,d })^\vee, \theta(\mathfrak{h}_{\Gamma, d})^\vee \right)}
   =\frac{d^2}{N_\mathfrak{h}}\cdot \frac{\left(\theta(\mathfrak{g}_{\Gamma}), \theta(\mathfrak{g}_{\Gamma}) \right)}{\left(\theta(\mathfrak{h}_{\Gamma,d }), \theta(\mathfrak{h}_{\Gamma, d}) \right)}
   =\frac{d}{N_\mathfrak{h}}.
\end{equation}
Here we have defined the integer $N_{\mathfrak{h}}$ as the ratio between the norm-squared of $\theta(\mathfrak{h}^\vee_{\Gamma,d })$ and that of $\theta(\mathfrak{h}_{\Gamma, d})^\vee$. The highest weight vector is always a long root which means that $\theta(\mathfrak{h}_{\Gamma, d})^\vee$ is a short coroot of $\mathfrak{h}$ or, equivalently a short root of $\mathfrak{h}^\vee$. So $N_{\mathfrak{h}}$ is simply the ratio between the norm-squares of the long and short roots of $\mathfrak{h}^\vee$ (which is the same ratio for $\mathfrak{h}$) so
\begin{equation}
    N_{\mathfrak{h}}=
\begin{cases}
1, \quad \textnormal{$\mathfrak{h}_{\Gamma,d}$ is simply-laced}\\
2,  \quad \textnormal{$\mathfrak{h}_{\Gamma,d}$ is a $B$- or $C$-type, or $\mathfrak{f}_4$ algebra}\\
3, \quad \textnormal{$\mathfrak{h}_{\Gamma,d}$ a $\mathfrak{g}_2$ algebra}
\end{cases}
\end{equation}

Therefore we arrive at the statement:
\begin{equation*}
    \boxed{\textnormal{Given a frozen algebra $\mathfrak{h}_{\Gamma, d}$, there exists a subalgebra $\mathfrak{h}^\vee_{\Gamma, d}\subset \mathfrak{g}_{\Gamma}$ of Dynkin index $d/N_{\mathfrak{h}}$}}
\end{equation*}

For a given frozen flux $r=n/d$, there may a priori be several possible $\mathfrak{h}^\vee_{\Gamma, d}\subset \mathfrak{g}_{\Gamma}$ which satisfy such a statement. We know that $\mathfrak{h}^\vee_{\Gamma, d}\subset \mathfrak{g}_{\Gamma}$ cannot be a regular subalgebra because a regular subalgebra of a simply-laced Lie algebra has Dynkin index 1 and is itself simply-laced. A subalgebra that is not regular is a special subalgebra (also referred to as an S-subalgebra) and in Table \ref{tab:dualcandidates} we list all possible special subalgebras of $\mathfrak{g}_\Gamma$ for the minimal $D$-type and the $E$-type cases. We see from this table that we have some spurious candidates. For instance, we see for the frozen $D_4$ singularity that, up to taking further subgroups, there are three candidate non-trivial (Langlands duals of) frozen algebras even though we suspect from dualities and compute from the freezing rule that $\mathfrak{h}_{\Gamma, 2}=\varnothing$ in this case. For the $\mathfrak{e}^{(1/2)}_{7}$ case, we have two candidate subalgebras from Table \ref{tab:dualcandidates} namely $\mathfrak{sp}(3)$ and $\mathfrak{f}_4$. The former is correct as $(\mathfrak{h}^\vee_{\Gamma, 2})^\vee=(\mathfrak{sp}(3))^\vee=\mathfrak{so}(7)$ while we still cannot eliminate the latter from our instanton fractionalization argument alone. Note that for the $\mathfrak{e}^{(1/4)}_8$ case, one can obtain the correct index-4 subgroup $\mathfrak{su}(2)\subset \mathfrak{e}_8$ because there is a Dynkin index-4 subgroup $\mathfrak{su}(2)\subset \mathfrak{g}_2$ and Dynkin indices are multiplicative under taking sequences of subgroups\footnote{For physicist-friendly resources on special subalgebras, see \cite{Esole:2020tby} and \cite{Yamatsu:2015npn}.}.

\begin{table}
\begin{equation*}
\begin{array}{|c|c|}
\hline
\textnormal{Frozen Singularity} & \textnormal{Relevant Maximal S-Subalgebras} \\
\hline
D^{(1/2)}_4 & \mathfrak{so}(7)^{[1]}, \; \mathfrak{so}(5)^{[1]}, \; \mathfrak{su}(2)^{[2]} \\
\mathfrak{e}^{(n/d)}_6 & \mathfrak{sp}(4)^{[1]}, \; \mathfrak{f}_4^{[1]}, \; \mathfrak{su}(3)^{[2]}\oplus \mathfrak{g}_2^{[1]}  \\
\mathfrak{e}^{(n/d)}_7 &  \mathfrak{su}(2)^{[3]}\oplus \mathfrak{f}_4^{[1]}, \; \mathfrak{sp}(3)^{[1]}\oplus \mathfrak{g}^{[1]}_2, \; \mathfrak{su}(2)^{[7]}\oplus \mathfrak{g}^{[1]}_2 \\
\mathfrak{e}^{(n/d)}_8 & \mathfrak{f}_4^{[1]}\oplus\mathfrak{g}^{[1]}_2 \\
\hline
\end{array}
\end{equation*}
\caption{Here we list the maximal special (S) subalgebras of $\mathfrak{g}_{\Gamma}$ relevant to our discussion, see \cite{Fuchs:1997jv}. The superscript denotes the Dynkin index of the subgroup. Simple subgroups of these with index $d/N_{\mathfrak{h}}$ are candidates for Langlands dual to the frozen algebras.}
\label{tab:dualcandidates}
\end{table}

\paragraph{Aside on D-brane Quivers}

Given the limitations on the constraint of having the correct instanton fractionalization on the frozen singularity, we now seek to derive a formula for the rank and dual Coxeter number of the frozen algebra. Our approach will be to reduce on a transverse circle to the IIA frozen background
\begin{equation}
    \mathrm{IIA}\left( X=\mathbb{C}^2/\Gamma, \; \; \int_{\partial X} C_3=\frac{n}{d}\right) \,,
\end{equation}
and study the quiver gauge theories living on BPS D$p$ probes of this frozen singularity. Recall that the low-energy physics of D$p$ probes of unfrozen $ADE$ singularities are described by $(p+1)$-dimensional gauge theories with 8 supercharges whose gauge/matter content are summarized by a quiver given by the affine $ADE$ Dynkin diagram \cite{Douglas:1996sw,Johnson:1996py}. While we expect in the frozen cases that the Douglas--Moore quivers of such theories will be in the shape of the affine Dynkin diagrams of $\mathfrak{h}_{\Gamma, d}$, such a derivation is outside of the scope of this work and would in principle require one to understand how open strings behave on non-trivial RR backgrounds. Our goal will be more modest. We will show that certain simple gauge theory factors that appear for the D$p$ probe branes for the unfrozen singularity develop a gauge anomaly in the presence of the frozen flux.

For concreteness, let us consider a D0 probing the $\mathfrak{e}^{(1/2)}_6$ IIA frozen singularity. Without the frozen flux, the 1D field theory content consists of the gauge group
\begin{equation}\label{eq:d0gaugegrp}
    G_{\mathrm{quiv.}}=U(1)^3\times U(2)^3\times U(3) \,,
\end{equation}
and $\mathcal{N}=8$ bifundamental hypermultiplets associated with each of the links of the affine $\mathfrak{e}_6$ quiver, see Figure \ref{fig:affine_dynkin_DE}. The Coulomb branch of this theory parameterizes the motion of fractional D0 branes along the $\mathbb{R}^6$ parallel to the singularity of the D0 brane along the $\mathbb{R}^6$ parallel to the singularity. In general the real dimension of this Coulomb branch is $5\times h_{\mathfrak{g}}^\vee$.\footnote{The factor of 5 is from the transverse directions, while the dual Coxeter number is related to the quiver gauge group $\prod^{\mathrm{rank}(\mathfrak{g})}_{i=0}U(n_i)$ as $h^\vee_{\mathfrak{g}}=\sum^{\mathrm{rank}(\mathfrak{g})}_{i=0}n_i$.} For instance, the scalars in the $U(k)$ vector multiplets correspond to positions of $\frac{k}{24}\text{D}0$ branes where the $1/24$ fraction is due to the fact that $|\Gamma_{E_6}|=24$. The $U(k)$ gauge factor in total then describes $k$ coincident $\frac{k}{24}\text{D}0$ branes.

Let us denote the 1-form gauge potentials for each factor of \eqref{eq:d0gaugegrp} by $a^{(i)}$, $b^{(j)}$, and $c$ where $i,j=1,2,3$. In the presence of the frozen flux $\int_{\mathbb{C}^2/\Gamma_{E_6}}G_4=1/2$, we claim that we have the additional topological terms
\begin{equation}\label{eq:1dCSterms}
    \frac{1}{2}\int_{L} \left( \sum_{i}a^{(i)}+2\sum_{i}\mathrm{Tr}(b^{(i)})+3\mathrm{Tr}(c) \right)
\end{equation}
where $L$ is the D0 brane worldvolume. These terms are neither invariant under $U(1)$ nor $U(3)$ large gauge transformations due to their fractional coefficients. To see how such terms can arise, consider the flux background in flat space
\begin{equation}
    \mathrm{IIA}\left( X=\mathbb{C}^2, \; \; \int_{X} G_4=12\right).
\end{equation}
Orbifolding such a background by the binary tetrahedral group $\Gamma_{E_6}$ produces the $\mathfrak{e}^{(1/2)}_{6}$ frozen singularity because of the normalization of the volume forms:
\begin{equation}
\int_{\mathbb{C}^2}=24\int_{\mathbb{C}^2/\Gamma_{E_6}}.
\end{equation}
Also, a single D$0$ brane at the origin of $\mathbb{C}^2$ becomes a $\frac{1}{24}\text{D}0$-brane at the origin of $\mathbb{C}^2/\Gamma_{E_6}$ upon oribifolding. Notice that a D$0$ brane can be formed by taking a D$4$ and $\overline{\text{D}4}$ pair along $\mathbb{C}^2\times L$ and turning on an Abelian instanton background localized at the origin\footnote{Note that one can more rigorously treat a $U(1)$ instanton localized in $\mathbb{R}^4$ by turning on a small B-field background, where the $U(1)$ instanton is smooth in a non-commutative deformation of $\mathbb{R}^4$ \cite{Gross:2000wc}. }. Explicitly, we have a Wess-Zumino coupling
\begin{equation}
    \int_{\mathbb{C}^2\times L} C_1\wedge (f_+-f_-)^2 \,,
\end{equation}
where $f_{\pm}$ are the field strengths for the $U(1)_+\times U(1)_-$ gauge group of the D$4$/$\overline{\text{D}4}$ stack. This coupling implies that a D$0$ brane embedded inside of the D$4$/$\overline{\text{D}4}$ stack can be engineered from a singular Abelian instanton background $(f_+-f_-)^2\sim \delta_{L}$ as this localizes to $\int_{L} C_1$ \cite{Douglas:1995bn}. As standard in realizing branes from higher-dimensional branes, the $U(1)$ gauge group associated with the D0 is a subgroup of $U(1)_+\times U(1)_-$ with potential $a:=(a_+-a_-)$. The key point is that the 4-brane stack also contains the term
\begin{equation}
    \int_{\mathbb{C}^2\times L} C_3\wedge (f_+-f_-)=\int_{\mathbb{C}^2\times L} G_4\wedge (a_+-a_-)=12\int_{L}a
\end{equation}
which localizes to a 1D Chern--Simons term on the D0 with level $12$. After orbifolding, the Wess-Zumino action of this D0 brane is multiplied by $1/24$ so its action is now
\begin{equation}\label{eq:124d0}
    \frac{1}{24}\int_{L}\left(C_1 +12a\right)
\end{equation}
where now the second term is not gauge invariant under the large gauge transformation\footnote{We take the worldvolume fluxes to be normalized as $\int f_{\pm}=2\pi n$.} $a\rightarrow a+\lambda$ where $d\lambda=0$ and $\int_L \lambda =2\pi$. This argument reproduces all of the $a^{(i)}$ terms in \eqref{eq:1dCSterms}. As for the $U(3)$ factor in $G_{\mathrm{quiv.}}$, we can consider going unto its Coulomb branch whereby the $3$ coincident $\frac{1}{8}\text{D}0$ branes are separated from each other at a generic point. Each of these $\frac{1}{8}\text{D}0$ branes is associated with a $U(1)_k\subset U(3)$, $k =1,2,3$. Now the only thing we change is that prior to taking the orbifold we place a charge-3 instanton in the D$4/\overline{\text{D}4}$ stack localized at the origin of $\mathbb{C}^2$ which engineers 3 D$0$ branes there. After orbifolding, the analog of \eqref{eq:124d0} is
\begin{equation}\label{eq:18d0}
    \frac{3}{24}\int_{L}\left(C_1 +12c^{(k)}\right)
\end{equation}
where $c^{(k)}:=a^{(k)}_+-a^{(k)}_-$ which is not gauge invariant due to the fractional level $12/8=3/2$. If we take these $\frac{1}{8}\text{D}0$ branes to be coincident, then we obtain the $3/2\mathrm{Tr}(c)$ term in \eqref{eq:1dCSterms} from the fact that $\mathrm{Tr}(c)=\sum^3_{k=1}c^{(k)}$. We note that for D2 and D4 fractional probes, we would obtain similar conclusions but with 3D and 5D Chern--Simons terms being non-gauge invariant due to a fractional level. One can derive these terms from realizing these branes as instantons inside a D$6/\overline{\text{D}6}$ pair and a $\text{D}8/\overline{\text{D}8}$ pair, respectively.

What can we conclude then from the fact that the $U(1)^3\times U(3)\subset G_{\mathrm{quiv.}}$ subgroup is inconsistent and thus must be projected out of the low-energy degrees of freedom of D$p$ probe of $\mathfrak{e}^{(1/2)}_6$? We know generally that the geometric deformations/resolutions of an $ADE$ singularity correspond to Fayet–Iliopoulos (FI) parameters of the quiver gauge theory living on the D$p$ probe \cite{Douglas:1996sw,Johnson:1996py}. There is a $SU(2)_R$ triplet of FI parameters associated to each factor of $G_{\mathrm{quiv.}}$ which of course correspond to the blow-ups of the of the $ADE$ singularity. Given that the $U(1)^3\times U(3)$ gauge factor cannot be present, we see that number of possible blow-up modes of $\mathfrak{e}^{(1/2)}_6$ is reduced by $4$. In 7D terms, this means that the rank is now $6-4=2$, which singles out the correct frozen gauge algebra $\mathfrak{h}_{E_6,2} = \mathfrak{su}(3)$. It is not difficult to see that if we repeat the above calculations for any frozen singularity we have the general statement
\begin{equation*}
    \boxed{\mathrm{rank}(\mathfrak{g}_{\Gamma})-\mathrm{rank}(\mathfrak{h}_{\Gamma, d})=\textnormal{\# of dual Coxeter labels of $\mathfrak{g}_{\Gamma}$ such that $d \not |\  d_i$}}
\end{equation*}
This statement is implied by the freezing rule of Section \ref{ssec:locfreezing}, and carries the general spirit of the dependence on the dual Coxeter labels which, from the point-of-view of gauge anomalies on the D$p$ brane probes, arises due to the ranks of the gauge factors in the quiver gauge group. These ranks ultimately arise from the dimensions of the irreducible representations of $\Gamma_{E_6}$ and it would be interesting to understand if there is a full derivation of the frozen quiver in terms of this data, similar to how discrete torsion affects quiver representations (see for instance \cite{Aspinwall:2000xs}). In particular, we do not see from this gauge anomaly argument why the non-anomalous $U(2)^3\subset G_{\mathrm{quiv.}}$ reduces to the expected $U(1)^3$ of the frozen quiver. It would also be desirable to have an anomaly argument directly in the M-theory frame which, however, seems to be hard to achieve since we heavily used the string theory specific branes within branes construction associated to gauge fields on D$p$ branes.

Summarizing our results, we have shown directly in M-theory that the frozen flux implies an instanton fractionalization condition which gives a restriction on the possible $\mathfrak{h}_{\Gamma,d}$, which can be compactly phrased in terms of the Langlands dual $\mathfrak{h}^\vee_{\Gamma,d}$, and that we can solve for the rank of $\mathfrak{h}_{\Gamma,d}$ using D$p$ probes after circle compactification. These arguments are ``duality-free" in the sense that the frozen flux $r=\int C_3$ and the orbifold geometry are not changed in any way. What is striking about these two conditions is that they are enough to completely determine $\mathfrak{h}_{\Gamma,d}$ in all cases. Given the rank reduction formula, the only cases that are ambiguous before applying the instanton fractionalization statement are when $\mathrm{rank}(\mathfrak{h}_{\Gamma,d})\geq 2$ and one can check explicitly that the ambiguity goes away. For example, in the case $\mathfrak{e}^{(1/2)}_8$ we know from the D$p$ probe anomaly statement that $\mathrm{rank}(\mathfrak{h}_{E_8, 2})=4$ and we know from the instanton fractionalization that $\mathfrak{h}^\vee_{\Gamma, 2}\subset \mathfrak{g}_{\Gamma}$ is a maximal subalgebra with index $1$ if doubly-laced and $2$ if simply-laced. From Table \ref{tab:dualcandidates} this uniquely fixes $\mathfrak{h}^\vee_{E_8, 2}=\mathfrak{f}_4$ which means $\mathfrak{h}_{E_8, 2}=\mathfrak{f}^\vee_4=\mathfrak{f}_4$. Finally, notice that the while this freezing leaves M2 branes wrapping exceptional 2-cycles out of the spectrum simply because they are in the same vector multiplet as the resolution scalars, we see that there is no such restriction on the lattice of M5 brane states, a behavior we will confirm in the twisted F-theory duality frame in the next section.

\section{Defect Groups of Frozen Singularities}\label{sec:defectgrps}

Having established how the $G_4$ flux reduces the gauge symmetry on $ADE$ singularities, we now turn our attention to defects charged under generalized symmetries in frozen backgrounds.
Specifically, we are interested in the group $\mathbb{D}$ of defects charged under 1- and 4-form symmetries of $\mathcal{T}^{(\mathrm{M})}[\mathfrak{g}^{(n/d)}]$ \cite{DelZotto:2015isa}.
In the non-frozen cases, $\mathcal{T}^{(\mathrm{M})}[\mathfrak{g}]$ has a defect group $\mathbb{D}= Z(G)^{(1)}\oplus Z(G)^{(4)}$ for the Wilson and 't Hooft operators, where $Z(G)$ is the center of the simply-connected Lie group $G$ associated to $\mathfrak{g}$.
Geometrically, if we let $\widetilde{X}_\Gamma\rightarrow \mathbb{C}^2/\Gamma$ denote the fully resolved space, these defects arise from M2 and M5 branes, respectively, wrapping relative 2-cycles in $H_2(\widetilde{X}_\Gamma, \partial \widetilde{X}_\Gamma) \cong \text{Hom} (H_2(\widetilde{X}_\Gamma), \mathbb{Z}) \cong \Lambda_\text{cowt}^\mathfrak{g} \cong \Lambda_\text{wt}^\mathfrak{g}$, and their charge (screened by M2/M5 branes wrapping compact 2-cycles) is precisely given by \cite{Morrison:2020ool, Albertini:2020mdx}
\begin{align}
    Z(G) = \Lambda_\text{wt}^\mathfrak{g} / \Lambda_\text{root}^\mathfrak{g} = \frac{H_2(\widetilde{X}_\Gamma, \partial \widetilde{X}_\Gamma)}{H_2(\widetilde{X}_\Gamma)} \cong H_1(\partial \widetilde{X}_\Gamma) \, .
\end{align}
We have listed the relevant details in Table \ref{table:relative 2-cycles}.

\begin{table}[ht]
\centering
\begingroup
\setlength{\tabcolsep}{10pt} % Default value: 6pt
\renewcommand{\arraystretch}{1.5} % Default value: 1
\begin{tabular}{c|c|c}
 Singularity & $Z(G_{\Gamma})$ & Generator(s) \\
 \hline\hline
 $A_{k-1}$ & $\mathbb{Z}_k$ &  $w_1 \equiv \frac{1}{k}\sum^{k-1}_{j=1} j \, \alpha_j \mod \langle \alpha_i \rangle_\mathbb{Z}$\\
 \hline
 $D_{4+k}$  ($k=4n$) & $\mathbb{Z}_2\times \mathbb{Z}_2$ &  $w_{k+3} \equiv \frac{1}{2} \sum^{k+2}_{j=1} j \, \alpha_{j} + \frac12 \alpha_{k+4} \mod \langle \alpha_i \rangle_\mathbb{Z}$\\
 & &  $w_{k+4} \equiv \frac{1}{2} \sum^{k+2}_{j=1} j\, \alpha_{j} + \frac12 \alpha_{k+3} \mod \langle \alpha_i \rangle_\mathbb{Z}$ \\
 \hline
 $D_{4+k}$  ($k=4n+2$) & $\mathbb{Z}_2\times \mathbb{Z}_2$ &  $w_{k+3} \equiv \frac{1}{2} \sum^{k+2}_{j=1} j \, \alpha_{j} + \frac12 \alpha_{k+3} \mod \langle \alpha_i \rangle_\mathbb{Z}$\\
 & &  $w_{k+4} \equiv \frac{1}{2} \sum^{k+2}_{j=1} j\, \alpha_{j} + \frac12 \alpha_{k+4} \mod \langle \alpha_i \rangle_\mathbb{Z}$ \\
 \hline
  $D_{4+k}$  ($k = 2n+1$) & $\mathbb{Z}_4$ &  $w_{k+3} \equiv \frac{1}{2} \sum^{2+k}_{i=1} j\, \alpha_{j} +\frac{k}{4}\alpha_{k+3}+\frac{k-2}{4}\alpha_{k+4} \mod \langle \alpha_i \rangle_\mathbb{Z}$ \\
 \hline
  $\mathfrak{e}_6$ & $\mathbb{Z}_3$ &  $w_1 \equiv \frac{1}{3}(\alpha_1+2\alpha_2+\alpha_4+2\alpha_5) \mod \langle \alpha_i \rangle_\mathbb{Z}$\\
 \hline
 $\mathfrak{e}_7$ & $\mathbb{Z}_2$ &  $w_1 \equiv \frac{1}{2}(\alpha_4+\alpha_6+\alpha_7) \mod\langle \alpha_i \rangle_\mathbb{Z}$\\
 \hline
  $\mathfrak{e}_8$ & $\varnothing$  &  $w_1 \equiv 0 \mod \langle \alpha_i \rangle_\mathbb{Z}$\\
\end{tabular}
\endgroup
\caption{We list the generators $w_\ell$ for the group $Z(G_\Gamma)$ for each of the $ADE$ singularities/Lie algebras as fractional linear combinations of the simple roots $\alpha_{i}$, which are identified up to integer linear combinations of the $\alpha_i$ (denoted as $\langle \alpha_i \rangle_\mathbb{Z}$). The $w_\ell$ are fundamental weights (satisfying $(\alpha_i, w_\ell) = \delta_{i\ell}$) known to generate the center, so their coefficients are given by rows of the inverse Cartan matrix (modulo 1). Our conventions for the enumeration of $\alpha_i$s for the $D$- and $E$-type cases follow from Figure \ref{fig:affine_dynkin_DE} and we use the obvious ordering for $A$-type.}
\label{table:relative 2-cycles}
\end{table}

For the frozen theories, the relationship between the various group-theoretic and geometric lattices are more delicate.
As we will see in detail, the freezing rule leads to a non-trivial identification of the Wilson and 't Hooft defects of the frozen gauge sector with elements of $H_1(\partial \mathbb{C}^2/\Gamma)$, which in turn will become important for the discussion of the SymTFT in Section~\ref{sec:symtft}.
This procedure leads to two surprising results.
First, we see that, while the line defects, i.e., wrapped M2 branes, generally suffers a reduction compatible with the rank reduction of the gauge algebra, the 4-manifold defects from wrapped M5 branes remain identical to the unfrozen case even though they may not have a clear description as 't Hooft defect of the frozen gauge theory. Second, in the specific case of ${\cal T}^{({\rm M})}[\mathfrak{e}_8^{(1/4)}]$ which has an $\mathfrak{h} = \mathfrak{su}(2)$ gauge algebra, there are \emph{neither} line \emph{nor} 4-manifold defects, as expected from geometry. We list these results in Table \ref{table:frozendessert}.

To corroborate these findings, we perform a geometric analysis of twisted circle compactifications of F-theory on Kodaira singularities that are (essentially) double-T-dual to the frozen M-theory backgrounds \cite{deBoer:2001wca,Tachikawa:2015wka} which verifies the direct application of the freezing map.
We will suggest a physical interpretation of these results in the next section.

\begin{table}[ht]
\centering
 \begin{tabular}{c|c|c|c|c}
  Frozen Singularity & $\mathbb{D}^{(1)}$ & $\mathbb{D}^{(4)}$ & $Z(G_\Gamma)$ & $Z(H_{\Gamma,d})$  \\
  \hline\hline
  $D^{(1/2)}_{4}$ & $0$ & $\mathbb{Z}^2_2$ & $\mathbb{Z}^2_2$ & 0  \\
  \hline
  $D^{(1/2)}_{k\geq 5}$  $(k=2n)$ & $\mathbb{Z}_2$ & $\mathbb{Z}^2_2$ & $\mathbb{Z}^2_2$ & $\mathbb{Z}_2$  \\
  \hline
  $D^{(1/2)}_{k\geq 5}$  $(k=2n+1)$ & $\mathbb{Z}_2$ & $\mathbb{Z}_4$ & $\mathbb{Z}_4$ & $\mathbb{Z}_2$  \\
  \hline
   $\mathfrak{e}^{(1/2)}_{6}$ & $\mathbb{Z}_3$ & $\mathbb{Z}_3$ & $\mathbb{Z}_3$ & $\mathbb{Z}_3$ \\
   \hline
 $\mathfrak{e}^{(1/3)}_{6}$ & $0$ & $\mathbb{Z}_3$ & $\mathbb{Z}_3$ & $0$ \\
   \hline
 $\mathfrak{e}^{(1/2)}_{7}$ and $\mathfrak{e}^{(1/3)}_{7}$ & $\mathbb{Z}_2$ & $\mathbb{Z}_2$ & $\mathbb{Z}_2$ & $\mathbb{Z}_2$ \\
   \hline
    $\mathfrak{e}^{(1/4)}_{7}$ & $0$ & $\mathbb{Z}_2$ & $\mathbb{Z}_2$ & $0$ \\
   \hline
     $\mathfrak{e}^{(r)}_{8}$ & $0$ & $0$ & $0$ & $0$
 \end{tabular}
 \caption{Our results for defect groups of M-theory frozen singularities, $\mathbb{D}=\mathbb{D}^{(1)}\oplus \mathbb{D}^{(4)}$.}
 \label{table:frozendessert}
 \end{table}

\subsection{Defect Groups from Freezing Map}\label{ssec:defectgrpsfreezemap}

To incorporate the coroot and (co-)weight lattices within the freezing map, it will be convenient to understand the freezing rule \eqref{eq:freezingrule} as follows.
First, we indicate the ``bad'' simple roots of $\mathfrak{g}$, i.e., those $\alpha_i$ with $d  \not | \, d_i$, with a tilde.
Then we construct the hyperplane $E$ via
\begin{align}
    H_2(\mathbb{C}^2/\Gamma; \mathbb{R}) = \Lambda^\mathfrak{g}_\text{root} \otimes \mathbb{R} \supset E := \left\{ v = \textstyle\sum_{i \geq 1} \lambda_i \alpha_i \, | \, \lambda_i \in \mathbb{R} \, , (v, \widetilde\alpha_j) =0 \right\} \, .
\end{align}
The arguments of the previous section imply that the M2 branes are only allowed to wrap 2-cycles in
\begin{align}
    H_2(\widetilde{X}_\Gamma) \cap E = \Lambda_\text{root}^\mathfrak{g} \cap E = \Lambda_\text{root}^\mathfrak{h} \, ,
\end{align}
the \emph{restriction} of the unfrozen roots to $E$.
As the line defects are M2's wrapping relative 2-cycles which can be interpreted as fractional linear combination of compact cycles, the same arguments can be extended to also restrict the weight lattice,
\begin{align}
    H_2(\widetilde{X}_\Gamma, \partial \widetilde{X}_\Gamma) \cap E = \Lambda_\text{wt}^\mathfrak{g} \cap E =: \Lambda_\text{w}^\mathfrak{h} \supset \Lambda_\text{root}^\mathfrak{h} \, .
\end{align}
Practically, this is most easily computed by first re-expressing the ``good'' simple roots of $\mathfrak{g}$, i.e., those $\alpha_i$ with $d \, |\,  d_i$, in terms of (possibly fractional) linear combinations of the ``bad'' simple roots $\widetilde\alpha_j$ and the frozen simple roots $\beta_i$, allowing us to write the $\mathfrak{g}$-weights as $w_\ell = \sum_i \lambda_i \beta_i + \sum_j \mu_j \widetilde\alpha_j$.
Then the orthogonal weights are integer linear combinations of these weights such that the coefficients of $\widetilde\alpha_j$ are integer.
We will see below that $\Lambda_\text{w}^\mathfrak{h} = \Lambda_\text{wt}^\mathfrak{h}$, the frozen weight lattice, for all cases except for ${\cal T}^{(\rm M)}(\mathfrak{e}_8^{(1/4)})$.

Conversely, the M5 branes wrapped on 2-cycles do not experience any restrictions from the freezing flux. Hence, we still expect the full set of four-dimensional defects that exist in the unfrozen theory (and fill out $\Lambda_\text{cowt}^\mathfrak{g} = \Lambda_\text{wt}^\mathfrak{g} \supset \Lambda_\text{(co)root}^\mathfrak{g}$) to be also present in ${\cal T}^{(\rm M)}(\mathfrak{g}^{(n/d)})$.
These are generally magnetically charged defects of $\mathfrak{h}$, with the charges determined by their pairing with the roots $\Lambda_\text{root}^\mathfrak{h}$.
Denoting by $\pi_E$ the orthogonal projection
\begin{equation}
    \pi_E: \Lambda_\text{wt}^\mathfrak{g} \twoheadrightarrow E,
\end{equation}
we clearly have
\begin{align}
    (u, v) = (\pi_E(u), v) \quad \forall \, u \in \Lambda_\text{wt}^\mathfrak{g} \, , \forall \, v \in \Lambda_\text{w}^\mathfrak{h} \subset E \, .
\end{align}
From this, we will verify case by case that
\begin{align}
    \pi_E(\Lambda_\text{root}^\mathfrak{g}) = \Lambda_\text{coroot}^\mathfrak{h} \, , \quad \text{and} \quad \pi_E(\Lambda_\text{wt}^\mathfrak{g}) = \Lambda_\text{cowt}^\mathfrak{h} \, \text{ (except for ${\cal T}^{(\rm M)}(\mathfrak{e}_8^{(1/4)})$)} \,,
\end{align}
and thus find that $\pi_E$ is equivalent to $\pi$ in \eqref{eq:proj}. Together, this will further allow us to determine the geometric representatives for
\begin{align}
    Z(H)^{(1)} = \Lambda_\text{wt}^\mathfrak{h}/ \Lambda_\text{root}^\mathfrak{h} \quad \text{and} \quad Z(H)^{(4)} = \Lambda_\text{cowt}^\mathfrak{h} / \Lambda_\text{coroot}^\mathfrak{h}
\end{align}
in the boundary homology $H_1(S^3/\Gamma) = \Lambda^\mathfrak{g}_\text{wt} / \Lambda^\mathfrak{g}_\text{root}$.

Note that while the lattices can be all thought of as embedded into one Euclidean vector space, it is important for the M-theory interpretation to separate the vectors that correspond to wrapped M2 vs.~M5 branes.
In our notation, the lattices $\Lambda_\text{wt}^\mathfrak{h}$ and $\Lambda_\text{root}^\mathfrak{h}$ are associated with M2 branes, while $\Lambda_\text{cowt}^\mathfrak{h}$ and $\Lambda_\text{coroot}^\mathfrak{h}$ are associated with M5 branes.

\subsubsection{\boldmath{${D^{(1/2)}_{4+k}}$}}

The root lattice of the frozen algebra $\mathfrak{h} = \mathfrak{sp}(k)$ is given in Table \ref{tab:frozen_algebras_roots}, with the long roots satisfying $(\beta_k, \beta_k) = 4 = 2d$ and $(\beta_k, \beta_j) = 2\delta_{k,j-1}$ for $j \neq k$.
It is easy to verify that we have
\begin{align}\label{eq:so_roots_as_sp}
\begin{split}
    & \alpha_2 = -\frac12 \left(\alpha_0 + \alpha_1 + 2\sum_{i=1}^{k-1} \beta_i + \beta_k \right) \, , \quad \alpha_{k+2} = \frac12 (\beta_k - \alpha_{k+3} - \alpha_{k+4}) \, ,\\
    & \alpha_{i+2} = \beta_{i} \, , \quad i = 1,..., k-1 \, ,
\end{split}
\end{align}
where the ``bad'' roots of $\mathfrak{g} = \mathfrak{so}(2k+8)$ (those with Coxeter label $d_j=1$) are $\widetilde\alpha \in \{\alpha_0, \alpha_1, \alpha_{k+3}, \alpha_{k+4}\}$ which are orthogonal to the $\beta_i$.
The orthogonal projection onto $E$, which is spanned by $\beta_i$ as a vector space, is obtained by simply dropping any term proportional to any $\widetilde\alpha$.
This gives $\pi_E (\alpha_2) = -\sum_{i=1}^{k-1} \beta_i - \frac12 \beta_k$, $\pi_E(\alpha_{k+2}) = \frac12 \beta_k$, and $\pi_E(\alpha_i) = \beta_{i-2}$ or $0$ for the others.
Thus we see that the projection of the unfrozen root lattice is indeed the coroot lattice $\Lambda_\text{coroot}^\mathfrak{h}$ with generators
\begin{align}
    \beta_i^\vee = \frac{2 \beta_i}{(\beta_i, \beta_i)} = \begin{cases}
        \beta_i \, , & i < k \\
        \frac12 \beta_k \, , & i = k \, .
    \end{cases}
\end{align}
To obtain the weight and coweight lattice, we orthogonalize and project the unfrozen weight lattice, respectively.

\paragraph{\boldmath{$k > 0$} even}
In this case $\Lambda_\text{wt}^\mathfrak{g}$ is generated by $w_{k+3}$ and $w_{k+4}$ (in Table \ref{table:relative 2-cycles}) together with the roots.
Using \eqref{eq:so_roots_as_sp}, we have for both
\begin{align}\label{eq:so_weights_as_sp-even}
\begin{split}
    w_{k+(3 \text{ or } 4)} & = \frac12 \left( \alpha_1 + 2 \alpha_2 +\sum_{j=1}^{k-1} (j+2) \, \beta_j +(k+2) \alpha_{k+2} \right) + \frac12 (\alpha_{k+3} \text{ or } \alpha_{k+4}) \\
    & = \frac12 \left( - \alpha_0 + \sum_{j=1}^{k-1} j \, \beta_j - \beta_k + \frac{k+2}{2}(\beta_k -\alpha_{k+3} - \alpha_{k+4})  \right) + \frac12 (\alpha_{k+3} \text{ or } \alpha_{k+4}) \, ,
\end{split}
\end{align}
which due to the $\alpha_0$ terms are not in the plane $E$.
To orthogonalize, observe that
\begin{align}
\begin{split}
    \Lambda_\text{wt}^\mathfrak{g} \ni &\, w_{k+3} + w_{k+4} - (k+1)\alpha_{k+2} + \alpha_0 \\
    = & \sum_{j=1}^{k-1} j \, \beta_j - \beta_k + \alpha_{k+2} + \frac12 (\alpha_{k+3} + \alpha_{k+4}) \\
    = & \sum_{j=1}^{k-1} j \, \beta_j - \frac12 \beta_k \in E \, .
\end{split}
\end{align}
So we have
\begin{align}
    \Lambda_\text{w}^\mathfrak{h} := \Lambda_\text{wt}^\mathfrak{g} \cap E = \langle \beta_i \rangle_{\mathbb{Z}} + \frac12 \beta_k =: \langle \beta_i \rangle_{\mathbb{Z}} + {\rm w}\, .
\end{align}
From \eqref{eq:so_weights_as_sp-even} the projections are:
\begin{align}
    \pi_E(w_{k+3}) = \pi_E(w_{k+4}) = \frac12 \left( \sum_{j=1}^{k-1} j\, \beta_j - \beta_k \right) \equiv \underbrace{\frac12 \sum_{j=1}^{k-1} j \, \beta_j^\vee}_{=: {\rm u}} \mod \langle \beta_i^\vee \rangle_\mathbb{Z} \, ,
\end{align}
and we define $\Lambda_\text{cw}^\mathfrak{h} := \pi_E(\Lambda_\text{wt}^\mathfrak{g}) = \langle \beta_i^\vee \rangle_\mathbb{Z} + u$.

It is easy to see that $\Lambda_\text{w}^\mathfrak{h} / \Lambda_\text{root}^\mathfrak{h} = \Lambda_\text{cw}^\mathfrak{h} / \Lambda_\text{coroot}^\mathfrak{h} = Z(Sp(k)) = \mathbb{Z}_2$.
Moreover, it is straightforward to verify that
\begin{align}
    \forall i =1 ,..., k : \quad ({\rm w}, \beta_i^\vee) \in \mathbb{Z} \, , \quad ({\rm u}, \beta_i) \in \mathbb{Z} \, .
\end{align}
This shows that we indeed have $\Lambda_\text{w}^\mathfrak{h} = \Lambda_\text{wt}^\mathfrak{h}$ and $\Lambda_\text{cw}^\mathfrak{h} = \Lambda_\text{cowt}^\mathfrak{h}$.

Notice that $w_{k+3} + w_{k+4} \in \Lambda_\text{wt}^\mathfrak{g} = H_2(\widetilde{X}_\Gamma, \partial \widetilde{X}_\Gamma)$ is still a valid relative cycle wrapped by M5 branes, which projects trivially onto $\Lambda_\text{cowt}^\mathfrak{h}/\Lambda_\text{coroot}^\mathfrak{h}$.
Therefore these (unscreened) four-dimensional defects are not charged under the magnetic center symmetry of the $\mathfrak{sp}$ gauge sector and also link trivially with M2 defects in $\Lambda_\text{wt}^\mathfrak{h}$ that survived the freezing, but nevertheless give rise to a ${\mathbb{Z}_2}^{(4)}$-symmetry independent of the center symmetries.
Geometrically this $\mathbb{Z}_2$ is the diagonal of $\mathbb{Z}_2 \times \mathbb{Z}_2 = H_1(\partial \widetilde{X}_\Gamma)$.

\paragraph{\boldmath{$k$} odd}

In this case $\Lambda_\text{wt}^\mathfrak{g}$ is generated by
\begin{align}
    w_{k+3} = \frac12 \left( -\alpha_0 +\sum_{j=1}^{k-1} j \, \beta_j - \beta_k + (k+2) \alpha_{k+2} \right) + \frac{k}{4} \alpha_{k+3} + \frac{k-2}{4} \alpha_{k+4} \, .
\end{align}
To obtain a weight vector lying in $E$, we need
\begin{align}
\begin{split}
    \Lambda_\text{wt}^\mathfrak{g} \ni & \, 2w_{k+3} + \alpha_0 \equiv k \alpha_{k+2} + \frac{k}{2}(\alpha_{k+3} +\alpha_{k+4}) \mod \langle \beta_i \rangle_\mathbb{Z} \\
    = & \, \frac{k}{2} (2\alpha_{k+2} + \alpha_{k+3} + \alpha_{k+4} ) \mod \langle \beta_i \rangle_{\mathbb{Z}} \\
    = & \, \frac12 \beta_k \mod \langle \beta_i \rangle_\mathbb{Z} \, \Rightarrow \Lambda_\text{w}^\mathfrak{h} = \langle \beta_i \rangle_\mathbb{Z} + \frac12 \beta_k \, .
\end{split}
\end{align}
On the other hand, the projection (using $\pi_E(\alpha_{k+2}) = \frac12 \beta_k$) becomes
\begin{align}
\begin{split}
    \pi_E(w_{k+3}) & = \frac12 \sum_{j=1}^{k-1} j \beta_j - \frac12 \beta_k + \frac{k+2}{2} \beta_k = \underbrace{\frac12 \sum_{j=1}^{k-1} j \, \beta_j^\vee}_{=: {\rm u}} + k \beta_k^\vee \\
    \Rightarrow \, \Lambda_\text{cw}^\mathfrak{h} := \pi_E(\Lambda_\text{cowt}^\mathfrak{g}) & = \langle \beta_i^\vee \rangle_\mathbb{Z} + {\rm u} \, .
\end{split}
\end{align}
We again find $\Lambda_\text{w}^\mathfrak{h}/\Lambda_\text{root}^\mathfrak{h} = \Lambda_\text{cw}^\mathfrak{h}/\Lambda_\text{coroot}^\mathfrak{h} = Z(Sp(k)) = \mathbb{Z}_2$, as well as
\begin{align}
    \forall i =1 ,..., k : \quad ({\rm w}, \beta_i^\vee) \in \mathbb{Z} \, , \quad ({\rm u}, \beta_i) \in \mathbb{Z} \, ,
\end{align}
so $\Lambda_\text{w}^\mathfrak{h} = \Lambda_\text{wt}^\mathfrak{h}$ and $\Lambda_\text{cw}^\mathfrak{h} = \Lambda_\text{cowt}^\mathfrak{h}$, as claimed.

Notice again that we have unscreened M5 brane defects that are uncharged under the $Z(Sp(k))^{(4)}$-symmetry, coming from M5 branes wrapping $2w_{k+3}$ which projects onto a coroot.
Thus, they again link trivially with M2 brane defects, and are uncharged under the magnetic center symmetry.
They are charged under a $\mathbb{Z}_2$ 4-form symmetry, which is the $\mathbb{Z}_2 \subset \mathbb{Z}_4 = H_1(S^3/\Gamma)$ subgroup.

\paragraph{\boldmath{$k=0$}}

In this special case, the freezing leaves no wrapped M2 branes, so there is no 1-form symmetry.
On the other hand, there are no restrictions on the M5 branes, so we again expect a $\mathbb{Z}_2 \times \mathbb{Z}_2 = H_1(\partial \mathbb{C}^2/\Gamma)$ 4-form symmetry without the presence of a gauge sector.

\subsubsection{\boldmath{${\mathfrak{e}^{(1/d)}_6}$}}

\paragraph{\boldmath{$d=2$}}
For ${\cal T}^\text{(M)}[\mathfrak{e}_6^{(1/2)}]$ the roots $\beta_i$ of $\mathfrak{h} = \mathfrak{su}(3)$ satisfy $(\beta_i, \beta_j) = d \, C^{\mathfrak{su}(3)}_{ij}$, where $C^{\mathfrak{su}(3)}$ is the standard $\mathfrak{su}(3)$ Cartan matrix.
The ``good'' roots of $\mathfrak{e}_6$, expressed in terms of the roots of $\mathfrak{h} = \mathfrak{su}(3)$ and the ``bad'' nodes $\widetilde\alpha \in \{\alpha_0, \alpha_1, \alpha_3,\alpha_5\}$ are
\begin{align}
    \alpha_2 = \frac12 (\beta_1 - \alpha_1 - \alpha_3) \, , \quad \alpha_4 = \frac12 (\beta_2 - \alpha_3 - \alpha_5) \, , \quad \alpha_6 = -\frac12 (\alpha_0 +\alpha_3 + \beta_1 + \beta_2) \, .
\end{align}
Thus we have
\begin{align}
    \pi_E(\alpha_2) = \frac12 \beta_1 = \beta_1^\vee \, , \quad \pi_E(\alpha_4) = \beta_2^\vee \, , \quad \pi_E(\alpha_6) = -\beta_1^\vee - \beta_2^\vee \, ,
\end{align}
which indeed span $\Lambda_\text{coroot}^\mathfrak{h}$.

Proceeding with the weight lattice, we have
\begin{align}
\begin{split}
    w_1 & \, = \frac13 (\alpha_1 + 2\alpha_2 + \alpha_4 + 2\alpha_5) = \frac16 (2\beta_1 + \beta_2) +\frac12 (\alpha_5 - \alpha_3) \\
    \Rightarrow \quad 2w_1 & - (\alpha_5-\alpha_3) = \frac13 (2\beta_1 + \beta_2) =: {\rm w} \in \Lambda_\text{wt}^\mathfrak{g} \cap E = \Lambda_\text{w}^\mathfrak{h} \, ,
\end{split}
\end{align}
and verify that this is indeed the weight lattice $\Lambda_{\text{wt}}^{\mathfrak{h}}$ of $\mathfrak{h} = \mathfrak{su}(3)$ using $({\rm w}, \beta_i) = \delta_{i,1} \in \mathbb{Z}$ and $\Lambda_\text{w}^\mathfrak{h} / \Lambda_\text{root}^\mathfrak{h} = \mathbb{Z}_3$.

The projection yields
\begin{align}
    \pi_E(w_1) = \frac13 \beta_1 + \frac16 \beta_2 = \frac13 (2 \beta_1^\vee + \beta_2^\vee) =: {\rm u} \quad \Lambda_\text{cw}^\mathfrak{h} = \langle \beta_i^\vee \rangle_\mathbb{Z} + {\rm u} \, ,
\end{align}
which is the $\mathfrak{su}(3)$ coweight lattice $\Lambda^{\mathfrak{h}}_{\text{cowt}}$, by $({\rm u}, \beta_i^\vee) = \delta_{i,1}) \in \mathbb{Z}$ and $\Lambda_\text{cw}^\mathfrak{h} / \Lambda_\text{coroot}^\mathfrak{h} = \mathbb{Z}_3$.

Note that in this case, there are no (unscreened) M5 branes wrapping coweights of $\mathfrak{g} = \mathfrak{e}_6$ which project onto something that is uncharged under the center of $\mathfrak{h}$, so there is no left-over 4-form symmetry.

\paragraph{\boldmath{$d=3$}}

In this case the freezing forbids M2 branes to wrap any 2-cycles, so there is no 1-form symmetry charges.
However, the M5 branes are still present, and form defects charged under a $\mathbb{Z}_3^{(4)}$ symmetry.

\subsubsection{\boldmath{${\mathfrak{e}^{(1/d)}_7}$}}

\paragraph{\boldmath{$d=2$}}

The frozen algebra in this case is $\mathfrak{h} = \mathfrak{so}(7)$ with $Z(Spin(7)) = \mathbb{Z}_2 = \Lambda_\text{wt}^\mathfrak{h}/\Lambda_\text{root}^\mathfrak{h} = \Lambda_\text{cowt}^\mathfrak{h} / \Lambda^{\mathfrak{h}}_\text{coroot}$.
The simple roots $\beta_{i=1,2,3}$ satisfy (see Table \ref{tab:frozen_algebras_roots})
\begin{align}
    (\beta_i, \beta_j) = \begin{pmatrix}
        4 & -2 & 0 \\
        -2 & 4 & -2 \\
        0 & -2 & 2
    \end{pmatrix} \quad \Rightarrow \quad \beta_1^\vee = \frac12 \beta_1 \, , \quad \beta_2^\vee = \frac12 \beta_2 \, , \quad \beta_3^\vee = \beta_3 \, .
\end{align}
In particular, the long roots have length-squared $2d$.
The ``bad'' $\mathfrak{g}$-roots are $\widetilde\alpha \in \{\alpha_0, \alpha_2, \alpha_4, \alpha_6\}$, and the others can be expressed as
\begin{align}
    \begin{split}
        & \alpha_1 = -\frac12 (\alpha_0 + \alpha_2 + \beta_1 + 2\beta_2 + 2\beta_3) \, , \quad \alpha_3 = \frac12 (\beta_2 - \alpha_2 - \alpha_4) \, ,\\
        & \alpha_5 = \frac12 (\beta_1 - \alpha_4 - \alpha_6) \, , \quad \alpha_7 = \beta_3 \, .
    \end{split}
\end{align}
From this we have the projections
\begin{align}
\begin{split}
    & \pi_E(\alpha_5) = \frac12 \beta_1 = \beta_1^\vee \, , \quad \pi_E(\alpha_3) = \beta_2^\vee \, , \quad \pi_E(\alpha_1) = \beta_1^\vee + 2\beta_2^\vee + \beta_3^\vee \\
    \Rightarrow \quad & \pi_E(\Lambda_\text{root}^\mathfrak{g}) = \Lambda_\text{coroot}^\mathfrak{h} \, .
\end{split}
\end{align}

The $\mathfrak{e}_7$ weight lattice is generated by the roots and $w_1$ from Table \ref{table:relative 2-cycles}.
Rewritten in the basis $\{\beta_i, \widetilde\alpha_j\}$, we have
\begin{align}
\begin{split}
    & w_1 = \frac12 (\alpha_4 + \alpha_6 + \beta_3) = \frac12 \beta_1 - \alpha_5 + \frac12 \beta_3 \\
    \Rightarrow \quad & {\rm w}:= w_1 + \alpha_5 = \frac12 (\beta_1 + \beta_3) \in E \cap \Lambda_\text{wt}^\mathfrak{g} \, .
\end{split}
\end{align}
So $2{\rm w} \in \Lambda_\text{root}^\mathfrak{h}$, and we easily verify $({\rm w}, \beta_j^\vee) = (1,-1,1)_{j}$, so $\Lambda_\text{root}^\mathfrak{h} + {\rm w} = \Lambda_\text{wt}^\mathfrak{h}$.
The projection produces
\begin{align}
    \pi_E(w_1) = \frac12 \beta_3 = \frac12 \beta^\vee =: {\rm u} \, ,
\end{align}
which is indeed an $\mathfrak{so}(7)$ coweight, since $2{\rm u} \in \Lambda_\text{coroot}^\mathfrak{h}$ and $({\rm u}, \beta_j) = (0,-1,1)_j$, so $\pi_E(\Lambda_\text{wt}^\mathfrak{g}) = \Lambda_\text{cowt}^\mathfrak{h}$.

Just as in the case of $\mathfrak{e}^{(1/2)}_6$, there are no unscreened M5 brane defects which are uncharged under the magnetic center symmetry of $\mathfrak{h}$.

\paragraph{\boldmath{$d=3$}}

The frozen algebra in this case is $\mathfrak{h} = \mathfrak{su}(2)$ with $Z(SU(2)) = \mathbb{Z}_2 = \Lambda_\text{wt}^\mathfrak{h}/\Lambda_\text{root}^\mathfrak{h} = \Lambda_\text{cowt}^\mathfrak{h} / \Lambda^{\mathfrak{h}}_\text{coroot}$.
$\Lambda_\text{root}^\mathfrak{h}$ and $\Lambda_\text{coroot}^\mathfrak{h}$ are generated by the simple (co-)roots (cf.~Table \ref{tab:frozen_algebras_roots})
\begin{align}
    \beta = 2\alpha_3 + 3 \alpha_4 + 2\alpha_5 + \alpha_6 + \alpha_7 \quad \text{with} \quad (\beta, \beta) = 6=2d \quad \Rightarrow \quad \beta^\vee = \frac13 \beta \, ,
\end{align}
which are orthogonal to the $\mathfrak{e}_7$ roots $\widetilde\alpha \in \{\alpha_0, \alpha_1, \alpha_3,\alpha_5, \alpha_6, \alpha_7\}$.
We then have
\begin{align}
    \alpha_2 = \frac13 (\alpha_0 - 2\alpha_1 - 2\alpha_3 - \alpha_7 - \beta) \, , \quad \alpha_4 = \frac13 (\beta - 2\alpha_3 -2\alpha_5 - \alpha_6 - \alpha_7) \, ,
\end{align}
which yields
\begin{align}
    \pi_E(\alpha_2) = - \pi_E(\alpha_4) = -\frac13 \beta = -\beta^\vee \quad \Rightarrow \quad \pi_E(\Lambda_\text{root}^\mathfrak{g}) = \Lambda_\text{coroot}^\mathfrak{h} \, .
\end{align}

Proceeding with the weights, we have
\begin{align}
\begin{split}
    w_1 & = \frac16 \beta + \frac13 (\alpha_6 + \alpha_7 - \alpha_3 - \alpha_5) \\
    \Rightarrow \quad {\rm w} &:= w_1 + \alpha_4 + \alpha_3+\alpha_5 = \frac12\beta \in E \cap \Lambda_\text{wt}^\mathfrak{g} \, ,
\end{split}
\end{align}
which is indeed an $\mathfrak{h} = \mathfrak{su}(2)$ weight, since $({\rm w}, \beta^\vee) = 1$, and $2{\rm w} \in \Lambda_\text{root}^\mathfrak{h}$.
The projection is
\begin{align}
    \pi_E(w_1) = \frac16 \beta = \frac12 \beta^\vee =: {\rm u} \, ,
\end{align}
which is also an $\mathfrak{h} = \mathfrak{su}(2)$ coweight since $({\rm u}, \beta) = 1$ and $2{\rm u} \in \Lambda_\text{coroot}^\mathfrak{h}$.
All unscreened M5 brane defects are charged under the magnetic $\mathbb{Z}_2$ center symmetry of $\mathfrak{h}$.

\paragraph{\boldmath{$d=4$}} This is again a completely frozen setting, so the theory has no gauge symmetry and consequently no 1-form symmetry charges from M2 branes on 2-cycles.
On the other hand, the unscreened M5 branes are charged under the $\mathbb{Z}_2 = H_1(\partial S^3/\Gamma)$ 4-form symmetry.

\subsubsection{\boldmath{$\mathfrak{e}_8^{(1/d)}$}}

\paragraph{\boldmath{$d=2$}}

Orthogonalizing the $\mathfrak{g} = \mathfrak{e}_8$ root lattice with respect to the ``bad'' roots $\widetilde\alpha \in \{ \alpha_0, \alpha_2, \alpha_4, \alpha_8\}$ gives the root lattice of $\mathfrak{h}= \mathfrak{f}_4$ with generators $\beta_{i=1,...,4}$ given in Table \ref{tab:frozen_algebras_roots}, satisfying
\begin{align}
  (\beta_i, \beta_j) = \begin{pmatrix}
    4 & -2 & 0 & 0 \\
    -2 & 4 & -2 & 0 \\
    0 & -2 & 2 & -1 \\
    0 & 0 & -1 & 2
  \end{pmatrix} \quad \Rightarrow \quad \beta_1^\vee = \frac12 \beta_1 \, , \quad \beta_2^\vee = \frac12 \beta_2 \, , \quad \beta_3^\vee = \beta_3 \, ,\quad \beta_4^\vee =\beta_4 \, ,
\end{align}
with the long roots satisfying $(\beta, \beta) = 2d$.
The ``good'' roots of $\mathfrak{g}$ are
\begin{align}
  \begin{split}
    & \alpha_1 = -\frac12 (\alpha_0 + \alpha_2 + 2\beta_1 + 3\beta_2 + 4\beta_3 + 2\beta_4) \, , \quad \alpha_3 = \frac12 (\beta_1 - \alpha_2 - \alpha_4) \, , \\
    & \alpha_5 = \frac12 (\beta_2 - \alpha_4 - \alpha_8) \, , \quad \alpha_6 = \beta_3 \, , \quad \alpha_7 = \beta_4 \, .
  \end{split}
\end{align}
From this we immediately see that
\begin{align}
\begin{split}
  & \pi_E(\alpha_3) = \beta_1^\vee \, , \quad \pi_E(\alpha_5) = \beta_2^\vee \, , \quad \pi_E(\alpha_6) = \beta_3^\vee \, , \quad \pi_E(\alpha_7) = \beta_4^\vee \, , \\
  & \pi_E(\alpha_1) = - (2\beta_1^\vee + 3 \beta_2^\vee +2 \beta_3^\vee + \beta_4^\vee) \\
  \Rightarrow \quad & \pi_E(\Lambda_\text{root}^{\mathfrak{g}}) = \Lambda_\text{coroot}^\mathfrak{h} \, .
\end{split}
\end{align}

Now, since $\Lambda_\text{wt}^\mathfrak{g} = \Lambda_\text{root}^\mathfrak{g}$ for $\mathfrak{g} = \mathfrak{e}_8$, this means
\begin{align}
  \Lambda_\text{w}^\mathfrak{h} := \Lambda_\text{wt}^\mathfrak{g} \cap E = \Lambda_\text{root}^\mathfrak{g} \cap E = \Lambda_\text{root}^\mathfrak{h} \, , \quad \Lambda_\text{cw}^\mathfrak{h} := \pi_E(\Lambda_\text{wt}^\mathfrak{g}) = \pi_E(\Lambda_\text{root}^\mathfrak{g}) =\Lambda_\text{coroot}^\mathfrak{h}\, .
\end{align}
This again agrees with the fact that $\mathfrak{h} = \mathfrak{f}_4$ has trivial center, so (co-)roots and (co-)weights are identical, i.e., $\Lambda_\text{w}^\mathfrak{h} = \Lambda_\text{wt}^\mathfrak{h}$ and $\Lambda_\text{cw}^\mathfrak{h} = \Lambda_\text{cowt}^\mathfrak{h}$.
So there are no unscreened defects charged under the 1- and 4-form symmetries at all, which is anyway the naive expectation for an $\mathfrak{f}_4$ gauge theory.

\paragraph{\boldmath{$d=3$}}

In this case the ``bad'' roots are $\widetilde{\alpha} = \{ \alpha_0, \alpha_1, \alpha_3, \alpha_4, \alpha_6, \alpha_7 \}$, leaving the roots $\beta_{i=1,2}$ of $\mathfrak{h} = \mathfrak{g}_2$ with
\begin{align}
  (\beta_i, \beta_j) = \begin{pmatrix}
    6 & -3\\
    -3 & 2
  \end{pmatrix} \quad \Rightarrow \quad \beta_1^\vee = \frac13 \beta_1 \, , \quad \beta_2^\vee = \beta_2 \, ,
\end{align}
with long root having length-squared $6 = 2d$.
For the ``good'' $\mathfrak{g}$-roots we have $\alpha_8 = \beta_2$ and
\begin{align}
\begin{split}
  & \alpha_2 = -\frac13 (\alpha_0 + 2\alpha_1 + 2\alpha_3 + \alpha_4 + 2\beta_1 + 3\beta_2) \, , \quad \alpha_5 = \frac13 (\beta_1 - \alpha_3 - 2 \alpha_4 - 2\alpha_6 -\alpha_7) \\
  \Rightarrow \quad & \pi_E(\alpha_8) = \beta_2^\vee \, , \quad \pi_E(\alpha_5) = \beta_1^\vee \, , \quad \pi_E(\alpha_2) = -2\beta_1^\vee - \beta_2^\vee \, .
\end{split}
\end{align}
So we have $\Lambda_\text{coroot}^\mathfrak{h} = \pi_E(\Lambda_\text{root}^\mathfrak{g}) = \pi_E(\Lambda_\text{wt}^\mathfrak{g}) = \Lambda_\text{cowt}^\mathfrak{h}$ and $\Lambda_\text{wt}^\mathfrak{h} = \Lambda_\text{wt}^\mathfrak{g} \cap E = \Lambda_\text{root}^\mathfrak{g} \cap E = \Lambda_\text{root}^\mathfrak{g}$, as expected for the centerless algebra $\mathfrak{h} = \mathfrak{g}_2$.

\paragraph{\boldmath{$d=4$}}

Applying the freezing rule to $\mathfrak{g} = \mathfrak{e}_8$ with flux $r=1/4$ gives rise to some peculiarities that we will now detail.
In this case the ``bad'' roots are $\widetilde\alpha \in \{ \alpha_0, \alpha_1, \alpha_2, \alpha_4, \alpha_5, \alpha_7, \alpha_8 \}$.
If we seek for integer linear combinations of all $\mathfrak{g}$ roots $\alpha_j$ such that they are orthogonal to $\widetilde\alpha$, then we find that they are all integer multiples of the vector $b = \alpha_4 + 2 \alpha_5 + 2\alpha_6 + \alpha_7 + \alpha_8$.
However, it turns out that $(b,b) = 2 \neq 2 d = 8$.

In order to confirm with the instanton fractionalization, as discussed in Section \ref{sec:freezemap}, we must therefore consider the lattice generated by $\beta = 2b$, as listed in Table \ref{tab:frozen_algebras_roots}, to be the frozen root lattice $\Lambda_\text{root}^\mathfrak{h}$.
Consequently, the coroot lattice is generated by $\beta^\vee = \frac14 \beta$.
The ``good'' $\mathfrak{e}_8$ roots are
\begin{align}
    \alpha_3 = -\frac14 (\alpha_0 + 2\alpha_1 + 3\alpha_2 + 3\alpha_4 + 2\alpha_5 + \alpha_8 + \beta) \, , \quad \alpha_6 = \frac14 (\beta - 2\alpha_4 - 4\alpha_5 -2\alpha_7 - 2\alpha_8) \, ,
\end{align}
which project to the coroot,
\begin{align}
    \pi_E(\alpha_3) = -\pi_E(\alpha_4) = -\frac14 \beta = -\beta^\vee \, .
\end{align}

However, given that the weight lattice of $\mathfrak{g}$ is the same as its root lattice, we cannot obtain the naively expected (co-)weights which would be generated by ${\rm w} = \frac12 \beta \in \Lambda_\text{wt}^\mathfrak{h}$ and ${\rm u} = \frac12 \beta^\vee \in \Lambda_\text{cowt}^\mathfrak{h}$.
This means that this $\mathfrak{h} = \mathfrak{su}(2)$ gauge theory does not have any unscreened defects from M2 or M5 branes wrapping relative 2-cycles which are charged under the electric or magnetic center symmetries.

We will present a bottom-up explanation of this somewhat surprising result in Section \ref{sec:symtft}.
For now, we would like to crosscheck this finding, as well as the ``additional'' 4-manifold defects charged under the 4-form symmetry we encountered in the other cases above, from a dual F-theory perspective.

\subsection{Review of Duality to Twisted F-theory Compactifications}

This subsection serves to briefly review the chain of dualities that relate M-theory frozen singularities to twisted compactifications of F-theory. Readers familiar with \cite{deBoer:2001wca} and \cite{Tachikawa:2015wka} can safely skip this subsection. For reviews on F-theory see \cite{Weigand:2018rez, Cvetic:2018bni}.

Up until this point, we have implicitly considered the M-theory frozen singularity geometries to have a metric asymptotic to the conical metric
\begin{equation}
    ds^2=dR^2+R^2 d\Omega^2_{S^3/\Gamma} \,,
\end{equation}
where $d\Omega^2_{S^3/\Gamma}$ is the metric of $S^3/\Gamma$ inherited from an $S^3$ metric of constant curvature. Such hyper-K\"ahler manifolds are known as asymptotically locally Euclidean (ALE). To perform the aforementioned duality chain, we first need to embed the ALE frozen singularities into
a non-compact elliptic K3 manifold, with so-called Kodaira singularities. For each $D$- and $E$-type ALE singularity there is essentially a unique way of doing this
\begin{equation}\label{eq:Kodairaemb}
    D_{4+k}\hookrightarrow I^*_{k} \,, \quad \mathfrak{e}_6\hookrightarrow IV^* \,, \quad \mathfrak{e}_7\hookrightarrow III^* \,, \quad \mathfrak{e}_8\hookrightarrow II^* \,,
\end{equation}
where we have used the standard notation for Kodaira singularity types (for a review aimed at physicists see for instance \cite{Weigand:2018rez}).

Let $X_{ell.}$ denote one of the Kodaira singularities appearing in \eqref{eq:Kodairaemb} which is elliptically-fibered $\mathbb{E}\rightarrow X_{ell.}\rightarrow \mathbb{C}_w$. Our starting point is M-theory with $\int_{\partial X_{ell.}} C_3=n/d$ which we will write as
\begin{equation}
    \mathrm{M}\left( X_{ell.}, \; \; \int_{\partial X_{ell.}}C_3=\frac{n}{d}\right).
\end{equation}
Here, the asymptotic boundary $\partial X_{ell.}$ is the elliptic fibration over the bounding circle of $\mathbb{C}_w$.
If we compactify on an $S^1$ transverse to $X_{ell.}$ then we have IIA string theory on the same background where now $C_3$ is the RR monodromy:
\begin{equation}
    \mathrm{M}\left( X_{ell.}, \; \; \int_{\partial X_{ell.}}C_3=\frac{n}{d}\right)\quad \xrightarrow[]{S^1 \text{ comp.}} \quad \mathrm{IIA}\left( X_{ell.}, \; \; \int_{\partial X_{ell.}}C_3=\frac{n}{d}\right).
\end{equation}
Then we can perform two T-dualities, one for each of the circles of the elliptic fiber of $X_{ell.}$:
\begin{equation}\label{eq:doubleT}
    \mathrm{IIA}\left( X_{ell.}, \; \; \int_{\partial X_{ell.}}C_3=\frac{n}{d}\right)\quad \xrightarrow[]{\textnormal{double T-duality}} \quad \mathrm{IIA}\left( \widehat{X}_{ell.} , \; \; \int_{S^1_\varphi}C_1=\frac{n}{d}\right).
\end{equation}
Here $\widehat{X}_{ell.}$, the mirror dual\footnote{From the CFT analysis of \cite{Nahm:2001kh}, one can conclude that the Kodaira singularities $I^*_0$, $IV^*$, $III^*$, $II^*$ are mirror dual to birationally equivalent K3 manifolds with the same $SL(2,\mathbb{Z})$ monodromy as before but with a central fiber of the form $T^2/\mathbb{Z}_k$ for $k=2,3,4$ and $6$ respectively. Such results were conjectured in Section 4.6.1 of \cite{deBoer:2001wca}.} of $X_{ell.}$, is also elliptically-fibered over $\mathbb{C}_w$ whose asymptotic boundary we denote by $S^1_{\varphi}=\partial \mathbb{C}_w$ where $\varphi:=\mathrm{arg}(w)$. We see that the background $C_3$ has turned into a $C_1$ background and we can in fact lift the RHS of \eqref{eq:doubleT} to M-theory where the $C_1$ potential implies a non-trivial fibration of the IIA/M-theory circle. This lift takes the form \cite{deBoer:2001wca, Tachikawa:2015wka}
\begin{equation}
     \mathrm{IIA}\left( \widehat{X}_{ell.} , \; \; \int_{S^1_\varphi}C_1=\frac{n}{d}\right)  \quad \xrightarrow[]{g_s\rightarrow \infty} \quad \mathrm{M}\left( (Y_{ell.}\times S_\theta^1)/\mathbb{Z}_d\right) \,,
\end{equation}
where $Y_{ell.}$ is a Kodaira singularity, i.e., a local K3, such that $Y_{ell.}/\mathbb{Z}_d=\widehat{X}_{ell.}$ and $g_s$ is the IIA string coupling. If we let $\mathbb{C}_z$ denote the base of $Y_{ell.}$, then the $\mathbb{Z}_d$ quotient is defined as
\begin{align}\label{eq:twistedtea}
  \theta \sim \theta +\frac{2\pi}{d}, \; \; \; \phi \sim \phi+\frac{2\pi n}{d}, \; \; \; \mathbb{E}_z \sim \rho(\mathbb{E}_z) \,.
\end{align}
where $\phi=\mathrm{arg}(z)$, $\mathbb{E}_z$ is the elliptic fiber of $Y_{ell.}$ at $z$, and $\rho$ acts by the $SL(2,\mathbb{Z})$ monodromy matrix of $\widehat{X}_{ell.}$. Finally, if we take a limit such that the volume of the generic elliptic fiber of $Y_{ell.}$ shrinks to zero, then we arrive at F-theory compactified on $(Y_{ell.}\times S_\theta^1)/\mathbb{Z}_d$:
\begin{equation}
    \mathrm{M}\left( (Y_{ell.}\times S_\theta^1)/\mathbb{Z}_d\right)  \quad \xrightarrow[]{\mathrm{Vol}(\mathbb{E}_z)\rightarrow 0} \quad \mathrm{F}\left( (Y_{ell.}\times S_\theta^1)/\mathbb{Z}_d\right).
\end{equation}
In summary, we have the following duality
\begin{equation}\label{eq:fmduality}
       \mathrm{M}\left( X, \; \; \int_{\partial X}C_3=\frac{n}{d}\right)       \quad \Leftrightarrow \quad \mathrm{F}\left( (Y\times S_{\theta}^1)/\mathbb{Z}_d\right)
\end{equation}
where we have dropped the $(ell.)$ subscript to not overload the notation. In Table \ref{table:frozendinner}, we list examples of Kodaira singularities $Y$ and their corresponding M-theory frozen singularity (for a full list see Table 3 of \cite{Tachikawa:2015wka}).

\begin{table}[t!]
\centering
\begin{tabular}{c|c|c|c|c|c|c|c}
 Frozen Singularity & $Y$ & $X$ & $\mathbb{Z}_d$ & $\rho_1|_{ell.}$ & Outer & $\mathfrak{g}_Y$ & $\mathfrak{h}_{\Gamma, d}$ \\
 \hline\hline
 $D^{(1/2)}_{4+k}$ & $I_{2k}$ & $I^*_{k}$ & $\mathbb{Z}_2$ & $\begin{pmatrix}
     -1 & -k \\
     0 & -1
 \end{pmatrix}$ & $\mathbb{Z}_2$ & $\mathfrak{su}(2k)$  & $\mathfrak{sp}(k)$  \\
 \hline
 $\mathfrak{e}^{(1/2)}_6$  & $IV$ & $IV^*$ & $\mathbb{Z}_2$ & $\begin{pmatrix}
     -1 & -1 \\
     1 & 0
 \end{pmatrix}$ & $\varnothing$ & $\mathfrak{su}(3)$ & $\mathfrak{su}(3)$ \\
 \hline
$\mathfrak{e}^{(1/3)}_6$  & $I_0$ & $IV^*$ & $\mathbb{Z}_3$ & $\begin{pmatrix}
     -1 & -1 \\
     1 & 0
 \end{pmatrix}$ & $\varnothing$ & $\varnothing$ & $\varnothing$ \\
 \hline
 $\mathfrak{e}^{(1/2)}_{7}$ & $I^*_{0}$ & $III^*$ & $\mathbb{Z}_2$ & $\begin{pmatrix}
     0 & -1 \\
     1 & 0
 \end{pmatrix}$ & $\mathbb{Z}_2$ & $\mathfrak{so}(8)$  & $\mathfrak{so}(7)$  \\
  \hline
 $\mathfrak{e}^{(1/3)}_{7}$ & $III$ & $III^*$ & $\mathbb{Z}_3$ & $\begin{pmatrix}
     0 & -1 \\
     1 & 0
 \end{pmatrix}$ & $\varnothing$ & $\mathfrak{su}(2)$  & $\mathfrak{su}(2)$  \\
  \hline
$\mathfrak{e}^{(1/4)}_7$ & $I_0$ & $III^*$ & $\mathbb{Z}_4$ &  $\begin{pmatrix}
     0 & -1 \\
     1 & 0
 \end{pmatrix}$ & $\varnothing$ & $\varnothing$ & $\varnothing$ \\
  \hline
 $\mathfrak{e}^{(1/2)}_{8}$ & $IV^*$ & $II^*$ & $\mathbb{Z}_2$ & $\begin{pmatrix}
     0 & -1 \\
     1 & 1
 \end{pmatrix}$ & $\mathbb{Z}_2$ & $\mathfrak{e}_6$  & $\mathfrak{f}_4$  \\
 \hline
 $\mathfrak{e}^{(1/3)}_{8}$ & $I^*_{0}$ & $II^*$ & $\mathbb{Z}_3$ & $\begin{pmatrix}
     0 & -1 \\
     1 & 1
 \end{pmatrix}$ & $\mathbb{Z}_3$ & $\mathfrak{so}(8)$  & $\mathfrak{g}_2$\\
 \hline
 $\mathfrak{e}^{(1/4)}_{8}$ & $IV$ & $II^*$ & $\mathbb{Z}_4$ & $\begin{pmatrix}
     0 & -1 \\
     1 & 1
 \end{pmatrix}$ & $\mathbb{Z}_2$ & $\mathfrak{su}(3)$  & $\mathfrak{su}(2)$\\
 \hline
  $\mathfrak{e}^{(1/5)}_{8}$ & $II$ & $II^*$ & $\mathbb{Z}_5$ & $\begin{pmatrix}
     0 & -1 \\
     1 & 1
 \end{pmatrix}$ & $\varnothing$ & $\varnothing$  & $\varnothing$\\
 \hline
 $\mathfrak{e}^{1/6}_{8}$ & $I_0$ & $II^*$ & $\mathbb{Z}_6$ & $\begin{pmatrix}
     0 & -1 \\
     1 & 1
 \end{pmatrix}$ & $\varnothing$ & $\varnothing$ & $\varnothing$
\end{tabular}
\caption{M-theory frozen singularities $X$ whose F-theory dual involves map $\rho_1|_{ell.}$ which non-trivially acts on the 1-cycles of the elliptic fiber. The column ``Outer" denotes the group action that it generates at the level of $H_1(\partial Y)$; it is the action on $\mathfrak{g}_Y$ whose quotient gives $\mathfrak{h}_{\Gamma, d}$.}
\label{table:frozendinner}
\end{table}

\subsection{Calculation of Defect Groups in Twisted F-theory Frame}\label{ssec:outerauto}

Our goal in this section is to compute the defect groups of the M-theory frozen singularities on the LHS of the F-/M-theory duality  \eqref{eq:fmduality} by appealing to the relation to twisted F-theory compactifications on the RHS of \eqref{eq:fmduality}. See Table \ref{table:frozendinner} for the details on the F-theory geometry for each frozen singularity.

\begin{figure}
\centering
\includegraphics[scale=0.30, trim = {0cm 0cm 0cm 0cm}]{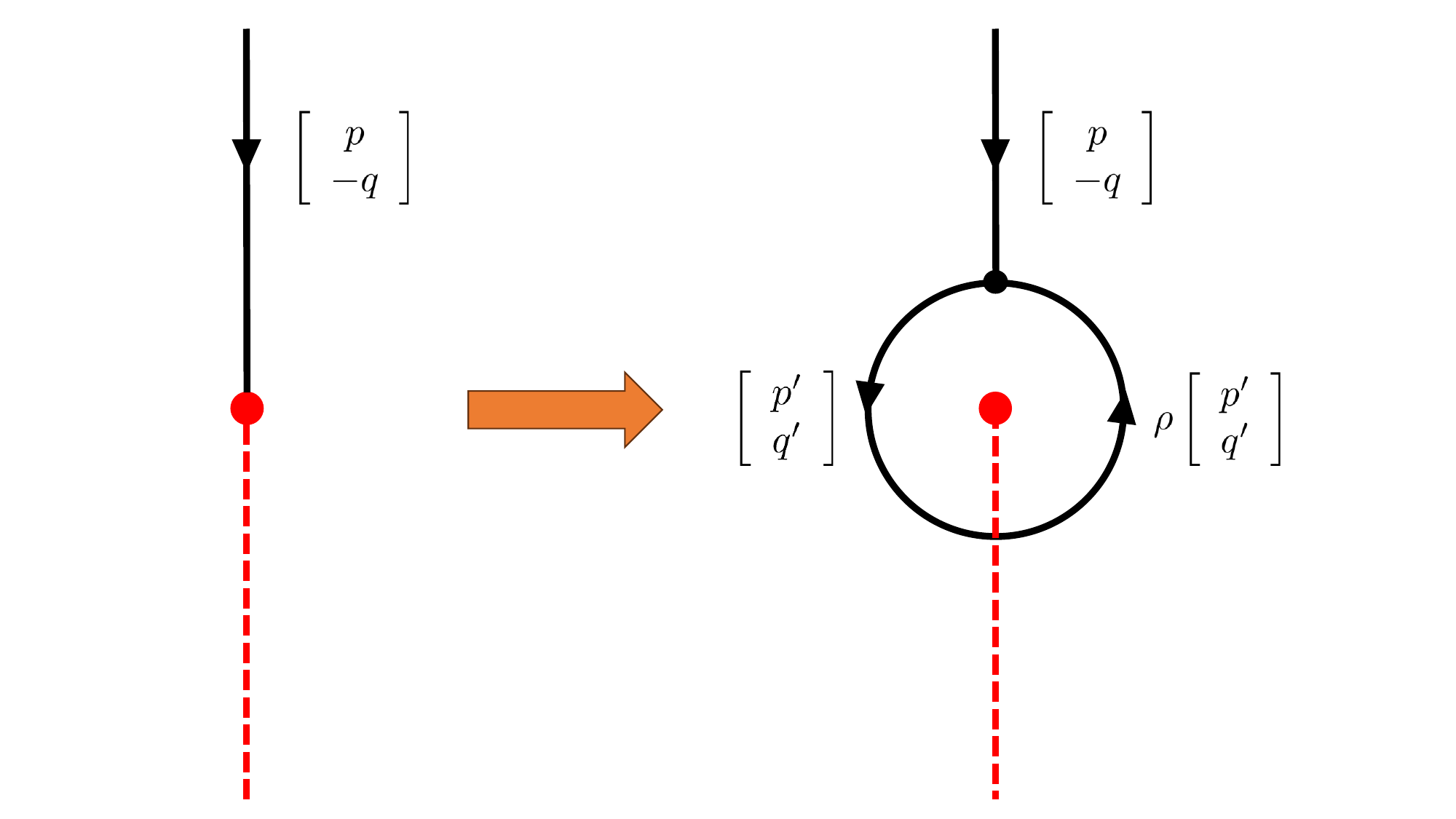}
\caption{On the left, we illustrate a $(p,q)$-string (black arrow) ending on a 7-brane (red dot) that generates monodromy $\rho$ (we signify a cut crossing which produces this monodromy action by the dotted red line). For certain $(p,q)$ charges, we can dynamically resolve this configuration to the right figure where $p', q'\in \mathbb{Z}$. Such $(p,q)$-strings are trivial in the defect group because this process implies that the line operators they generate in the 8D gauge theory can terminate in the 8D 7-brane worldvolume. Similar remarks apply for $(p,q)$-5-branes.}
\label{fig:stringjunc}
\end{figure}

To understand how to calculate the defect groups of F-theory on $Z=(Y\times S^1)/\mathbb{Z}_d$, first we recall how one may calculate the defect group for F-theory on some Kodaira singularity $Y$. This engineers an 8D SYM with some gauge algebra $\mathfrak{g}_Y$ and has a defect group made up of the electric 1-form symmetry and the magnetic 5-form symmetry $\mathbb{D}_{\rm 8D}=\mathbb{D}^{(1)}\oplus \mathbb{D}^{(5)}$. While it turns out that simply $\mathbb{D}_{\rm 8D}=(Z(G_Y))^{(1)}\oplus (Z(G_Y))^{(5)}$, the string theoretic approach to deriving this was systematized in \cite{Cvetic:2021sxm} whereby the defect groups of the 7-brane specified by $Y$ are related to the charged of string and 5-brane states that can end on the 7-brane localized at the fiber singularity. These string/5-brane charges are valued in a freely generated lattice, and the defect group is equivalent to this lattice modulo string/5-brane states ending on the 7-branes that can be resolved into so-called \textit{integer null junctions}\footnote{In precise terms, a $(p,q)$-string/5-brane admits an integer null junction if there exists solutions to $\rho_{11}p'+\rho_{12}q'+p=p'$ and $\rho_{21}p'+\rho_{22}q'+q=q'$ such that $p', q'\in \mathbb{Z}$.}, see Figure \ref{fig:stringjunc}. The reason for this is that such states can be dynamically radiated away from the 7-brane, and thus do not realize a defect in the 8D gauge theory charged under a global symmetry.

\begin{figure}[t!]
\centering
\includegraphics[scale=0.40, trim = {0cm 0cm 0cm 0cm}]{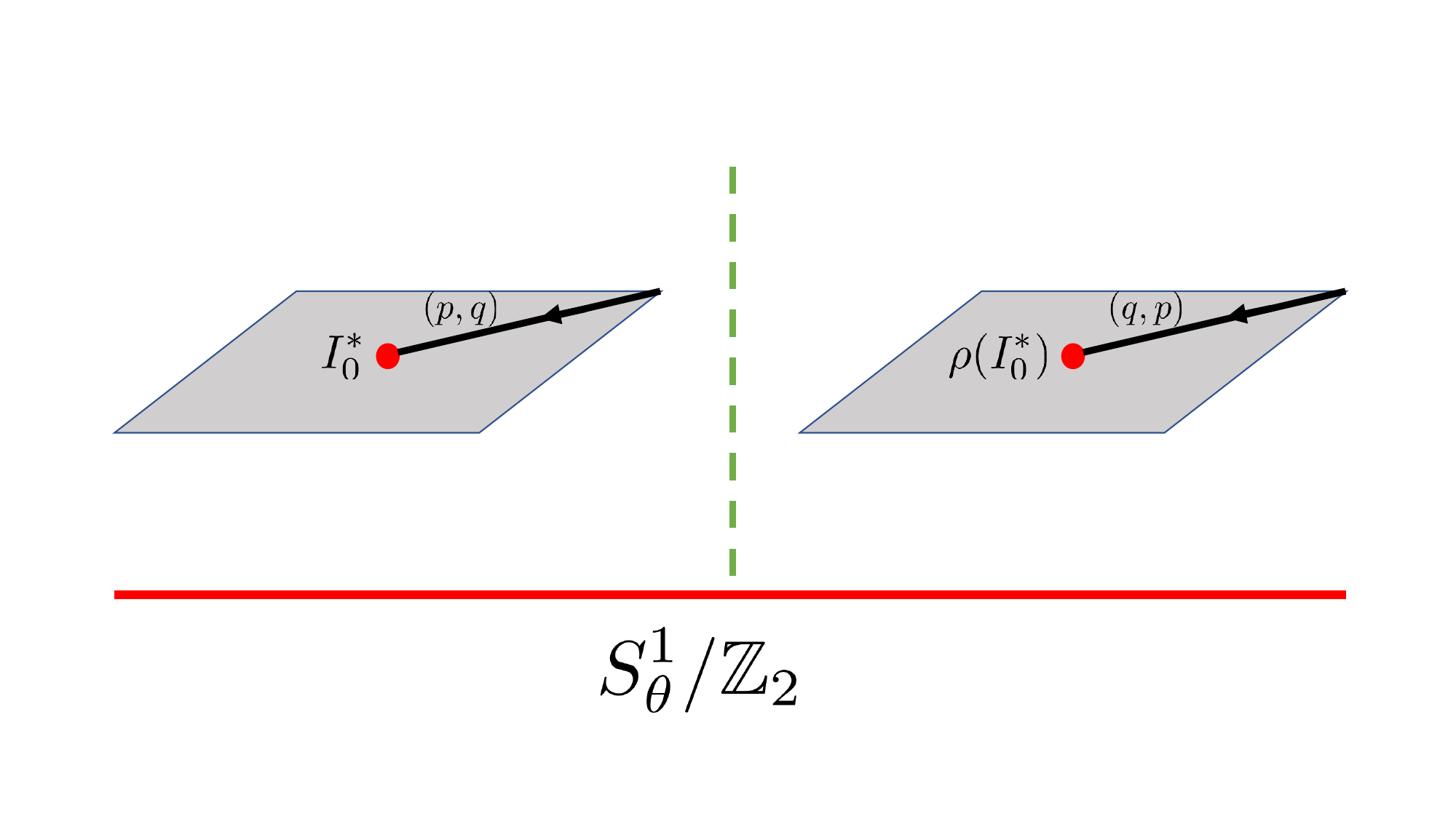}
\caption{F-theory compactified on $Z=(Y_{I^*_0}\times S_{\theta}^1)/\mathbb{Z}_2$ which is dual to M-theory on an $III^*$ singularity with $\int_{\partial X} C_3$=1/2 frozen flux. The gray planes are the bases of $Y_{I^*_0}$ with the $I^*_0$ 7-brane located at the origin. The green line indicates a $\rho$ monodromy action cut and acts on long string junctions as shown. Note that the $(p,q)$ long strings have a defect group charge $(p\; \mathrm{mod}\; 2, q\; \mathrm{mod}\; 2)$ which is well-defined up to screening of dynamical states on the 7-brane. The monodromy action $\rho$ then enforces that $(1,0)$ and $(0,1)$ strings are in the same equivalence class of the defect group (i.e. they have indistinguishable charges at the level of the 7D KK theory). This reduces the defect group from $\mathbb{Z}_2\times \mathbb{Z}_2$ to $\mathbb{Z}_2$ with the generator $(1,1)=(1,0)+(0,1)$ being equivalent to $(0,0)$.}
\label{fig:e7onehalf}
\end{figure}

For the case of F-theory on $Z=(Y\times S_\theta^1)/\mathbb{Z}_d$, states charged under the 1-form symmetry of the 7D KK theory are strings ending on the 7-brane which are localized at a specific value of $\theta_0\in S^1$. For a given $(p,q)$-string which is able to end on the 7-brane, we have a charge vector\footnote{Using the conventions of \cite{Weigand:2018rez} where the minus sign is due to the $SL(2,\mathbb{Z})$-invariant epsilon tensor. } $[p,-q]^\mathrm{T}$ and the monodromy around $S_{\theta}^1/\mathbb{Z}_d$ identifies
\begin{equation}\label{eq:stringcoker}
   \rho \begin{bmatrix}
        p \\
        -q
    \end{bmatrix} \sim \begin{bmatrix}
        p \\
        -q
    \end{bmatrix}.
\end{equation}
This means that the 1-form piece of the defect group is given by
\begin{equation}\label{eq:d1}
    \mathbb{D}^{(1)}=\mathrm{coker}(\rho-1).
\end{equation}
See Figure \ref{fig:e7onehalf} for an illustration/explanation in the case of the twisted compactification $Z=(Y_{I^*_0}\times S_{\theta}^1)/\mathbb{Z}_2$. As for 5-branes, these give rise to the 4-form part of the defect group by wrapping\footnote{This is due to the familiar fact that the 't Hooft operators of a $(D+1)$-dimensional gauge theory reduce to 't Hooft operators in a $D$-dimensional gauge theory upon circle reduction by wrapping the circle. Meanwhile Wilson operators do not wrap the circle.} $S^1_\theta/\mathbb{Z}_d$. Such 5-branes must be invariant under the monodromy (which now acts as an automorphism on $(Z(G_Y))^{(5)}$) to consistently wrap $S_\theta^1/\mathbb{Z}_d$ so therefore we have that
\begin{equation}\label{eq:d4}
    \mathbb{D}^{(4)}=\mathrm{ker}(\rho-1) \,.
\end{equation}
Similar to \eqref{eq:d1}, the $(\rho-1)$ operator is understood to act on $(p,q)$ charges of 5-branes which are able to end on the 7-brane. Notice also that we can consider the consistent charges of strings wrapped on $S_\theta^1/\mathbb{Z}_d$ to generate 0-dimensional defects in a 0-form defect group, $\mathbb{D}^{(0)}$ or similarly 5-branes not wrapped on the circle to generate 5-manifold defect operators in $\mathbb{D}^{(5)}$ for the 7D theory. Since these operators do not play a role in the global structure of the 7D gauge group, we will not consider them further in this work but their presence may be interesting to study in future work.

\begin{figure}
\centering
\includegraphics[scale=0.40, trim = {0cm 4cm 0cm 6cm}]{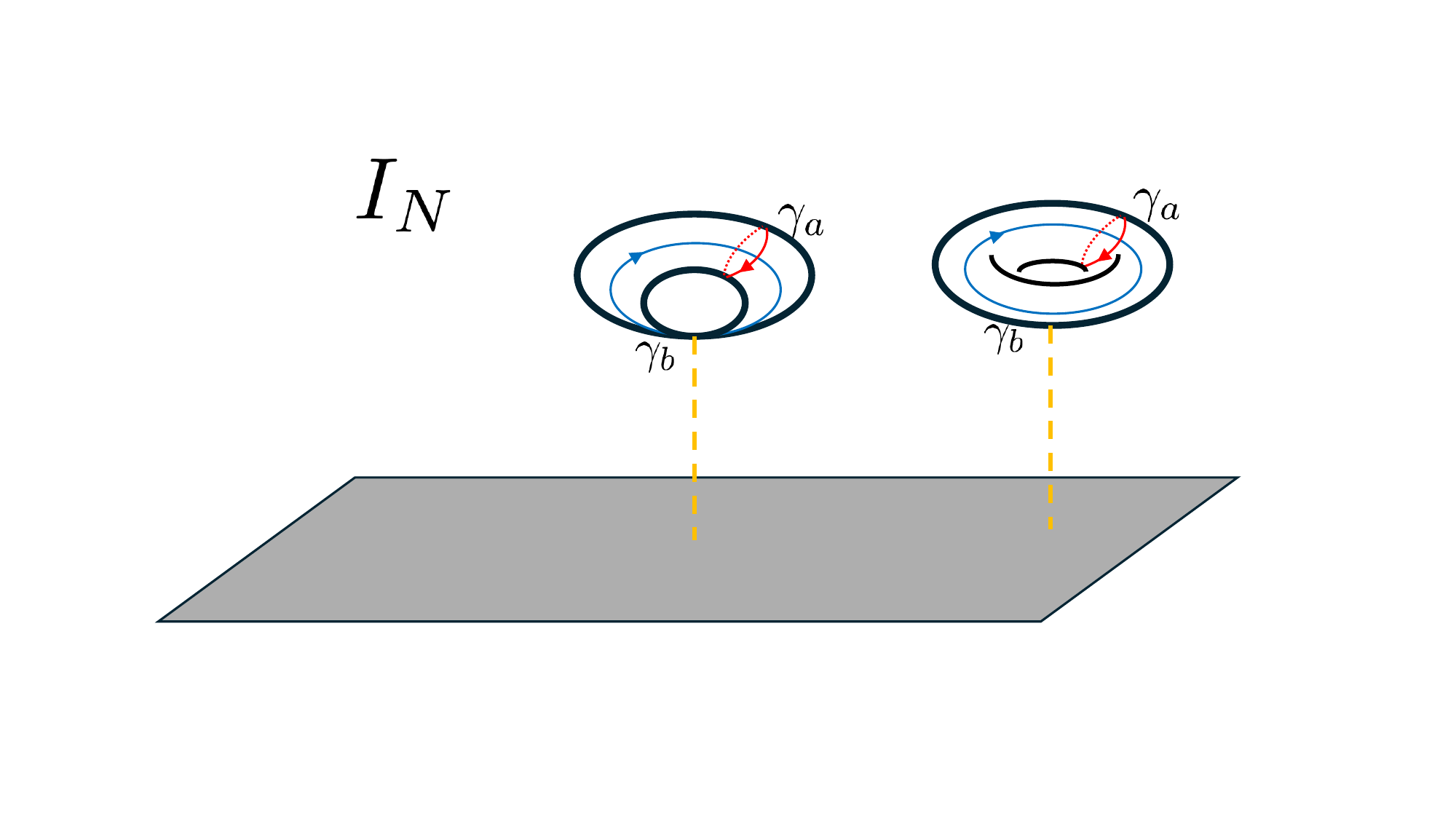}
\caption{Illustration of a $I_N$ Kodaira surface. The singular, central fiber is depicted on the left while the generic smooth fiber is shown on the right. At the central fiber the a-cycle, $\gamma_a$, of the elliptic fiber degenerates while $\gamma_b$ does not. This mean that the $\gamma_b$ cycle on the asymptotic boundary $\partial Y_{I_N}$ does not trivialize. This means that D1/D5 branes dual to M2/M5 branes wrapping $\gamma_b$ cannot end on a stack of $N$ D7s.}
\label{fig:h1nontriv}
\end{figure}

A convenient way to compute the defect group $\mathbb{D}=\mathbb{D}^{(1)}\oplus \mathbb{D}^{(4)}$ is done by compactifying the twisted F-theory compactifcation on a further $S^1_{\mathrm{extra}}$ to arrive at M-theory compactified on $Z=(Y\times S_{\theta}^1)/\mathbb{Z}_d$:
\begin{equation}\label{eq:fmduality6d}
         \mathrm{F}\left( (Y\times S_{\theta}^1)/\mathbb{Z}_d\times S^1_{\mathrm{extra}}\right)     \quad \Leftrightarrow \quad \mathrm{M}\left( (Y\times S_{\theta}^1)/\mathbb{Z}_d\right) \,.
\end{equation}
This allows us to easily package \eqref{eq:d1} and \eqref{eq:d4} in terms of homology groups of $Z$ relative to its asymptotic boundary $\partial Z$ as
\begin{align}
\mathbb{D}_{\text{7D}}^{(1)}=\mathbb{D}_{\text{6D}}^{(1)}=\mathrm{Tor}\left(\frac{H_2(Z,\partial Z)}{H_2(Z)} \right)_{\text{M}2} \,, \\ \mathbb{D}_{\text{7D}}^{(4)}=\mathbb{D}_{\text{6D}}^{(3)}=\mathrm{Tor}\left(\frac{H_3(Z,\partial Z)}{H_3(Z)} \right)_{\text{M}5} \,,\label{eq:6ddefectgrps}
\end{align}
where the 6D subscript refers to the 6D system in \eqref{eq:fmduality6d} and the M$p$ subscripts indicate which M-theory brane we are wrapping on these relative cycles. We restrict to torsion in order to isolate the charges that can arise from strings/5-branes. From the long exact sequence which defines relative homology, we arrive at another presentation of these groups
\begin{equation}
    \mathrm{Tor}\left(\frac{H_i(Z,\partial Z)}{H_i(Z)} \right)\simeq  \mathrm{Tor}\big(\mathrm{Ker}\left( H_{i-1}(\partial Z)\rightarrow H_{i-1}(Z)\right) \big)=: \mathrm{Tor} \big( H_{i-1}(\partial Z) \big)|_{\mathrm{triv.}}
\end{equation}
where the subscript `$_{\text{triv.}}$' indicates that $\mathrm{Tor} \big(H_{i-1}(\partial Z)\big)|_{\mathrm{triv.}}$ is the subgroup of $\mathrm{Tor} \big(H_{i-1}(\partial Z)\big)$ that trivializes when embedded as $(i-1)$-cycles into the bulk $Z$. Intuitively, the groups $H_i(\partial Z)$ (of appropriate degree) encode the possible string/5-brane charges that may consistently be measured at spatial infinity while the restriction $|_{\mathrm{triv.}}$ identifies those charges that may actually end on the 7-brane. Such a restriction is non-trivial when $Y$ is a type $I_N$ 7-brane because $H_1(Y)\neq 0$ in this case which physically can be understood as the inability for D1 strings to end on D7 branes, see Figure \ref{fig:h1nontriv} for a illustration of the F-theory geometry. In Section \ref{sec:symtft}, we will see that in cases where this restriction is non-trivial and give a SymTFT picture of what is going on.

\begin{figure}[t!]
\centering
\includegraphics[scale=0.40, trim = {0cm 0cm 0cm 0cm}]{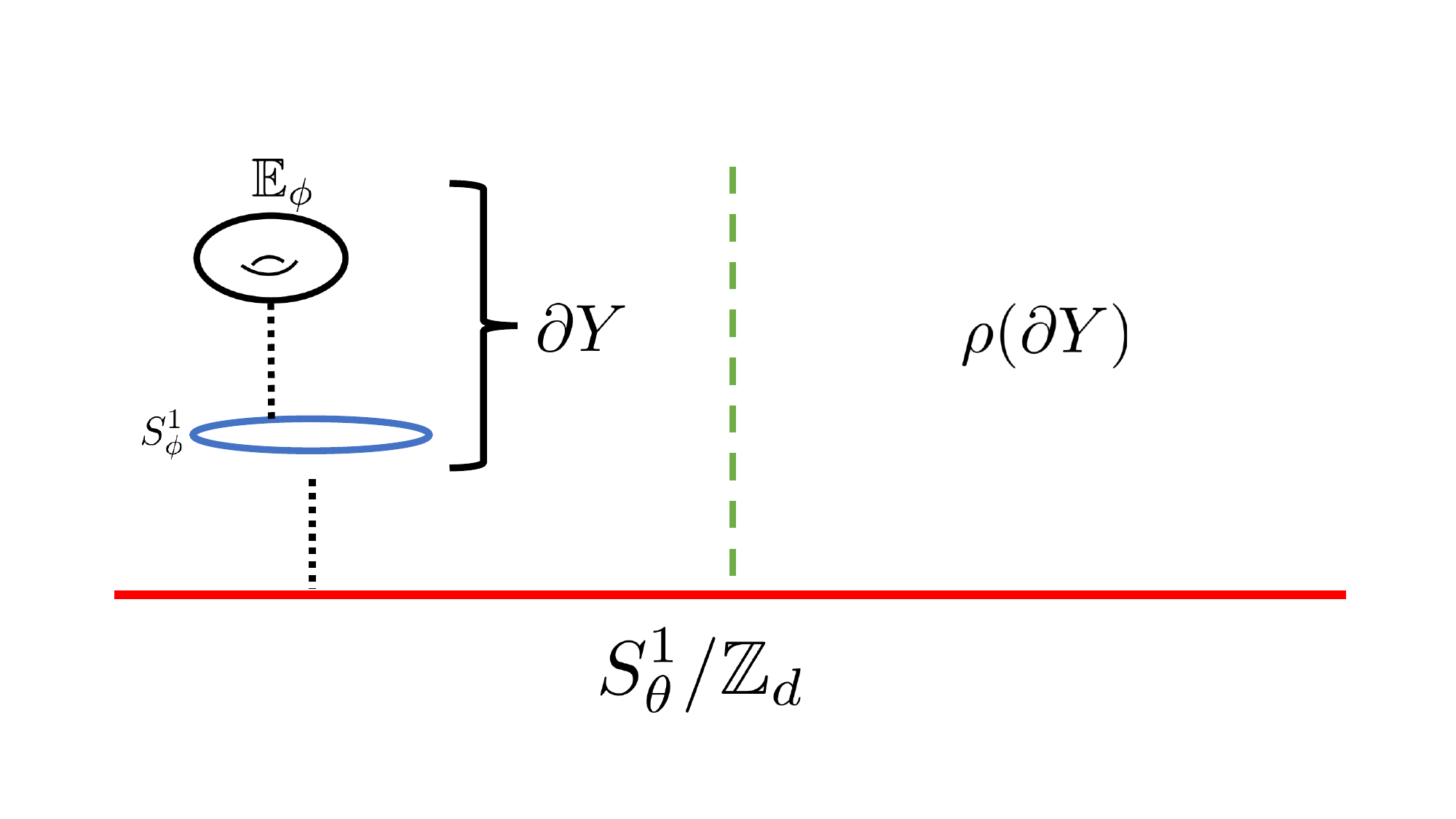}
\caption{Illustration of $\partial Z$ regarded a fibration of $\partial Y$ over $S^1_\theta/\mathbb{Z}_d$. We also show the fibration structure of $\partial Y$ itself over a circle $S^1_\phi$ with elliptic fiber $\mathbb{E}_\phi$. The green dotted line denotes a cut for the monodromy action $\rho$.}
\label{fig:partialZ}
\end{figure}

Calculating the groups in \eqref{eq:6ddefectgrps} is a fairly straightforward task as $Z$ and $\partial Z$ have the structure of a fibration over a circle and the action of $\mathbb{Z}_d$ on the cycles of $H_*(Z)$ and $H_*(\partial Z)$ can be derived as follows. From \eqref{eq:twistedtea} we have a fibration structure of $Z$ and $\partial Z$ over a circle:
\begin{align}
    Y \hookrightarrow Z \xrightarrow{\pi} S^1/\mathbb{Z}_d \\
    \partial Y \hookrightarrow \partial Z \xrightarrow{\pi} S^1/\mathbb{Z}_d.\label{eq:partialZfib}
\end{align}
See Figure \ref{fig:partialZ} for an illustration of this fibration structure for $\partial Z$. We immediately see that the 1-cycle associated to the $\phi$ direction in $\partial Y$ is not acted upon when encircling the base of \eqref{eq:partialZfib} since the quotient \eqref{eq:twistedtea} simply acts by a rotation. However, cycles in $Y$ and $\partial Y$ will generically transform due to the action of $\rho$ in \eqref{eq:twistedtea} which we will detail in the examples below. Notice from Table \ref{table:frozendinner} that the action of $\rho$ on $H_1(Y)$ and $H_2(Y)$ is trivial for the $\mathfrak{e}^{(1/2)}_6$ and $\mathfrak{e}^{(1/3)}_7$ cases so their defect groups will be identical to that of $Y$, so we omit these trivial cases from our calculations.

Another geometric detail that will be useful in what follows are the homology groups for $Y$ and $\partial Y$. The former homology groups are given as follows ($*=0,...,4$)
\begin{align}
    H_*(Y)&=\{ \mathbb{Z}, H_1(Y), \mathbb{Z},0,0\} \,, \\
    H_1(Y)&=\left\{
    \begin{array}{lr}
      \mathbb{Z}^2 &  Y=I_0\\
      \mathbb{Z} &  Y=I_{N>0}\\
      0 &  \mathrm{Otherwise}
    \end{array}
  \right\}.\label{eq:h1y}
\end{align}
which can be derived from deformation retracting $Y$ to the central fiber $\mathbb{E}_0$ whose topology follows from the Kodaira type of $Y$. Meanwhile, the homology groups for $\partial Y$ depend crucially on the $ADE$-type of the singularity located in the central fiber $\mathbb{E}_0$. A local patch of the $ADE$ is diffeomorphic to $\mathbb{C}^2/\Gamma$ for a finite subgroup $\Gamma\subset SU(2)$. $H_*(\partial Y)$ is then given as follows (see \cite{Cvetic:2021sxm} for more details)
\begin{align}
    H_*(\partial Y) &=\{ \mathbb{Z}, H_1(\partial Y), H_2(\partial Y), \mathbb{Z}\} \,, \\
    H_1(\partial Y) &=\left\{
    \begin{array}{lr}
      \mathbb{Z}\oplus \mathbb{Z}^2 &  Y=I_0\\
      \mathbb{Z} \oplus \mathbb{Z}\oplus \mathbb{Z}_{N}&  Y=I_{N>0}\\
       \mathbb{Z} \oplus \mathrm{Ab}(\Gamma_Y) &  \mathrm{Otherwise}
    \end{array}
  \right\} \,, \\
H_2(\partial Y) &=\left\{
    \begin{array}{lr}
      \mathbb{Z}\oplus \mathbb{Z}^2 &  Y=I_0\\
      \mathbb{Z} \oplus \mathbb{Z}&  Y=I_{N>0}\\
       \mathbb{Z} &  \mathrm{Otherwise}
    \end{array}\right\} \,,
\end{align}
where $\Gamma_Y$ denotes the $ADE$ subgroup of $SU(2)$ for which $\mathbb{C}^2/\Gamma_Y$ is the local $ADE$ singularity of $Y$. The first $\mathbb{Z}$ summand for $H_1(\partial Y)$ and $H_2(\partial Y)$ are, respectively, the asymptotic circle $S_{\phi}^1\equiv \partial \mathbb{C}_z$ and the generic $\mathbb{E}_z\simeq T^2$ fiber of $Y$ along $S^1_{\phi}$, which exist for all geometries.

For readers that wish to skip the details of the geometric computations, we list in Table~\ref{table:frozendessert} our results for the defect groups. These follow from the homologies for $Z$ and $\partial Z$ whose results we state below:
\begin{equation}
     H_*(Z) =\{ \mathbb{Z}, H_1(Z), \mathbb{Z}, \mathbb{Z}, 0, 0\} \,,
\end{equation}
\begin{align}
    H_1(Z)=\left\{
    \begin{array}{lr}
      \mathbb{Z} \oplus \mathbb{Z}_{2}&  Z=(I_{N>0}\times S^1)/\mathbb{Z}_2\\
       \mathbb{Z} \oplus \mathrm{Ab}(\Gamma_X) &  \mathrm{Otherwise}
    \end{array}
  \right\} \,,
\end{align}
\begin{equation}
    H_*(\partial Z)=\{ \mathbb{Z}, H_1(\partial Z), H_2(\partial Z), \mathbb{Z}^2, \mathbb{Z}\} \,,
\end{equation}
\begin{align}
H_1(\partial Z)=H_2(\partial Z)=\left\{
    \begin{array}{lr}
      \mathbb{Z}^2\oplus \mathbb{Z}_2^2 &  Z=(I_{2k}\times S^1)/\mathbb{Z}_2 \; \; \textnormal{($k$ even)}\\
      \mathbb{Z}^2\oplus \mathbb{Z}_4 &  Z=(I_{2k}\times S^1)/\mathbb{Z}_2 \; \; \textnormal{($k$ odd)}\\
      \mathbb{Z}^2\oplus \mathbb{Z}_3 & Z=(I_0\times S^1)/\mathbb{Z}_3\\
       \mathbb{Z}^2\oplus \mathbb{Z}_2 &  Z=(I^*_0\times S^1)/\mathbb{Z}_2 \\
       \mathbb{Z}^2\oplus \mathbb{Z}_2 & Z=(I_0\times S^1)/\mathbb{Z}_4\\
       \mathbb{Z}^2 & \mathrm{Otherwise}
    \end{array}\right\} \,.
\end{align}
Here $\Gamma_X$ denotes the $ADE$ subgroup of $SU(2)$ for which $\mathbb{C}^2/\Gamma_X$ is the local $ADE$ singularity of $X$, see Table \ref{table:frozendinner}.

\paragraph{D-Type Frozen Singularities}
We now study the topology of $Z$ and $\partial Z$ relevant to the $D$-type singularities with $\int_{\partial X} C_3=1/2$. We find it best to separate the cases $D_4$ and $D_{4+k\geq 5}$.

Starting with $D_4$, then $Y$ is has an $I_0$ Kodaira fiber, i.e. is topologically $\mathbb{C}\times T^2$, and $\partial Y\simeq T^3$. We have a fibration
\begin{equation}\label{eq:fibagaind4}
    \partial Y\hookrightarrow \partial Z\xrightarrow{\pi} S^1/\mathbb{Z}_2 \,,
\end{equation}
so the first and second homologies can be calculated from the short exact sequences
\begin{align}\label{eq:ses}
  & 0\rightarrow \mathrm{coker}(\rho_1-1)\rightarrow H_1(\partial Z)\rightarrow \mathrm{ker}(\rho_0-1)\rightarrow 0  \,,\\
  & 0\rightarrow \mathrm{coker}(\rho_2-1)\rightarrow H_2(\partial Z)\rightarrow \mathrm{ker}(\rho_1-1)\rightarrow 0 \,,
\end{align}
where $\rho_k:H_k(\partial Y)\rightarrow H_k(\partial Y)$ is the action of the monodromy on $H_k(\partial Y)$ which is induced from the action $\rho$ appearing in \eqref{eq:twistedtea}. From Table \ref{table:frozendinner}, we see that the action $\rho_1$ on $H_1(\partial Y)=\mathbb{Z}\oplus \mathbb{Z}^2$ is given as
\begin{equation}\label{eq:rhod4partial}
    \rho_1=1\oplus \begin{pmatrix}
        -1 & 0\\
        0 & -1 \\
    \end{pmatrix} \,.
\end{equation}
Additionally, the action of $\rho_2$ on $H_2(\partial Y)=\mathbb{Z}\oplus \mathbb{Z}^2$ is exactly the same as in \eqref{eq:rhod4partial} since the action of $\rho$ preserves the 2-cycle of the generic fiber\footnote{This amounts to the statement that an $SL(2,\mathbb{Z})$ matrix is determinant 1 so preserves the volume form of the elliptic fiber.} $[\mathbb{E}_z]$, while the $\mathbb{Z}^2$ factor in $H_2(\partial Y)$ are generated by $[(\textnormal{A cycle of $\mathbb{E}_z$})\times S^1_\phi]$ and $[(\textnormal{B cycle of $\mathbb{E}_z$})\times S^1_\phi]$.
We see then that
\begin{align}\label{eq:dtypehomology1}
H_1(\partial Z) &=\mathbb{Z}\oplus \mathbb{Z}_2\oplus \mathbb{Z}_2\oplus \mathbb{Z} \,, \\
  H_2(\partial Z) &=\mathbb{Z}\oplus \mathbb{Z}_2\oplus \mathbb{Z}_2\oplus \mathbb{Z} \,,
\end{align}
where now the last $\mathbb{Z}$ factor for $H_1$ is generated by the base circle of the fibration \eqref{eq:fibagaind4}, and the last $\mathbb{Z}$ factor for $H_2$ is generated by the product of this circle with $S^1_\phi$. Similarly, we can use the following short exact sequences to solve for $H_1(Z)$ and $H_2(Z)$:
\begin{align}\label{eq:h1z}
 & 0\rightarrow \mathrm{coker}(\rho_1-1)\rightarrow H_1(Z)\rightarrow \mathrm{ker}(\rho_0-1)\rightarrow 0 \,, \\
 &  0\rightarrow \mathrm{coker}(\rho_2-1)\rightarrow H_2(Z)\rightarrow \mathrm{ker}(\rho_1-1)\rightarrow 0 \,.\label{eq:h2z}
\end{align}
Now $\rho_1$ is acts as the $(2 \times 2)$ matrix appearing in \eqref{eq:rhod4partial} on $H_1(Y)=\mathbb{Z}^2$ while $\rho_2$ acts on $H_2(Y)=\mathbb{Z}$ by the identity. This implies that
\begin{align}\label{eq:dtypehomology111}
H_1(Z)=\mathbb{Z}_2\oplus \mathbb{Z}_2 \oplus \mathbb{Z} \,, \quad \quad H_2(Z)=\mathbb{Z} \,.
\end{align}
To finally compute the 1-form symmetry piece of the defect group, $\mathbb{D}^{(1)}_{\text{7D}}$, we just need to understand the kernel of the inclusion map $H_1(\partial Z)\rightarrow H_1(Z)$ which is an isomorphism on the $\mathbb{Z}^2_2$ factor as well as the $\mathbb{Z}$ factor generated by $[S^1/\mathbb{Z}_2]$. Therefore
\begin{equation}\label{eq:defectgrpd4}
    D^{(1/2)}_4: H_1(\partial Z)|_{\mathrm{triv.}}=\mathbb{Z} \quad \implies \quad \mathbb{D}^{(1)}_{\text{7D}}=0 \,,
\end{equation}
after taking the torsion part. Additionally, we see that the kernel of $H_2(\partial Z)\rightarrow H_2(Z)$ must be $\mathbb{Z}\oplus \Z^2_2$ which means that
\begin{equation}
     D^{(1/2)}_4: \mathbb{D}^{(4)}_{\text{7D}}=\Z^2_2 \,.
\end{equation}
The fact that this group is non-trivial will have impications for our application to O$7^+$ planes in Section \ref{ssec:O7plus}.

Moving on to $D_{4+k}$ with $k\geq 1$, we again first solve for $H_1(\partial Z)$ and
$H_2(\partial Z)$. Recall that for an $I_{2k}$ singularity we have $H_1(\partial Y)=\mathbb{Z}\oplus \mathbb{Z}\oplus\mathbb{Z}_{2k}$ where the second two factors follow from the fact that
\begin{equation}
  \mathrm{coker}\begin{pmatrix}
      0 & 2k \\
      0 & 0
  \end{pmatrix}  \simeq \mathbb{Z}_{2k}\oplus \mathbb{Z} \,,
\end{equation}
since the above matrix is the monodromy matrix for the $I_{2k}$ singularity minus the identity. This motivates explicitly presenting the elements of $\mathbb{Z}_{2k}\oplus \mathbb{Z}$ as
\begin{equation}\label{eq:xy}
    \begin{pmatrix}
        x \; \mathrm{mod}\; 2k\\
        y
    \end{pmatrix}, \quad \quad x,\; y \in \mathbb{Z} \,.
\end{equation}
From Table \ref{table:frozendinner}, the action of $\rho_1$ on $H_1(\partial Y)=\mathbb{Z}\oplus \mathbb{Z}_{2k}$ is then given as multiplication by
\begin{equation}\label{eq:rho1}
    \rho_1=1\oplus \begin{pmatrix}
        -1 & -k \\
        0 & -1
    \end{pmatrix} \,.
\end{equation}
If $k$ is even, then we can perform a coordinate change $x'\; \mathrm{mod}\; 2k=(x-\frac{k}{2}y)\; \mathrm{mod}\; 2k$, $y'=-y$ after which
\begin{equation}\label{eq:coordchangeabove}
    \rho_1-1=0\oplus \begin{pmatrix}
        2 & 0 \\
        0 & 2
    \end{pmatrix} \,.
\end{equation}
Then it is straightforward to see that
\begin{equation}
    \textnormal{($k$ even):} \quad  H_1(\partial Z)=\mathbb{Z}\oplus\mathbb{Z}^2_2\oplus \mathbb{Z} \,.
\end{equation}
Similarly, if $k$ is odd then we can perform coordinate changes\footnote{These follow from the Smith normal decomposition $\begin{pmatrix}
        1 & 0 \\
        0 & 4
    \end{pmatrix}=\begin{pmatrix}
        -1 & (k-1)/2 \\
        2 & -k
    \end{pmatrix}\begin{pmatrix}
        -2 & -k \\
        0 & -2
    \end{pmatrix}\begin{pmatrix}
        0 & -1 \\
        1 & 2
    \end{pmatrix}$ when $k$ is odd.} in both the domain and codomain of $(\rho_1-1)$ to arrive at
\begin{equation}
    \rho_1-1=0\oplus \begin{pmatrix}
        1 & 0 \\
        0 & 4
    \end{pmatrix} \,,
\end{equation}
which implies that
\begin{equation}
    \textnormal{($k$ odd):} \quad  H_1(\partial Z)=\mathbb{Z}\oplus\mathbb{Z}_4\oplus \mathbb{Z} \,.
\end{equation}
Now solving for $H_2(\partial Z)$, we see first that $\mathrm{ker}(\rho_1-1)=\mathbb{Z}\oplus \mathbb{Z}_2$. The $\mathbb{Z}$ factor is generated by $[S^1_\phi\times S_{\theta}^1/\mathbb{Z}_2]$ while the torsion factor is generated by $k$ times the generator of $\mathbb{Z}_{2k}\subset H_1(\partial Y)$ fibered over $S^1_{\theta}$. Meanwhile, we have that the action of $\rho_2$ on $H_2(\partial Y)=\mathbb{Z}\oplus \mathbb{Z}=\langle [\mathbb{E}_z], [(\textnormal{B-cycle of $\mathbb{E}_z$})\times S^1_\phi] \rangle$ is given by
\begin{equation}\label{eq:someaction}
    \begin{pmatrix}
        1 & 0 \\
        0 & -1
    \end{pmatrix} \,,
\end{equation}
where we recall from \eqref{eq:xy} that the B-cycle of $\mathbb{E}_z$ has coordinate $y$ so the minus sign in \eqref{eq:someaction} follows from the bottom-right entry of the $I^*_k$ monodromy matrix \eqref{eq:rho1}. We then have that $\mathrm{coker}(\rho_2-1)=\mathbb{Z}\oplus \mathbb{Z}_2$ leading to
\begin{equation}
    H_2(\partial Z)=\mathbb{Z}\oplus \mathrm{Ext}(\mathbb{Z}_2,\mathbb{Z}_2)\oplus \mathbb{Z} \,,
\end{equation}
where $\mathrm{Ext}(\mathbb{Z}_2,\mathbb{Z}_2)$ is an extension of $\mathbb{Z}_2$ by $\mathbb{Z}_2$ that we a a priori do not know from the short exact sequence alone. We can fix this extension using Poincar\'e duality because $\partial Z$ is smooth. This implies $\mathrm{Tor} \big(H_i(\partial Z) \big) =\mathrm{Tor} \big(H_{\mathrm{dim}(\partial X)-i-1}(\partial Z) \big)$ which means that $\mathrm{Tor} \big(H_1(\partial Z) \big)=\mathrm{Tor} \big(H_2(\partial Z)\big)$. Therefore, the extension is trivial for $k$ even and non-trivial for $k$ odd. Since the integer factors also match we have that $H_2(\partial Z)=H_1(\partial Z)$.

We now show that
\begin{equation}
    H_1(Z)=\mathbb{Z}_2\oplus \mathbb{Z} \,, \quad H_2(Z)=\mathbb{Z} \,,
\end{equation}
where the $\mathbb{Z}_2$ factor follows from the previously mentioned effect of the $I^*_k$ monodromy matrix flipping the B-cycle, and we have that $\mathbb{Z}$ in $H_1(Z)$ is generated by $[S_{\theta}^1/\mathbb{Z}_2]$ while the $\mathbb{Z}$ in $H_2(Z)$ is generated by $[\mathbb{E}_z]$.

We are finally in a position to derive $H_1(\partial Z)|_{\mathrm{triv.}}$ and $H_2(\partial Z)|_{\mathrm{triv.}}$. We show the former is
\begin{equation}\label{eq:dkclaim}
    H_1(\partial Z)|_{\mathrm{triv.}}=\mathrm{ker}(H_1(\partial Z)\rightarrow H_1(Z))=\mathbb{Z}\oplus \mathbb{Z}_2 \,.
\end{equation}
The $\mathbb{Z}$ is simply the extra factor of $\mathbb{Z}$ that $H_1(\partial Z)$ has compared to $H_1(Z)$ which is generated by $[S^1_\phi]$. As for the torsion factors, we more interestingly have that
\begin{equation}\label{eq:dkclaim1}
    \mathbb{Z}_2=\left\{
    \begin{array}{lr}
      \mathrm{ker}(\mathbb{Z}^2_2\rightarrow \mathbb{Z}_2) &  \textnormal{$k$ even}\\
      \mathrm{ker}(\mathbb{Z}_4\rightarrow \mathbb{Z}_2) &  \textnormal{$k$ odd}
    \end{array}
  \right\} \,,
\end{equation}
where the map in the top line is simply
\begin{equation}
    \begin{pmatrix}
        x \; \mathrm{mod}\; 2 \\
        y \; \mathrm{mod}\; 2
    \end{pmatrix}\mapsto y \; \mathrm{mod}\; 2 \,,
\end{equation}
while the map in the bottom line is
\begin{equation}
    \begin{pmatrix}
        0 \\
        y \; \mathrm{mod}\; 4
    \end{pmatrix}\mapsto y \; \mathrm{mod}\; 2 \,,
\end{equation}
which clearly satisfy the claimed \eqref{eq:dkclaim1}. In summary then, the 1-form part of the defect group for the $D^{(1/2)}_{4+k}$ frozen singularity for any $k$ is
\begin{equation}\label{eq:defectdtype}
  \mathbb{D}^{(1)}_{\text{7D}}\big(D^{(1/2)}_{4+k}\big)=\left\{
    \begin{array}{lr}
       0, \; \; \; k=0\\
        \mathbb{Z}_{2}, \; \; \; k\geq 1
        \end{array}\right\} \,.
\end{equation}

We additionally have that
\begin{equation}\label{eq:dkmagclaim}
    H_2(\partial Z)|_{\mathrm{triv.}}=\mathrm{ker}(H_2(\partial Z)\rightarrow H_2(Z))=\left\{
    \begin{array}{lr}
       \mathbb{Z}\oplus \mathbb{Z}^2_2, \; \; \; \textnormal{$k$ even}\\
        \mathbb{Z}\oplus \mathbb{Z}_4, \; \; \; \textnormal{$k$ odd}
        \end{array}\right\} \,,
\end{equation}
which is immediately clear given that $\mathrm{Tor} \big(H_2(Z)\big)=0$. We then conclude that
\begin{equation}\label{eq:defectdtype1}
  \mathbb{D}^{(4)}_{\rm 7D}\big(D^{(1/2)}_{4+k}\big)= \left\{
    \begin{array}{lr}
        \mathbb{Z}^2_2, \; \; \; \textnormal{$k$ even}\\
        \mathbb{Z}_4, \; \; \; \textnormal{$k$ odd}
        \end{array}\right\} \,.
\end{equation}
Notice that we see that the 4-manifold defect charges, arising from M5 branes wrapped on relative 2-cycles, are indeed more numerous in than the line defect charges as we found in Section \ref{ssec:defectgrpsfreezemap}. Confirmation of this feature will persist for the fully frozen examples below.

\paragraph{$\mathbf{\mathfrak{e}^{(1/3)}_6}$ Frozen Singularity}
In this case $Y$ is a type $I_0$ surface, so $H_1(\partial Y)=\mathbb{Z}\oplus\mathbb{Z}^2$ and the monodromy action around $S^1_\theta/\mathbb{Z}_3$ is
\begin{equation}
   \rho_1= 1\oplus \begin{pmatrix}
        -1 & -1\\
        1 & 0
    \end{pmatrix} \,.
\end{equation}
We can calculate the kernel and cokernel of $(\rho_1-1)$ from its Smith normal form
\begin{equation}\label{eq:asthetwobytwomatrixin}
  \mathrm{SNF}( \rho_1-1)= 0\oplus \begin{pmatrix}
        1 & 0\\
        0 & 3
    \end{pmatrix} \,,
\end{equation}
which makes it clear that $\mathrm{ker}(\rho_1-1)=\mathbb{Z}$ and $\mathrm{coker}(\rho_1-1)=\mathbb{Z}_3$. From this we have that $H_1(\partial Z)=H_2(\partial Z)=\mathbb{Z}\oplus\mathbb{Z}_3$. This is not quite the defect group because $H_1(Z)$ is non-trivial. This follows from the fact that the action of $\rho_1$ on $H_1(Y)=\mathbb{Z}^2$ has the same Smith normal form as the $(2 \times 2)$ matrix in \eqref{eq:asthetwobytwomatrixin} which means that $H_1(Z)=\mathbb{Z}_3$. This means that the 1-form piece of the defect group is
\begin{equation}
    \mathbb{D}^{(1)}_{\rm 7D}(\mathfrak{e}^{(1/3)}_6) = H_2(Z,\partial Z)/H_2(Z) = (\mathbb{Z}\oplus \mathbb{Z}_3)/(\mathbb{Z}\oplus \mathbb{Z}_3)=0 \,,
\end{equation}
while the 4-form piece is
\begin{equation}
    \mathbb{D}^{(4)}_{\rm 7D}(\mathfrak{e}^{(1/3)}_6)=H_3(Z,\partial Z)/H_3(Z)=(\mathbb{Z}\oplus\mathbb{Z}_3)/\mathbb{Z}=\mathbb{Z}_3 \,.
\end{equation}

\paragraph{\boldmath{$\mathfrak{e}^{(1/2)}_7$} Frozen Singularity}
In this case, $Y$ is an $I^*_0$ surface so we have $H_1(\partial Y)=\mathbb{Z}\oplus \mathbb{Z}^2_2$, with the monodromy action around $S^1_{\theta}/\mathbb{Z}_2$ being given by
\begin{equation}
    \rho_1=1\oplus \begin{pmatrix}
        0 & 1\\
        1 & 0
    \end{pmatrix} \,,
\end{equation}
with the $(2 \times 2)$ matrix being the modulo-2 reduction of the monodromy matrix of an $III^*$ singularity, see Table \ref{table:frozendinner}. It is straightforward to calculate\footnote{Explicitly deriving the $\mathbb{Z}_2$ part of the cokernel, we indeed find that the $\mathbb{Z}^2_2$ vector $\begin{pmatrix}
    x \; \mathrm{mod} \; 2\\
     y \; \mathrm{mod} \; 2
\end{pmatrix}$ is now considered modulo $\begin{pmatrix}
    1\\
     1
\end{pmatrix}$ which means we are left with the quotient $\mathbb{Z}^2_2/\mathbb{Z}_2=\mathbb{Z}_2$ generated by $(x+y)\; \mathrm{mod}\; 2$. These considerations also make it clear that $\mathrm{ker}(\rho_1-1)=\mathbb{Z}\oplus \mathbb{Z}_2$.} that $\mathrm{coker}(\rho_1-1)=\mathbb{Z}\oplus\mathbb{Z}_2$ which gives
\begin{equation}
    H_1(\partial Z)=\mathbb{Z}\oplus\mathbb{Z}_2 \oplus \mathbb{Z} \,,
\end{equation}
after combining with $\mathrm{ker}(\rho_0-1)=\mathbb{Z}$ whose generator is given by $[S_{\theta}^1/\mathbb{Z}_2]$. Similar considerations yield $H_2(\partial Z)=H_1(\partial Z)$. Meanwhile from \eqref{eq:h1y} we see that since $H_1(Y)=0$, \eqref{eq:h1z} implies that $H_1(Z)=\mathrm{ker}(\rho_0-1)=\mathbb{Z}$ and \eqref{eq:h2z} implies that $H_2(Z)=\mathrm{coker}(\rho_2-1)=\mathbb{Z}$. We therefore have that
\begin{equation}
    H_1(\partial Z)|_{\mathrm{triv.}}=\mathbb{Z}\oplus\mathbb{Z}_2 \,.
\end{equation}
which after taking the torsion part gives the 1-form part of the defect group
\begin{equation}
  \mathbb{D}_{\text{7D}}^{(1)}(\mathfrak{e}^{(1/2)}_7)=\mathbb{Z}_2 \,.
\end{equation}
As for the 4-form part, this follows from homology groups quoted above to simply be
\begin{equation}\label{eq:defectetype4}
  \mathbb{D}_{\text{7D}}^{(4)}(\mathfrak{e}^{(1/2)}_7)=\mathbb{Z}_2 \,.
\end{equation}

\paragraph{\boldmath{$\mathfrak{e}^{(1/4)}_7$} Frozen Singularity}
This case is similar to the $\mathfrak{e}^{(1/3)}_6$ above, where again $H_1(\partial Y)=\mathbb{Z}\oplus\mathbb{Z}^2$. The monodromy action on $H_1(Y)$ around $S^1_\theta/\mathbb{Z}_4$ is
\begin{equation}
   \rho_1= 1\oplus \begin{pmatrix}
        0 & -1\\
        1 & 0
    \end{pmatrix} \,,
\end{equation}
and $(\rho_1-1)$ has a Smith normal form
\begin{equation}\label{eq:asthetwobytwomatrixin1}
  \mathrm{SNF}( \rho_1-1)= 0\oplus \begin{pmatrix}
        1 & 0\\
        0 & 2
    \end{pmatrix} \,.
\end{equation}
It follows that $\mathrm{ker}(\rho_1-1)=\mathbb{Z}$ and $\mathrm{coker}(\rho_1-1)=\mathbb{Z}_2$ which means that $H_1(\partial Z)=H_2(\partial Z)=\mathbb{Z}\oplus\mathbb{Z}_2$. Once more, this is not quite the defect group because $H_1(Z)$ is non-trivial since the action of $\rho_1$ on $H_1(Y)=\mathbb{Z}^2$ has the same Smith normal form as the $(2\times 2)$ matrix in \eqref{eq:asthetwobytwomatrixin1}. We see then that $H_1(Z)=\mathbb{Z}_2$ which makes 1-form piece of the defect group
\begin{equation}
    \mathbb{D}^{(1)}_{\rm 7D}\big(\mathfrak{e}^{(1/4)}_7\big)=H_2(Z,\partial Z)/H_2(Z)=(\mathbb{Z}\oplus\mathbb{Z}_2)/(\mathbb{Z}\oplus\mathbb{Z}_2)=0 \,,
\end{equation}
while the 4-form piece is
\begin{equation}
    \mathbb{D}^{(4)}_{\rm 7D}\big(\mathfrak{e}^{(1/4)}_7\big)=H_3(Z,\partial Z)/H_3(Z)=(\mathbb{Z}\oplus\mathbb{Z}_2)/\mathbb{Z}=\mathbb{Z}_2 \,.
\end{equation}

\paragraph{\boldmath{$\mathfrak{e}^{(r)}_8$} Frozen Singularities with \boldmath{$r=1/2,1/3$} and \boldmath{$1/4$}}
Following similar remarks from the previous paragraph, $H_1(Z)=H_2(Z)=\mathbb{Z}$ for all of the $\mathfrak{e}^{(r)}_8$ cases as well which, as for the $\mathfrak{e}^{(1/2)}_7$, gives the relations
\begin{align}
&\mathbb{D}^{(1)}_{\text{7D}}=\mathrm{Tor} \big(H_1(\partial Z)\big)|_{\mathrm{triv.}}=\mathrm{Tor} \big(H_1(\partial Z)\big) \,, \\
&\mathbb{D}^{(4)}_{\text{7D}}=\mathrm{Tor} \big(H_2(\partial Z)\big)|_{\mathrm{triv.}}=\mathrm{Tor} \big(H_2(\partial Z)\big) \,, \\
& \implies \quad \quad \mathbb{D}^{(1)}_{\rm 7D}=\mathbb{D}^{(4)}_{\rm 7D} \,,
\end{align}
where the last line follows from Poincar\'e duality of $\partial Z$. This means that we are left with the simpler task of just calculating $\mathrm{Tor}\big(H_1(Z)\big)$ to derive the higher-form symmetries. Note that in all of these examples, we will have $H_2(Z)=\mathbb{Z}=\langle [\mathbb{E}_z] \rangle$ since $H_1(Y)$ is trivial and the monodromy action $\rho_2$ on $H_2(Y)$ is the identity.

For $r=1/2$, $Y$ is a $IV^*$ surface which satisfies $\mathrm{Tor}\big(H_1(\partial Y)\big)=\mathbb{Z}_3$. Ones sees this from
\begin{equation}
    \mathrm{coker}(\rho_1 - 1)=\mathrm{coker}\begin{pmatrix}
        -2 & -1 \\
        1 & -1
    \end{pmatrix}=\mathbb{Z}_3 \,,
\end{equation}
whereby we can choose to represent the generator of the $\mathbb{Z}_3$ in any of the following ways:
\begin{equation}
    \begin{pmatrix}
        1\\
        0
    \end{pmatrix}\sim \begin{pmatrix}
        0\\
        -1
    \end{pmatrix}\sim \begin{pmatrix}
        -1\\
        1
    \end{pmatrix}\sim \begin{pmatrix}
        1+3M\\
        0
    \end{pmatrix}\equiv 1 \; \mathrm{mod} \; 3 \in \mathbb{Z}_3 \,.
\end{equation}
The action of $\rho_1$ is then
\begin{equation}
    \begin{pmatrix}
        0 & -1\\
        1 & 1
    \end{pmatrix}\begin{pmatrix}
        1 \\
        0
    \end{pmatrix}=\begin{pmatrix}
        0 \\
        1
    \end{pmatrix}\sim \begin{pmatrix}
        2 \\
        0
    \end{pmatrix}\equiv 2 \; \mathrm{mod} \; 3 \,,
\end{equation}
which means that the $\mathbb{Z}_2$ in the quotient $Z=(Y \times S_{\theta}^1)/\mathbb{Z}_2$ acts by an automorphism on $\mathbb{Z}_3$. This means that $(\rho_1-1)$ is the identity map which has trivial cokernel therefore
\begin{equation}
  \mathbb{D}_{\text{7D}}^{(1)}\big(\mathfrak{e}^{(1/2)}_8 \big)=\mathbb{D}_{\text{7D}}^{(4)}\big(\mathfrak{e}^{(1/2)}_8\big)=0 \,.
\end{equation}

When $r=1/3$, $Y$ is an $I^*_0$ and $\mathrm{Tor}\big(H_1(Y)\big)=\mathbb{Z}^2_2$. The $\mathbb{Z}_3$ action of $\rho_1$ is given by
\begin{equation}
    \begin{pmatrix}
        0 & -1\\
        1 & 1
    \end{pmatrix}\begin{pmatrix}
    x \; \mathrm{mod} \; 2\\
     y \; \mathrm{mod} \; 2
\end{pmatrix}=\begin{pmatrix}
    -y \; \mathrm{mod} \; 2\\
     x+y \; \mathrm{mod} \; 2
\end{pmatrix} \,,
\end{equation}
which is none other than the well-known triality automorphism on the Kleinian group. One can easily see that $\mathrm{im}(\rho_1-1)=\mathbb{Z}^2_2$ so $\mathrm{coker}(\rho_1-1)=0$ and thus
\begin{equation}
  \mathbb{D}_{\text{7D}}^{(1)}\big(\mathfrak{e}^{(1/3)}_8\big)=\mathbb{D}_{\text{7D}}^{(4)}\big(\mathfrak{e}^{(1/3)}_8\big)=0 \,.
\end{equation}

When $r=1/4$, $Y$ is a $IV$ surface which means that $\mathrm{Tor}\big(H_1(\partial Y)\big)=\mathbb{Z}_3$. The discussion is identical with that of the $r=1/2$ case since the monodromy matrices satisfy $\rho_{IV}\rho_{IV^*}=1$ which in particular implies that $\rho_{IV}$ and $\rho_{IV*}$ are related by similarity transformations so all of the linear algebra analysis in the $r = 1/2$ case carries over. Therefore we have that
\begin{equation}
  \mathbb{D}_{\text{7D}}^{(1)}\big(\mathfrak{e}^{(1/4)}_8\big)=\mathbb{D}_{\text{7D}}^{(4)}\big(\mathfrak{e}^{(1/4)}_8\big)=0 \,.
\end{equation}
The only geometric caveat is that $\mathbb{Z}_2\simeq \mathbb{Z}_4/\mathbb{Z}_2$ of the $\mathbb{Z}_4$ quotient in $Z=(Y\times S_{\theta}^1)/\mathbb{Z}_4$ acts effectively on $\mathrm{Tor}\big(H_1(\partial Y)=\mathbb{Z}_3\big)$ but this does not affect the topology at all\footnote{I.e. we can get an identical topological space by instead quotienting the $IV$ singularity $Y$ by the relations $\theta \sim \theta +\frac{2\pi}{2}, \; \; \; \phi \sim \phi+\frac{2\pi }{2}, \; \; \; \mathbb{E}_z \sim \rho(\mathbb{E}_z)$}.

Finally, when $r=1/6$, $Y$ is a type $I_0$ surface and our defect group calculation is similar to the $\mathfrak{e}^{(1/3)}_6$ and $\mathfrak{e}^{(1/4)}_7$ cases above. The monodromy action on $H_1(Y)$ around $S^1_\theta/\mathbb{Z}_6$ is
\begin{equation}
   \rho_1= 1\oplus \begin{pmatrix}
        0 & -1\\
        1 & 1
    \end{pmatrix} \,,
\end{equation}
and $(\rho_1-1)$ has a Smith normal form
\begin{equation}
  \mathrm{SNF}( \rho_1-1)= 0\oplus \begin{pmatrix}
        1 & 0\\
        0 & 1
    \end{pmatrix} \,.
\end{equation}
It follows that $\mathrm{ker}(\rho_1-1)=\mathbb{Z}$ and $\mathrm{coker}(\rho_1-1)=0$ which means that $H_1(\partial Z)=H_2(\partial Z)=\mathbb{Z}$. The defect group is thus trivial in this case:
\begin{equation}
\mathbb{D}^{(1)}_{\rm 7D}\big(\mathfrak{e}^{(1/6)}_8\big)=\mathbb{D}^{(4)}_{\rm 7D}\big(\mathfrak{e}^{(1/6)}_8\big)=0 \,.
\end{equation}

We see that the geometrical calculation in the dual F-theory frame perfectly reproduces the 1- and 4-form symmetries predicted by an application of the local freezing map in Section \ref{ssec:defectgrpsfreezemap}.

\section{SymTFTs of Frozen Singularities}
\label{sec:symtft}

Having given a detailed analysis of defects charged under 1-form and 4-form symmetries of 7D theories on frozen singularities, $\mathcal{T}^{(\mathrm{M})}[\mathfrak{g}^{(n/d)}]$, we now turn to the more refined information present in their 8D SymTFTs. Such a topological theory of one higher dimension, whereby $\mathcal{T}^{(\mathrm{M})}[\mathfrak{g}^{(n/d)}]$ lives on a boundary, captures the 't Hooft anomalies of the higher-form symmetries. Indeed, such a SymTFT analysis for M-theory on unfrozen $ADE$ singularities was carried out in \cite{Apruzzi:2021nmk} and leads to the usual mixed electric 1-form/magnetic 4-form anomaly (or, mutual non-locality of Wilson and 't Hooft operators) expected of 7D gauge theories (coupled to adjoint gauginos), whose results we will review below.

When applied to frozen singularities, the discrepancy between the topological defects present in the SymTFT and the charged defects we have computed above signals that the boundary degrees of freedom $\mathcal{T}^{(\mathrm{M})}[\mathfrak{g}^{(n/d)}]$ comprises more than just a simple SYM sector.
The novelties we will see come in two flavors:
\begin{enumerate}
    \item A modification in the usual 8D SymTFT boundary conditions along the 7D $\mathcal{T}^{(\mathrm{M})}[\mathfrak{g}^{(n/d)}]$ boundary.
    \item Additional 7D counterterms\footnote{A perhaps modern way of phrasing the procedure of adding topological counterterms is tensoring the 7D theory by a symmetry protected topological (SPT) phase.} localized on the $\mathcal{T}^{(\mathrm{M})}[\mathfrak{g}^{(n/d)}]$ boundary required for consistency with 't Hooft anomalies.
\end{enumerate}
The first type of modification can be derived explicitly in the twisted F-theory frame, using the geometric description on $Z$. There, the mismatch between $H_*(\partial
Z)|_{\mathrm{triv.}}$ and $H_*(\partial Z)$ resulting from a non-trivial ${\rm Tor}\big(H_*(Z)\big)$ can be translated into a topological term on the physical boundary which imposes Neumann boundary conditions for some or all of the SymTFT's 1-form symmetry operators. We will find that these cases of the first type also have additional counterterms which is explained in Appendix \ref{app:fullyfrozen}.

We propose topological modifications of the second type for the $\mathfrak{e}^{(1/2)}_6$ and $\mathfrak{e}^{(1/4)}_8$ singularities. The former is due to the fact that the term in the SymTFT action which characterizes the mixed 1-/4-form 't Hooft anomaly has the opposite sign from what would naively be expected from the 7D gauge theory. As for the latter case, such a topological modification for the 7D theory $\mathcal{T}^{(\mathrm{M})}[\mathfrak{e}_8^{(1/4)}]$ appears to have a trivial 1-form/4-form 't Hooft anomaly, and in fact no notion of global structure for its gauge group despite hosting a $\mathfrak{su}(2)$ gauge multiplet. Such a gauge theory with no notion of global structure is a new field theory phenomenon as far as we are aware.

\begin{figure}
\centering
\includegraphics[scale=0.40, trim = {0cm 0cm 0cm 0cm}]{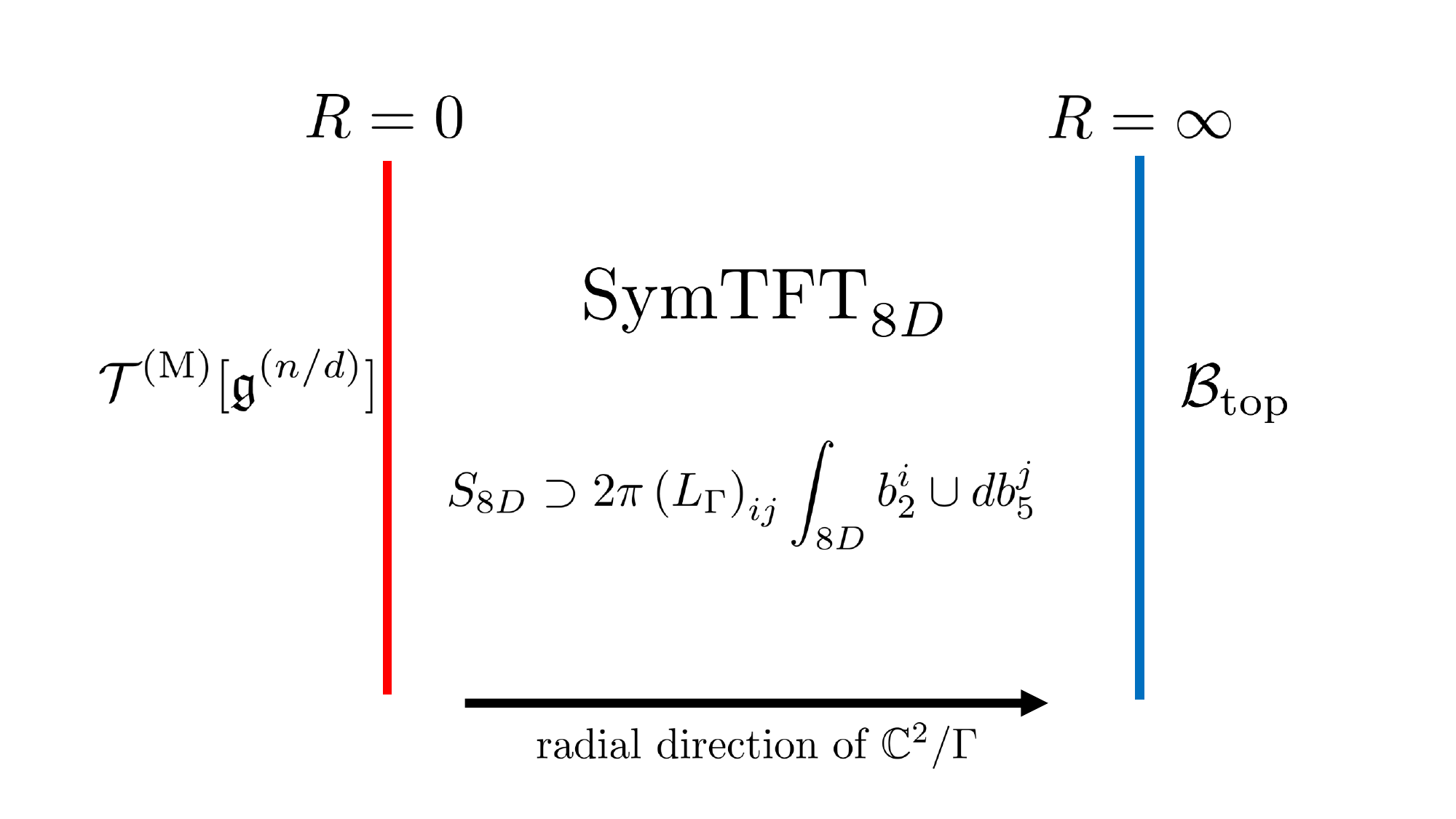}
\caption{8D SymTFT sandwich construction for the 7D theories localized on the $\mathfrak{g}^{(n/d)}$ frozen singularity.}
\label{fig:8Dsymtftgeneral}
\end{figure}

The SymTFTs considered here are topological theories in 8D with one boundary hosting the 7D theory $\mathcal{T}^{(\mathrm{M})}[\mathfrak{g}^{(n/d)}]$, and a gapped topological boundary which we call $\mathcal{B}_{\mathrm{top}}$. The $\mathcal{T}^{(\mathrm{M})}[\mathfrak{g}^{(n/d)}]$ boundary is of course gapless when $\mathfrak{g}^{(n/d)}$ is a partially frozen singularity and is gapped when it is fully frozen. The worldvolume of this 8D theory is $X_7\times [0,1]$ where we take\footnote{The precise relation is immaterial since the 8D SymTFT is topological. } the interval direction to be related to the physical radial direction by
\begin{equation}
    \pi T=2 \,\mathrm{arctan}(R) \,,
\end{equation}
see Figure \ref{fig:8Dsymtftgeneral} for an illustration. One obtains an absolute 7D theory by compactifying the SymTFT on the interval direction, whose partition function is given schematically as
\begin{equation}
    Z^{(\mathcal{B}_{\rm top})}_{\rm 7D}(X_7)=\braket{\mathcal{T}^{(\mathrm{M})}[\mathfrak{g}^{(n/d)}]|\mathcal{B}_{\rm top}} \,.
\end{equation}
Roughly speaking, the data of $\mathcal{B}_{\mathrm{\rm top}}$ determines the global structure of the 7D gauge group which, as we will review shortly, is precisely true for unfrozen singularities.

The 8D bulk topological theory is in principle derived by compactifying M-theory on $\partial X_{\Gamma}$ and truncating to the topological terms which are the leading contributions in the large volume limit\footnote{There is a small neighborhood of the singularity where this large volume limit of $\partial X_{\Gamma}$ breaks down, but these corrections, which include for instance 11D graviton scattering with the 7D degrees of freedom, vanish in the deep IR limit of the 7D field theory.} of $\partial X_{\Gamma}$. When $X_{\Gamma}$ is elliptically-fibered over $\mathbb{C}$, one can identify the radial direction with the radial direction of the base but this will not matter for our purposes because in cases where we need to appeal to the twisted F-theory dual, the volume of the generic fiber of $X_{\Gamma}$ is taken to infinity so it is essentially an $ADE$ singularity.

\subsection{Unfrozen SymTFTs}

Let us first analyze what we expect from a bottom-up field theory perspective for the frozen singularities. In other words, let us just assume we have a 7D SYM theory with some simple gauge algebra $\mathfrak{g}$. This will doubly serve as a review of the SymTFT construction for 7D gauge theories. If we let $G$ denote the associated simply-connected group, then this gauge theory is a priori a relative theory with a defect group \cite{Freed:2012bs, Aharony:2013hda, DelZotto:2015isa, Albertini:2020mdx}
\begin{equation}
\mathbb{D}=Z(G)^{(1)} \oplus Z(G)^{(4)} \,,
\end{equation}
whose charged objects are Wilson/'t Hooft defects whose representations have non-trivial highest weights/coweights in the quotients $\Lambda^{\mathfrak{g}}_{\mathrm{wt}}/\Lambda^{\mathfrak{g}}_{\mathrm{root}}$ and $\Lambda^{\mathfrak{g}}_{\mathrm{cowt}}/\Lambda^{\mathfrak{g}}_{\mathrm{coroot}}$, respectively. These objects are charged under 1-form/4-form symmetries which can be coupled to background $Z(G)$-valued gauge fields $b_2$ and $b_5$, respectively. These symmetries have a mixed 't Hooft anomaly which is captured by a term in the 8D SymTFT action of the form
\begin{equation}\label{eq:usualmixedanom}
S_{\text{BF}} = 2\pi (L_{\Gamma})_{ij} \int_{\rm 8D} b^i_2 \cup \delta b^j_5 \,.
\end{equation}
Note that for $\mathfrak{g}=\mathfrak{so}(4k)$ the center symmetry has two factors and one needs two linearly independent background fields with a $(2 \times 2)$ matrix $(L_{\Gamma})_{ij}$. In all other cases, to which we will restrict not to clutter notation, $i = j$ and $L_{\Gamma}$ becomes a number.

\begin{figure}
\centering
\includegraphics[scale=0.40, trim = {0cm 0cm 0cm 0cm}]{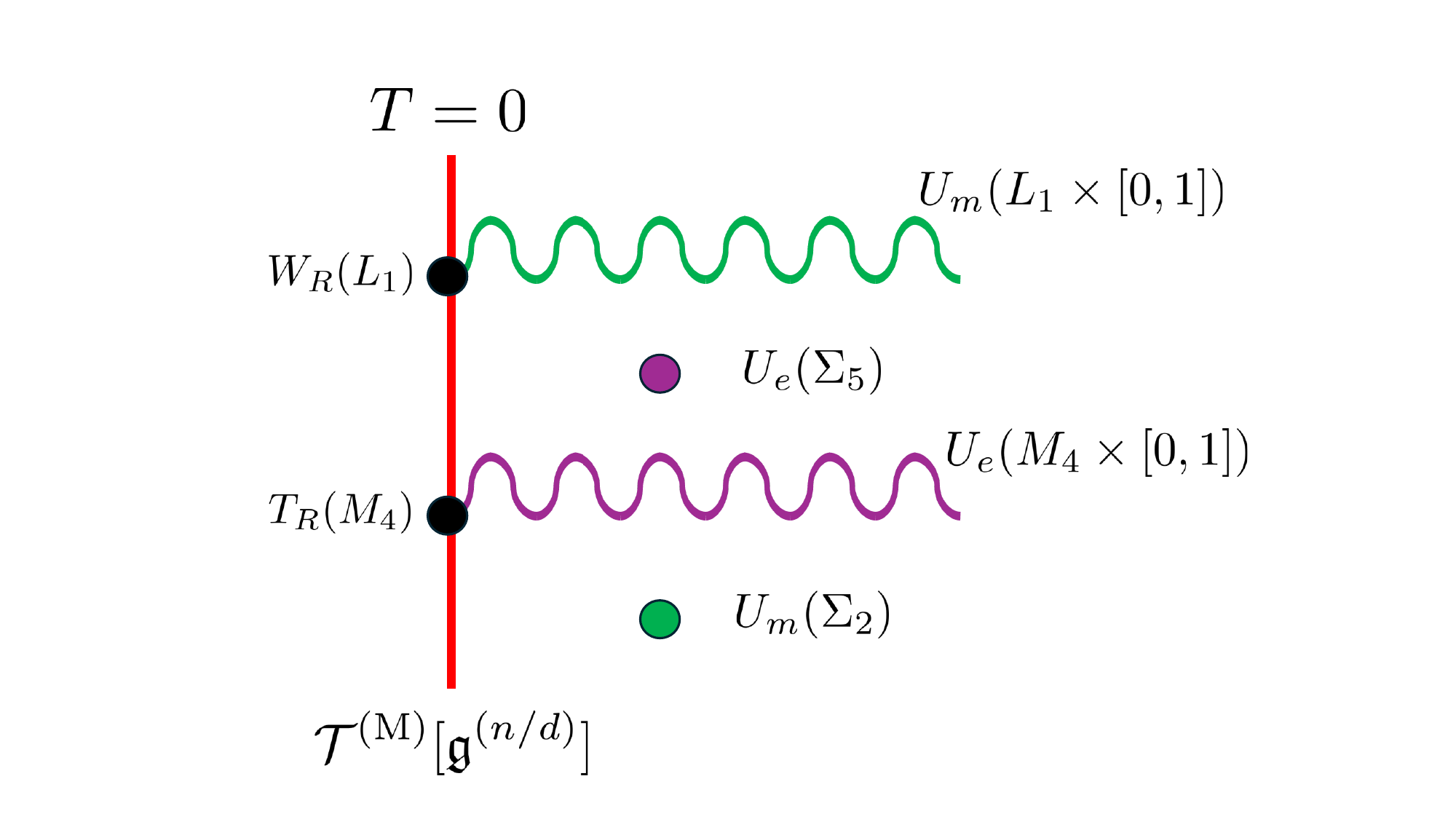}
\caption{Illustration of the $\mathcal{T}^{(\mathrm{M})}[\mathfrak{g}^{(n/d)}]$ boundary at $T=R=0$. The $U_m$ (green) and $U_e$ (purple) operators of the SymTFT can both end on this boundary on a Wilson line and 't Hooft defect, respectively. When $U_m$ / $U_e$ are placed parallel to the $T=0$ boundary they appear as topological symmetry operators for the 4-form/1-form global symmetries. }
\label{fig:8Dsymtfttphys}
\end{figure}

The coefficients $L_{\Gamma}$ have a natural interpretation in the 8D SymTFT as the link pairing of Wilson surface operators for the $b_2$ and $b_5$ dynamical gauge fields. These are defined by
\begin{equation}
U_{m} (\Sigma_2) = \text{exp} \bigg( i \int_{\Sigma_2} b_2 \bigg) \,, \quad U_e (\Sigma_5) = \text{exp} \bigg( i \int_{\Sigma_5} b_5 \bigg) \,,
\end{equation}
and have the algebraic relation
\begin{equation}\label{eq:usualtopops}
U_m (\Sigma_2) U_e (\Sigma_5)= \text{exp} \big( 2 \pi i \,  L_{\Gamma} \, \text{Link}(\Sigma_2, \Sigma_5) \big) \, U_e (\Sigma_5)U_m (\Sigma_2) \,,
\end{equation}
with $\text{Link}(\Sigma_2,\Sigma_5)$ defined as the linking number of $\Sigma_2$ and $\Sigma_5$ in the eight-dimensional theory with worldvolume $X_7\times [0,1]$. A key feature of the SymTFT paradigm is that these Wilson surface operators can appear as either topological symmetry operators or (generally non-topological) defect operators depending on whether $\Sigma_2$/$\Sigma_5$ are along the interval direction or not, see Figure \ref{fig:8Dsymtfttphys}. For example, when $\Sigma_2=L_1\times [0,1]\subset X_7\times [0,1]$ and $\Sigma_5\subset X_7\times \{ T_0\}$, then the linking amounts to the statement that a codimension-two topological symmetry operators acts on the Wilson line charged under $Z(G)^{(1)}$.

\begin{figure}
\centering
\includegraphics[scale=0.40, trim = {0cm 0cm 0cm 0cm}]{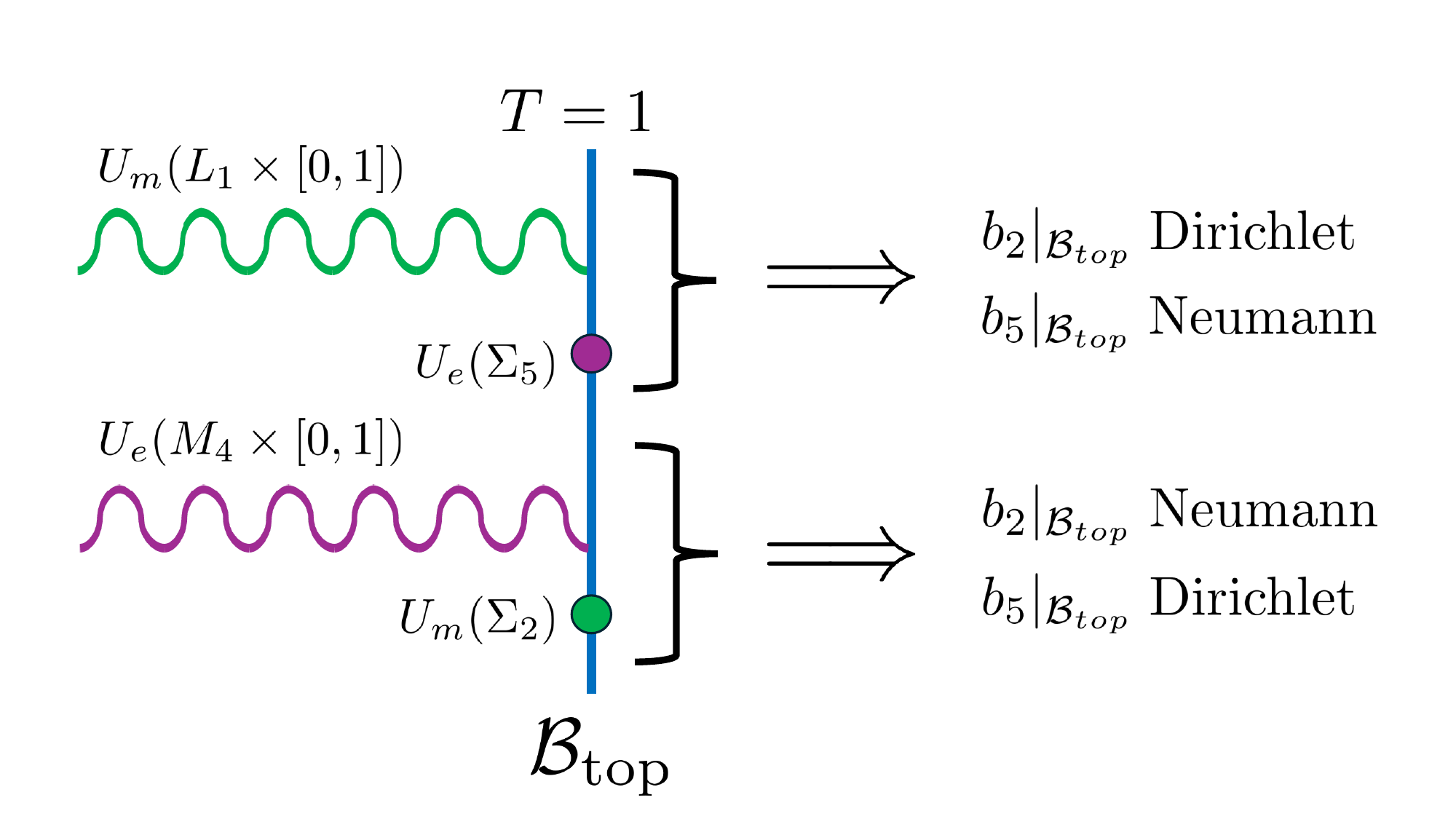}
\caption{Here we show two possible boundary conditions on $\mathcal{B}_{\rm top}$. The top-half showcases ``fully electric" boundary conditions while the bottom-half shows ``fully magnetic" boundary conditions. For all unfrozen SymTFTs these respectively lead to a simply-connected gauge group or 7D gauge group where the full center 1-form symmetry is gauged.}
\label{fig:8Dsymtftbtop}
\end{figure}

While the $U_e$ and $U_m$ operators are both free to end on the $T=0$ boundary because they end on gauge theoretic defect operators of the 7D gauge theory, on the $\mathcal{B}_{\rm top}$ boundary at $T=1$, one must choose a consistent combination of Neumann/Dirichlet boundary conditions for $b_2$ and $b_5$. This is easiest to see if we quantize the SymTFT on constant $T$ slices (or equivalently constant $R$ slices) whereby the possible boundary states $\ket{\mathcal{B}_{\rm top}}$ transform under the algebra \eqref{eq:usualtopops}. In particular one cannot simultaneously diagonalize $U_e$ and $U_m$ but rather a combination of $U^\ell_e$ and $U^k_m$ for some powers $\ell$ and $k$ such that their linking vanishes. For simplicity, we illustrate in Figure \ref{fig:8Dsymtftbtop} the cases where $(\ell,k)=(1,0)$ and $(0,1)$. When $U_m$ is diagonalized on $\ket{\mathcal{B}_{\rm top}}$, then $b_2|_{\mathcal{B}_{\rm top}}$ has a definite value (up to gauge transformations) so we identify this with Dirichlet boundary conditions, moreover the algebra \eqref{eq:usualtopops} implies that $b_5|_{\mathcal{B}_{\rm top}}$ has Neumann boundary conditions. Such a scenario is specifies an ``electric polarization" for the gauge algebra which means that the gauge \textit{group} is simply-connected.

From a top-down point of view, these gauge theoretic objects can also be constructed directly from M-theory for frozen and unfrozen $ADE$ singularities. The $U_e$ and $U_m$ operators are associated with M5 branes and M2 branes, respectively, and they correspond to defects/symmetry operators in the gauge theory depending on whether or not the branes have support along the radial direction or remain far away from the singularity \cite{Apruzzi:2021nmk, Heckman:2022muc}, see also \cite{GarciaEtxebarria:2022vzq, Apruzzi:2022rei}. Indeed the SymTFT action follows from reducing the 11D action of M-theory on $S^3/\Gamma$ and truncating to the topological degrees of freedom, see Appendix B of \cite{Baume:2023kkf}. The coefficients $L_{\Gamma}$ are then deduced from the geometry of the M-theory background, namely by the link pairing on $H_1(S^3/\Gamma, \mathbb{Z})$, which is summarized for the $ADE$ cases in Table \ref{tab:Linkings} which we borrow from \cite{Apruzzi:2021nmk}.

\subsection{Frozen SymTFTs}\label{ssec:fullyfrozen}

With the review material of the previous subsection in place, we are now ready to address a puzzle that arose in our defect group calculations for some of the frozen singularities:

\vspace{0.2cm}
\noindent Whenever $\mathfrak{g}^{(n/d)}=\{ D^{(1/2)}_{k}, \mathfrak{e}^{(1/3)}_6, \mathfrak{e}^{(1/4)}_7\}$, we have the expected line defects from the frozen gauge algebra $\mathbb{D}^{(1)}=Z(H_{\Gamma,d})^{(1)}$, but an ``excess" of four-dimensional defects, i.e.,
\begin{equation}
|\mathbb{D}^{(4)}|>|Z(H_{\Gamma,d})^{(4)}| \,.
\end{equation}

\begin{figure}
\centering
\includegraphics[scale=0.40, trim = {0cm 0cm 0cm 0cm}]{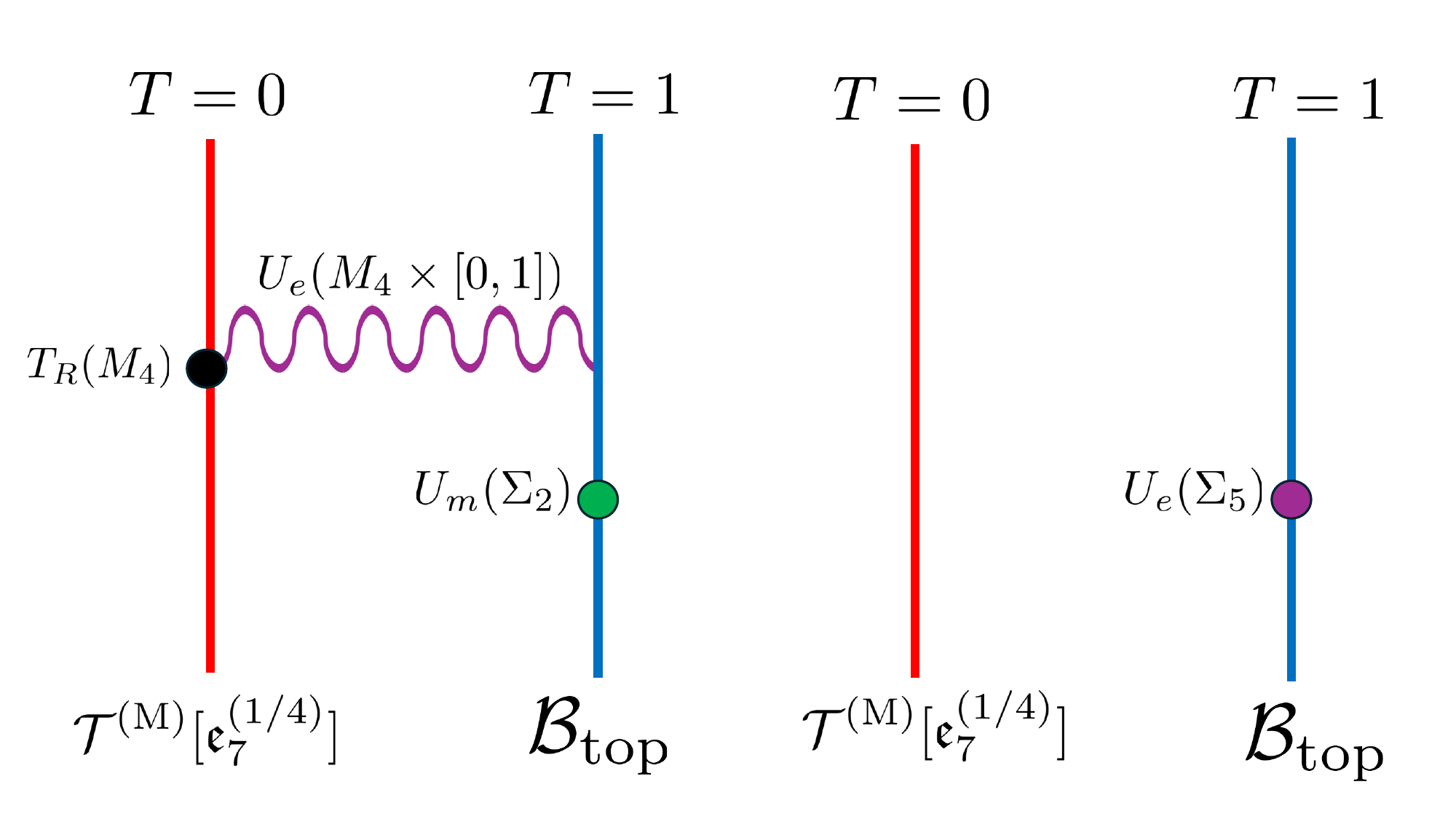}
\caption{The two possible $\mathcal{B}_{\rm top}$ boundary conditions for the $\mathfrak{e}^{(1/4)}_7$ SymTFT. For the fully magnetic boundary condition (left), the 7D theory has a magnetic defect and 4-form symmetry operator, while the fully electric boundary condition (right) admits 1-form symmetry operators which do not act on any defects.}
\label{fig:8Dsymtftexotic}
\end{figure}

To understand the SymTFT interpretation of what is going on, let us for concreteness focus on the $\mathfrak{e}^{(1/4)}_7$ case. Similar remarks will carry over for $D^{(1/2)}_{4}$ and $\mathfrak{e}^{(1/3)}_6$, and we discuss the $D^{(1/2)}_{k>4}$ cases at the end. In this case then, $\mathfrak{h}_{\mathfrak{e}_7, 4}=\varnothing$ so clearly the defect group would be trivial if one were only considering the gauge degrees of freedom. However, as calculated in Section \ref{ssec:outerauto}, the defect group is
\begin{equation}
\mathbb{D}\big(\mathfrak{e}^{(1/4)}_7\big)=\mathbb{D}\big(\mathfrak{e}^{(1/4)}_7\big)^{(1)}\oplus \mathbb{D}\big(\mathfrak{e}^{(1/4)}_7\big)^{(4)}=0^{(1)}\oplus \mathbb{Z}^{(4)}_2 \,.
\end{equation}
We can understand this mismatch by tabulating the symmetry operators and defect operators in the twisted F-theory frame. Recall that the twisted F-theory background further reduced on an $S^1$ leads to M-theory on $Z=(I_0\times S_{\theta}^1)/\mathbb{Z}_4$ where $\mathbb{Z}_4$ acts on the $I_0$ fiber as an $S\in SL(2,\mathbb{Z})$ transformation. An M5 brane on a relative 3-cycle $\Sigma_3\in H_3(Z,\partial Z)$ whose boundary $\partial \Sigma_3$ is the generator of $\mathrm{Tor} \big( H_2(\partial Z) \big)=\mathbb{Z}_2$ engineers the defect operator charged under the 4-form symmetry $\mathbb{Z}^{(4)}_2$ (albeit reduced on the M-/F-theory circle). From the homology computations in Section \ref{ssec:outerauto}, we can conclude that that this relative 3-cycle has a boundary in $H_2(\partial Z)$ which wraps the base circle $S^1_{\phi}$ of $\partial I_0$ and a 1-cycle in the boundary. The precise 1-cycle is only important modulo $(\rho_2-1)$ acting on the 2-cycles of $H_2(\partial I_0)$ with one leg along the base which span a sublattice $\mathbb{Z}^2\subset H_2(\partial I_0)$. In this case
\begin{equation}
    \rho_2=\begin{pmatrix}
    0 & -1 \\
    1 & 0
\end{pmatrix}.
\end{equation}
In F-theory language this defect is a $(p,q)$-fivebrane wrapping the $\mathbb{C}$ base of $I_0$ whose $(p,q)$ charge is non-trivial modulo\footnote{$(p,q)=(1,1)$ is one representative one can use for the $\mathbb{Z}_2$ generator.}
\begin{equation}
    \begin{pmatrix}
        -1 & -1 \\
        1 & -1
    \end{pmatrix} \,.
\end{equation}
Meanwhile, as mentioned in Section \ref{ssec:outerauto}, there is no electric defect operator because the putative one would arise from wrapping M2 branes on $H_2(Z,\partial Z)/H_2(Z)=(\mathbb{Z}\oplus \mathbb{Z}_2)/(\mathbb{Z}\oplus \mathbb{Z}_2)=0$. All together this means for the SymTFT that while $U_e(\Sigma_5)$ can end on the $\mathcal{T}^{(\mathrm{M})}[\mathfrak{e}^{(1/4)}_7]$ boundary in Figure \ref{fig:8Dsymtfttphys} as expected, the $U_m(\Sigma_2)$ operator cannot.
In fact, because the $\mathcal{T}^{(\mathrm{M})}[\mathfrak{e}^{(1/4)}_7]$ boundary is gapped, in perfect analogy to $\mathcal{B}_{\rm top}$ boundary at $T=1$, we simply have another set of topological boundary conditions and the geometry enforces that $b_2|_{T=0}$ is Neumann and $b_5|_{T=0}$ is Dirichlet, see Figure \ref{fig:8Dsymtftexotic}.

Such boundary conditions at $T=0$ are less surprising when we explicitly perform the dimensional reduction of M-theory on $Z$ which is a purely geometrical background; this yields a 6D theory which is the $S^1$-reduction of the 7D theory of interest.
The main ingredient is the fact that $\mathrm{Tor} \big(H^2(Z)\big)= {\rm Tor}\big( H^1(Z; U(1)) \big) = \mathbb{Z}_2 \neq 0$, which leads to a dynamical $\mathbb{Z}_2$-valued 2-form $b^{\rm 6D}_2$ from expanding $C_3$ on ${\rm Tor}\big( H^1(Z; U(1)) \big)$. From the long exact sequence of relative (co-)homology, we have a map ${\rm Tor}\big(H^2(Z)\big) \rightarrow {\rm Tor}\big(H^2(\partial Z)\big)$ from restriction of forms, so this 2-form field of the 6D theory is identified with the 2-form field $b^{\rm 7D}_2$ of the 7D SymTFT which is the reduction of $C_3$ on ${\rm Tor}\big(H^2(\partial Z)\big) = {\rm Tor}\big(H^1(\partial Z, U(1))\big)$, i.e. $b^{\rm 6D}_2 = b^{\rm 7D}_2 |_{T=0}$. We drop the 7D / 6D superscripts below whenever the context is clear.

On the other hand since $\mathrm{Tor}\big(H^3(Z)\big)=0$, we cannot expand $C_6$ to get a dynamical $\mathbb{Z}_2$-valued 4-form which would have been the F-/M-theory circle reduction of a $b^{\rm 7D}_5|_{T=0}$ with Neumann boundary conditions if that cohomology were non-vanishing.
We leave more technical details of this reduction to Appendix \ref{app:fullyfrozen}, where we also explicitly derive a 7D ``counterterm''
\begin{equation}\label{eq:fullyfrozencounterterm}
    S^{\mathrm{c.t.}}_{\rm 7D}=\pi \int_{T =0} b_2\cup b_5 \,,
\end{equation}
localized on the $\mathcal{T}^{(\mathrm{M})}[\mathfrak{e}^{(1/4)}_7]$ boundary.
In fact, it is easy to see that the equations of motion for the bulk-boundary system $S = S_{\rm 7D}^{\rm c.t.} + S_{\rm BF}$, with $S_{\rm BF} = \pi \int_{\rm 8D} b_2 \cup \delta b_5$ the SymTFT derived from the geometry of $\partial Z$, sets $b^{\rm 8D}_5|_{T=0} = b^{\rm 7D}_5$.
So this counterterm ``implements'' the boundary conditions at $T=0$ given the Neumann boundary conditions of $b_2$.\footnote{Analogous conclusions also apply to unfrozen M-theory models on complex surfaces $X^\circ$ with ${\rm Tor}\big(H_1(X^\circ)\big) \neq 0$, which was relevant in understanding the SymTFT in the context of 7D supergravity models and their global structure \cite{Cvetic:2023pgm, Gould:2023wgl}.}

Addressing now the symmetry operators, in terms of M-theory on $Z$, the electric 1-form symmetry operators are given by wrapping M5 branes on $\mathrm{Tor}\big(H_2(\partial Z)\big)=\mathbb{Z}_2$ and the magnetic 6D 3-form (which lifts to a 7D 4-form in F-theory) symmetry is given by an M2 branes wrapping $\mathrm{Tor}\big(H_2(\partial Z)\big)=\mathbb{Z}_2$. These operators are topological from the point of view of the field theory because they are formerly infinitely far away from the interior of $Z$ \cite{Heckman:2022muc}. These respectively lift to $(p,q)$-fivebranes wrapping the base of $\partial I_0$ and $(p,q)$-strings placed on a point in in the circle base of $\partial I_0$. This picture of symmetry operators is not qualitatively different from the unfrozen singularities: the SymTFT implies that we still have a algebra \eqref{eq:usualtopops} acting on the boundary state $\ket{\mathcal{B}_{\rm top}}$ which means we have to pick a consistent set of boundary conditions for the topological operators. In this case, because $\mathbb{Z}_2$ has no non-trivial subgroups, there are only two either fully electric, or fully magnetic. See Figure \ref{fig:8Dsymtftexotic}, where we illustrate the two possible boundary conditions for $\mathcal{B}_{\rm top}$. Notably, the fully electric boundary conditions implies that there is a 1-forms symmetry defect which does not faithfully act on any operators in the 7D theory.

Our discussion of the $\mathfrak{e}^{(1/4)}_7$ singularity applies equally well to $D^{(1/2)}_{4}$ and $\mathfrak{e}^{(1/3)}_6$ cases if one replaces $\mathbb{Z}_2$ with $\mathbb{Z}^2_2$ and $\mathbb{Z}_3$ throughout and with the counterterm in \eqref{eq:fullyfrozencounterterm} generally taking the form
\begin{equation}\label{eq:7Dcounterterm2}
    S^{\mathrm{c.t.}}_{\text{7D}}=2\pi (L_{\Gamma})_{ij} \int_{\text{7D}} b^i_2\cup b^j_5 \,.
\end{equation}
As for the $D$-type cases with a non-trivial gauge algebra, $D^{(1/2)}_{k>4}$, recall that the defect groups are
\begin{equation}
    \mathbb{D}(D^{(1/2)}_{k>4})=(\mathbb{Z}_2)^{(1)}\oplus (\mathbb{Z}^2_2)^{(4)} \; \;  (\textnormal{$k$ even}) \,, \quad \mathrm{or}\quad (\mathbb{Z}_2)^{(1)}\oplus (\mathbb{Z}_4)^{(4)} \; \; (\textnormal{$k$ odd}) \,.
\end{equation}
For $k$ even or odd, there is a $(\mathbb{Z}_2)^{(4)}$ subgroup which behaves similarly to the fully frozen cases: for one choice of $\mathcal{B}_{\rm top}$ boundary condition, there exists a 4-manifold defect charged under this 4-form symmetry group and for another choice there exists a $(\mathbb{Z}_2)^{(1)}$ symmetry operator which does not act on any defect. Additionally, there is a $(\mathbb{Z}_2)^{(1)}\oplus (\mathbb{Z}_2)^{(4)}$ quotient subgroup of $\mathbb{D}\big(D^{(1/2)}_{k>4}\big)$ for which the expect fundamental Wilson / 't Hooft defects for a $\mathfrak{sp}(k)$ gauge theory are valued in. Generally speaking, this subtlety is required when the 5-manifold $Z$ has non-trivial torsion in first homology, $\mathrm{Tor}\big(H_1(Z)\big)\neq 0$.

\subsection{\texorpdfstring{\boldmath{$\mathfrak{e}^{(1/2)}_6$}}{e6 with r=1/2}}\label{ssec:e126}

One interesting feature of \eqref{eq:usualmixedanom} for the $\mathcal{T}^{(\mathrm{M})}[\mathfrak{e}^{(1/2)}_6]$ theory, which is an $\mathfrak{su}(3)$ gauge theory, is that the 't Hooft anomaly coefficient differs from the $\mathfrak{su}(3)$ gauge theory engineered from M-theory on $\mathbb{C}^2/\mathbb{Z}_3$. We can denote the latter theory by $\mathcal{T}^{(\mathrm{M})}[A^{(0)}_2]$, and we see from Table~\ref{tab:Linkings} that its coefficient for the mixed electric 1-form/magnetic 4-form 't Hooft anomaly is $1/3$ while $\mathcal{T}^{(\mathrm{M})}[\mathfrak{e}^{(1/2)}_6]$ has a coefficient of $-1/3$.

\begin{table}
\begin{center}
\renewcommand{\arraystretch}{1.25}
\begin{tabular}{|| c | c |  c  ||}
\hline
   $\Gamma$ & $Z(G_\Gamma)=H_1(S^3/\Gamma)$ & $L_\Gamma$  \\ [0.5ex]
 \hline\hline $\mathbb{Z}_{k}$ & $\mathbb{Z}_{k}$ & $1/k$ \\
\hline $D_{2k}$ & $\mathbb{Z}_2\times \mathbb{Z}_2$ & $\frac{1}{2}\begin{pmatrix}
  k & k-1 \\
  k-1 & k
\end{pmatrix} $\\
\hline $ D_{2k+1}$ & $\mathbb{Z}_4$ & $\frac{2k-1}{4}$ \\
\hline $\Gamma_{E_6}$ & $\mathbb{Z}_3$ & $-1/3$\\
\hline $\Gamma_{E_7}$ & $\mathbb{Z}_2$ & $1/2$\\
\hline $\Gamma_{E_8}$ & $\varnothing$ & 0\\
\hline
\end{tabular}
\end{center}
 \caption{Linking invariants for the 3-manifolds $S^3/\Gamma$. The numerical values in the $L_\Gamma$ column are well-defined in $\mathbb{Q}/\mathbb{Z}$.}\label{tab:Linkings}
\end{table}

To account for this difference, $\mathcal{T}^{(\mathrm{M})}[\mathfrak{e}^{(1/2)}_6]$ must differ from $\mathcal{T}^{(\mathrm{M})}[A^{(0)}_2]$ by some additional topological degrees of freedom. Note that a field theorist would say that the 't Hooft anomaly coefficient of D-dimensional $\mathfrak{su}(3)$ gauge theory would canonically be $1/3$ since this fixes the action a generator, say, the generator of $(\mathbb{Z}_3)^{(1)}$ which is $U_e(M_{D-2})$ in our notation, to act on a Wilson line in the fundamental representation by a phase $\exp(2\pi i /3)$. A phase of $\exp(-2\pi i /3)$ would mean that the fundamental Wilson line appears as the anti-fundamental Wilson line\footnote{One can of course act by charge conjugation (i.e. $\mathfrak{su}(3)$ outer automorphism) to go between these two scenarios. Considering the remarks on the counterterm \eqref{eq:e6counter}, performing a charge conjugation would then appear to shift the theory by this term. This means that these theories have a mixed 't Hooft anomaly between charge conjugation, the electric 1-form, and magnetic 4-form symmetries. Such an anomaly can be derived directly from M-theory from the considerations of \cite{Cvetic:2021maf}.}. This difference just amounts to a shift in the counterterm
\begin{equation}\label{eq:e6counter}
    S^{\text{c.t.}}_{\text{7D}}=\frac{2\pi}{3}\int_{\rm 7D}b_2\cup b_5 \,.
\end{equation}
which indeed inflows into the 8D bulk to give the correct difference in SymTFTs between the $\mathcal{T}^{(\mathrm{M})}[\mathfrak{e}^{(1/2)}_6]$ theory and the $\mathcal{T}^{(\mathrm{M})}[A^{(0)}_2]$. In more modern field theory language, two theories that differ by a local topological counterterm differ by a stacking of a symmetry-protected topological (SPT) phase so our conclusion can then be written schematically as
\begin{equation}\label{eq:spt}
    \mathcal{T}^{(\mathrm{M})}[A^{(0)}_2]=\mathcal{T}^{(\mathrm{M})}[\mathfrak{e}^{(1/2)}_6]\otimes \big(\mathrm{SPT}(b_2,b_5)\big)\otimes (...) \,,
\end{equation}
where $\mathrm{SPT}(b_2,b_5)$ has the action \eqref{eq:e6counter}\footnote{Note that the fact that an SPT is by definition an invertible topological theory, one could invert this equation to $\mathcal{T}^{(\mathrm{M})}[A^{(0)}_2]\otimes \big(\mathrm{SPT}(b_2,b_5)\big)^*=\mathcal{T}^{(\mathrm{M})}[\mathfrak{e}^{(1/2)}_6] \otimes (...)$. }. We have added $(...)$ to  \eqref{eq:spt} to signify that we are ignoring other possible topological sectors which may appear on the (un)frozen singularities such as 7D fractional quantum Hall phases \cite{Heckman:2017uxe}.

\subsection{\texorpdfstring{\boldmath{$\mathfrak{e}^{(1/4)}_8$}}{e8 with r=1/4}}
\label{ssec:e148}

Since the frozen singularity hosts a 7D SYM theory with gauge algebra $\mathfrak{su}(2)$ one expects a SymTFT capturing the mixed 't Hooft anomaly between $\mathbb{Z}_2$ 1-form and 4-from symmetries. However, the geometrical and F-theory calculation shows that the actual SymTFT is trivial, see Table \ref{table:frozendessert}. This means that the 7D $\mathfrak{su}(2)$ theory localized at the singularity must somehow forbid the existence of genuine Wilson and 't Hooft operators in half-integer spin representations of $SU(2)$. We can reconcile these two observations by the inclusion of an additional topological theory in the 8D bulk of the form
\begin{equation}
\Delta S = \pi \int b_2 \cup \delta b_5 \,.
\label{eq:e8counter}
\end{equation}
This term combines with the SymTFT of the $\mathfrak{su}(2)$ gauge theory to produce
\begin{equation}
S_{\text{BF}} + \Delta S =  \pi \int b_2 \cup \delta b_5 +  \pi \int b_2 \cup \delta b_5 = 2 \pi  \int b_2 \cup \delta b_5 \,.
\end{equation}
But this term is trivial, in the sense that it does not produce any phases in linking the extended operators \eqref{eq:usualtopops}. At the same time it produces a boundary term on the physical boundary, that takes the form of a counterterm only depending on background fields
\begin{equation}
S^{\text{c.t.}}_{\text{7D}} = \pi \int_{\text{7D}} b_2 \cup b_5 \,.
\label{eq:counterterm}
\end{equation}
This counterterm is responsible for breaking the 1- and 4-form symmetries. For that one simply realizes, that the system of boundary theory and SymTFT is invariant under 1-form and 4-form gauge transformations, thus after the inclusion of $S^{\text{c.t.}}_{\text{7D}}$ the variation of the complete system can be deduced by the properties of the counterterm alone. With the variations given by
\begin{equation}
b_2 \rightarrow \delta \lambda_1 \,, \quad b_5 \rightarrow \delta \lambda_4 \,,
\end{equation}
we see that \eqref{eq:counterterm} is not invariant under 4-form symmetries for non-trivial $b_2$ and not invariant under 1-form symmetries for non-trivial $b_5$.

To better understand the effect of adding such a counterterm to the $\mathfrak{su}(2)$ gauge theory, let us consider a similar such counterterm in a $U(2)$ gauge theory as well as what happens after decoupling the overall $U(1)$\footnote{This is similar to the approach of \cite{Kapustin:2014gua} and \cite{Gaiotto:2014kfa} which also studied gauge theories with a $\mathfrak{su}(N)$ Lie algebra via a $U(N)$ gauge theory.}. We parameterize the $U(2)$ connection in terms of its $SU(2)$ and $U(1)$ parts as\footnote{In terms of the equivalence $U(2)=(SU(2)\times U(1))/\mathbb{Z}_2$, $a/2$ is the connection for the $U(1)$ factor in the numerator.}
\begin{equation}
    \mathcal{A}=A+\tfrac{1}{2}a \mathbbm{1}_{2\times 2} \,,
\end{equation}
where $f=da=\mathrm{Tr}(\mathcal{F})$ is quantized such that $\int_{\Sigma_2} f\in \mathbb{Z}$. The (bosonic) action for the SYM theory is
\begin{equation}\label{eq:suymact}
    S_{\rm YM}= 2\pi \int \mathrm{Tr}(\mathcal{F}\wedge *\mathcal{F})+2 C_5\wedge f+2 B_2\wedge *f \,,
\end{equation}
where we have turned on the background fields for the $U(1)^{(1)}_e$ electric and $U(1)^{(4)}_m$ magnetic symmetries. The former acts on the fields as
\begin{align}\label{eq:el1}
    a &\rightarrow a-2\lambda_1 \,, \\
    \label{eq:el2} B_2 &\rightarrow B_2+d\lambda_1 \,.
\end{align}
The latter meanwhile can be expressed in terms of the electromagnetic dual 4-form of the $U(1)$ connection, $d\widetilde{a}\equiv *f$, as
\begin{align}
    \widetilde{a} &\rightarrow \widetilde{a}-2\lambda_4 \,, \\
    C_5 &\rightarrow C_5+d\lambda_4 \,.
\end{align}
The electric and magnetic topological symmetry operators are $U^{(e)}_{\eta}=\exp(i\eta  \int *f)$ and $U^{(m)}_{\theta}=\exp(i\theta \int f)$, respectively.

The fact that the Lagrangian \eqref{eq:suymact} is neither invariant under $U(1)^{(1)}_e$ gauge transformations when $C_5\neq 0$ nor under $U(1)^{(4)}_m$ gauge transformations when $B_2\neq 0$ is of course due to the mixed $U(1)_e^{(1)}$-$U(1)^{(4)}_m$ anomaly with a 9-form anomaly polynomial proportional to $dB_2\wedge dC_5$. In 8D SymTFT terms, this means we can consider a coupled 7D-8D system with action
\begin{equation}
    S_{\rm YM}+ \int_{M_7\times [0,1]} \left( 2 B_2\wedge dC_5\right) \,,
\end{equation}
which is simultaneously invariant under $U(1)^{(1)}_e$ and $U(1)^{(4)}_m$ gauge transformations. The action of the 8D piece reproduces the 9-form anomaly polynomial after taking an exterior derivative and contributes to anomaly in-flow via a 7D term
\begin{equation}
    S_{\mathrm{inflow}}=2\int_{M_7}B_2\wedge C_5 \,,
\end{equation}
which means $S_{\rm YM}+S_{\mathrm{inflow}}$ is invariant. Adding a counterterm
\begin{equation}
    S_{\text{c.t.}} =-2\int_{M_7}B_2\wedge C_5 \,,
\end{equation}
would mean that $S_{\rm YM}+S_{\mathrm{inflow}}+S_{\text{c.t.}}$ is no longer invariant under generic gauge transformations of the $U(1)$ higher-form global symmetries.

If we now make $B_2$ dynamical, i.e., the transformations \eqref{eq:el1} and \eqref{eq:el2} become gauged, we obtain an $SO(3)$ gauge theory because $\frac{1}{2}a$ can be now be trivialized. The theory with action $S_{\rm YM}+S_{\mathrm{sym}}$ now has a $\mathbb{Z}_2$ valued background field $C_5$ since the equations of motion for the, now dynamical field, $B_2$ is $2dC_5=0$. $C_5$ is of course the background field form the usual magnetic $\mathbb{Z}^{(4)}_2$ symmetry for a $SO(3)$ gauge theory. The generating topological operator for this symmetry is given by
\begin{equation}\label{eq:magso3topop}
    U^{(m)}_{-1}=\exp\left(\frac{i}{2} \int f\right)=\exp\left(i\int B_2\right) \,,
\end{equation}
where the last equality follows from from the $*f$ equations of motion $df=2dB_2$. On the other hand, the theory with action $S_{\rm YM}+S_{\mathrm{sym}}+S_{\text{c.t.}}$ is not invariant under $C_5$ gauge transformations and therefore does not have a magnetic $\mathbb{Z}^{(4)}_2$ symmetry.

A natural question is then: what defects are allowed in the $SU(2)$ 7D SYM theory with the counterterm $S_{\mathrm{c.t.}}$? Without the counterterm, the $SU(2)$ gauge theory has a genuine Wilson line in the fundamental representation: $W_\mathbf{2}$. We claim that this operator cannot appear in the theory with the added counterterm, and the lattice of genuine Wilson lines is instead generated by $W_{\mathbf{3}}$. In terms of the $U(2)$ gauge fields, the fundamental Wilson line can be expressed as
\begin{equation}
    W^{U(2)}_{\mathbf{2}_{1/2}}=W^{SU(2)}_{\mathbf{2}}\exp\left(\frac{i}{2}\int a\right) \,,
\end{equation}
so we see that under $U(1)^{(1)}_e$:
\begin{equation}
    W^{U(2)}_{\mathbf{2}_{1/2}}\rightarrow W^{U(2)}_{\mathbf{2}_{1/2}}\exp\left(-i\int\lambda_1\right) \,.
\end{equation}
On the other hand, the bosonic action with the counterterm transforms as
\begin{equation}\label{eq:noninv}
    \delta_m(S_{\rm YM}+S_{\mathrm{sym}}+S_{\mathrm{c.t.}})=-2B_5\wedge d\lambda_1 \,,
\end{equation}
which introduces an extra sign in the braiding between $\exp(i\int B_5)$ and $W^{U(2)}_{\mathbf{2}_{1/2}}$. More precisely, recall that the term $B_5\wedge f=B_5\wedge da$ gives the braiding relation\footnote{We have now restored notation indicating the manifolds that the operators are supported on.}
\begin{equation}
    \exp\left(i\int_{\Sigma_5} B_5\right) W^{U(2)}_{\mathbf{2}_{1/2}}(\Sigma_1)=e^{i\pi \mathrm{Link}(\Sigma_5,\Sigma_1)}W^{U(2)}_{\mathbf{2}_{1/2}}(\Sigma_1)\exp\left(i\int_{\Sigma_5} B_5\right) \,,
\end{equation}
where $\mathrm{Link}(\sigma_5,\Sigma_1)\in \mathbb{Z}$ is the linking number. However, the gauge non-invariance \eqref{eq:noninv} implies that we cannot fix the overall sign:
\begin{equation}
    \exp\left(i\int_{\Sigma_5} B_5\right) W^{U(2)}_{\mathbf{2}_{1/2}}(\Sigma_1)=\pm e^{i\pi \mathrm{Link}(\Sigma_5,\Sigma_1)}W^{U(2)}_{\mathbf{2}_{1/2}}(\Sigma_1)\exp\left(i\int_{\Sigma_5} B_5\right)
\end{equation}
From \eqref{eq:magso3topop}, we see that we cannot unambiguously define a linking between $\exp((i/2)\int *f)$ and $W^{U(2)}_{\mathbf{2}_{1/2}}$, but this is contradictory since $W^{U(2)}_{\mathbf{2}_{1/2}}$ should satisfy $\int_{S^5} *f=1/2$ for any $S^5$ that links with it. Therefore we see that $W^{U(2)}_{\mathbf{2}_{1/2}}$ cannot be consistently included in the spectrum of extended operators of the theory.\footnote{One could equivalently phrase this in terms of the quantization of this 7D theory on a spatial worldvolume $S^5\times \mathbb{R}^1$, where we cannot define a $\int_{S^5} *f=1/2$ sector of the Hilbert space in the presence of the counterterm.}

Similar considerations follow for the non-genuine minimal charge 't Hooft defects of the $SU(2)$ theory, as well as the non-genuine fundamental Wilson/genuine minimal charge 't Hooft operators of the $SO(3)$ theory with the added counterterm. In fact, we expect an equivalence of theories
\begin{equation}   \mathcal{T}^{(\mathrm{M})} [{\mathfrak{e}^{(1/4)}_8}]=\left(\textnormal{$SO(3)$ 7D SYM $+$ $S_{\mathrm{c.t.}}$}\right)= \left(\textnormal{$SU(2)$ 7D SYM $+$ $S_{\mathrm{c.t.}}$}\right) \,,
\end{equation}
as they cannot be distinguished either at the level of their extended operators in flat space or their global symmetry operators. We leave the understanding this equivalence on more general manifolds as well as how one proper treatment of the difference in topological sum over different principle bundles to future work.

Finally, we comment that we can apply this same procedure of adding a counterterm to pure non-Abelian gauge theories to completely break their higher-form symmetries. This applies regardless of the dimension or the amount of supersymmetry. Explicitly, consider a $D$-dimensional pure gauge theory (possibly coupled to adjoint matter) with gauge group $G$. We have the higher-form symmetries
\begin{equation}\label{eq:thesesymmetries}
    Z(G)^{(1)}\oplus \pi_1(G)^{(D-3)} \,.
\end{equation}
If we denote $\widetilde{G}$ as the simply-connected cover of $\widetilde{G}$, then the $(D+1)$-dimensional SymTFT has an Lagrangian term proportional to $(L_\mathfrak{g})_{ij} \, b^i_2\cup \delta b^j_{D-2}$ where $b^i_p$ are $p$-forms valued in $Z(\widetilde{G})$ with the index $i$ running over the simple factors of $Z(\widetilde{G})$. Here $(L_\mathfrak{g})_{ij}$ are physically the 1-/4-form mixed 't Hooft anomaly coefficients which depends solely on $\mathfrak{g}$. The symmetries \eqref{eq:thesesymmetries}
are completely broken if we add a counterterm
\begin{equation}
    (L_\mathfrak{g})_{ij}\int_{X_D} b^i_2\cup b^j_{D-2} \,.
\end{equation}
As before this makes the lattice of genuine Wilson / 't Hooft defects such that they have trivial charge under higher-form symmetries, or more specifically, their (co)weight labels are (co)roots. Such a situation could be called a ``trivial polarization" of the gauge algebra.

\section{Applications}\label{sec:applications}

Having understood the defect groups, higher-form symmetries, and SymTFTs of M-theory frozen singularities, we now turn to applying this data to better understand more complicated M-theory backgrounds.

As our first application, we will study M-theory on a $I^*_{4+k}$ Kodaira surface with two frozen $D$-type singularities. Such a geometry, when lifted to F-theory by taking the elliptic fiber to zero size, becomes equivalent to a $k$ D$7$-branes probing a O$7^+$ singularity. For $k=0$, it was conjectured in \cite{Cvetic:2022uuu}, that there is a so-called ``evenness condition" whereby a $(p,q)$-string can end on the O$7^+$ plane only if $p$ and $q$ are even. This condition was crucial in \cite{Cvetic:2022uuu} for providing consistent lattices of electric/magnetic charges for 8D $\mathcal{N}=1$ supergravity theories with non-maximal rank. We will see how this condition arises from the defect group calculation of $D^{(1/2)}_4$ and how this condition is relaxed when $k>0$.

As a second application, we study M-theory on compact K3 manifolds with frozen singularities. From understanding the defect groups, one can understand the global structure of the 7D $\mathcal{N}=1$ gauge group, or equivalently, what representations electric particles and magnetic 3-branes can take with respect to the 7D gauge algebra
\begin{equation}\label{eq:7Dalg}
    \mathfrak{g}_{\mathrm{full}}=\mathfrak{g}_{ADE}\oplus \mathfrak{u}(1)^{b_{2}} =\left(\oplus_i \mathfrak{g}_i\right)\oplus \mathfrak{u}(1)^{b_{2}}.
\end{equation}
Here each $\mathfrak{g}_i$ corresponds to a singularity in the K3 of the form $\mathbb{C}^2/\Gamma_{\mathfrak{g}_i}$, $\mathfrak{g}_{ADE}$ is the non-Abelian part of $\mathfrak{g}_{\mathrm{full}}$, and $b_2$ is the second Betti number of the K3. The corresponding 7D gauge group can then generally be of the form
\begin{equation}\label{eq:7Dgrp}
    G_{\mathrm{full}}=\frac{G_{ADE}/\mathcal{Z} \times U(1)^{b_2}}{\mathcal{Z}'}
\end{equation}
for some simply-connected non-Abelian group $G_{ADE}$, and finite groups $\mathcal{Z}$ and $\mathcal{Z}'$. Concretely we will discuss M-theory on $T^4/\Gamma$ for all cases of $\Gamma$ where frozen singularities appear, as well as an elliptically-fibered K3 with two $II^*$ singularities.

Note that in both of these examples we will requires that the total flux vanishes, i.e.,
\begin{equation}\label{eq:fluxvanishes}
    \int_{X}G_4=0 \,.
\end{equation}
For the first application, this is required in order to have a well-defined F-theory uplift since otherwise the $G_4$ flux becomes a ``KK-flux" with respect to the F-theory circle direction whose radius is inversely proportional to the volume of the M-theory elliptic fiber. For the second this is required in order to have a 7D $\mathcal{N}=1$ background which has a vanishing cosmological constant\footnote{Relaxing this condition would also relax the Ricci-flatness condition of the internal four-manifold.}. Therefore given some collection of $N_{sing.}$ frozen singularities each with their own fractional flux labels $r_i$, we require\footnote{We implicitly consider rational lifts $\hat{r}_i$ which satisfy $\sum^{N_{sing.}}_{i=1} \hat{r}_i = 0$.}
\begin{equation}
    \sum^{N_{sing.}}_{i=1} r_i=0\; \mathrm{mod}\;  1 \,,
\end{equation}
such that the total flux can vanish.

\subsection{Only Even Charge Strings Can End on an O\texorpdfstring{$\mathbf{7^+}$}{7+} Plane}\label{ssec:O7plus}

\begin{figure}
\centering
\includegraphics[scale=0.40, trim = {0cm 1cm 0cm 3cm}]{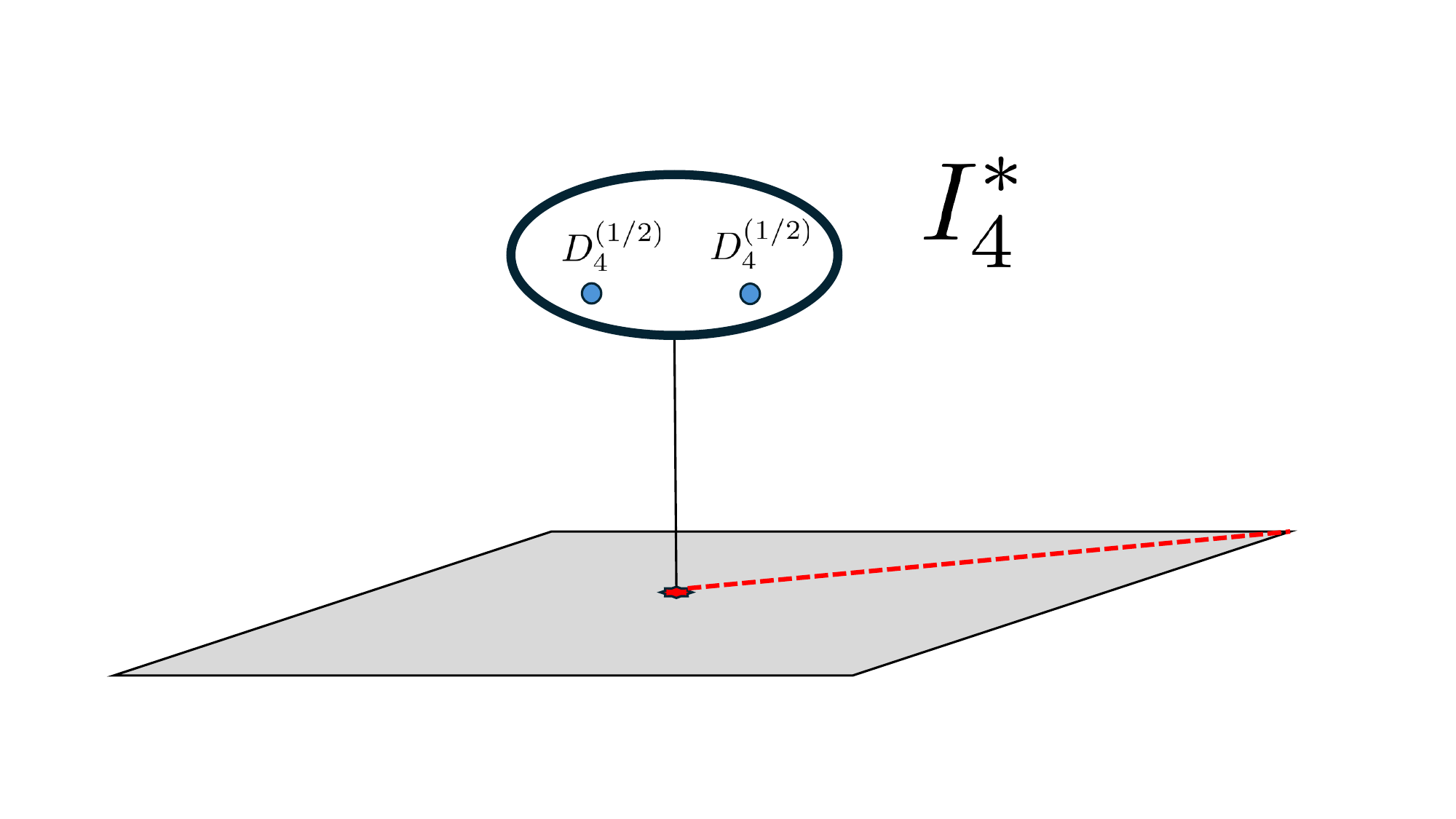}
\caption{Illustration of the elliptic F-theory geometry corresponding to an O$7^+$. Here the $I^*_4$ fiber is such that its central $\mathbb{P}^1$ is resolved (the $\alpha_4$ cycle in the notation of Figure \ref{fig:affine_dynkin_DE}). There are then two $D_4$-type singularities, each with a frozen flux.}
\label{fig:O7plus}
\end{figure}

As mentioned above, an O$7^+$ in F-theory compactified on a circle is equivalent to M-theory on a type $I^*_4$ Kodaira surface with two $D^{(1/2)}_4$ frozen singularities as shown in Figure \ref{fig:O7plus}. Recall from Table \ref{table:frozendessert} that the defect group of a frozen $D^{(1/2)}_4$ singularity is
\begin{equation}
    \mathbb{D}(D^{(1/2)}_4)=(\mathbb{Z}_2 \times \mathbb{Z}_2)^{(4)} \,,
\end{equation}
which only has a magnetic 4-form piece while the electric 1-form symmetry is trivial. This means that while M$5$ branes are free to wrap the two generating relative 2-cycles in Table~\ref{table:relative 2-cycles}, M$2$ branes are forbidden to wrap such relative 2-cycles.

We already see intuitively that for the $I^*_4$ fiber in Figure \ref{fig:O7plus}, there can be no M2 branes wrapping the relative 2-cycles of $I^*_4$ fiber which generate the group $\mathbb{Z}^2_2$ since they are made of linear combination of the two $D_4$ relative 2-cycles. Upon lifting to F-theory, this then reproduces the statement that $(p,q)$-strings of odd $p$ or $q$ cannot end on the O$7^+$ plane. More rigorously, the relative 2-cycles of the $I^*_4$ fiber which generate $\mathbb{Z}^2_2$ are given by (using the notation of Figure \ref{fig:affine_dynkin_DE}):
\begin{align}
    \mathfrak{T}_{s} &= \tfrac{1}{2}(\alpha_1+\alpha_3+\alpha_5+\alpha_7) \,,\\
    \mathfrak{T}_c &=\tfrac{1}{2}(\alpha_1+\alpha_3+\alpha_5+\alpha_8) \,.
\end{align}
Considering the decomposition of the $D_8$ affine Dynkin diagram into two $D_4$ Dynkin diagrams connected by a node (whose corresponding cycle is $\alpha_4$), we have that the relative 2-cycles for one of the $D_4$ singularities is\footnote{Note that $\alpha_0$ can be expressed as a linear combination of the other $\alpha_i$ because the elliptic fiber class vanishes: $[\mathbb{E}]\sim 0\in H_2(X_{I^*_4}, \partial X_{I^*_4})/H_2(X_{I^*_4})$.}
\begin{align}
    \mathfrak{T}^{(1)}_{s} &=\tfrac{1}{2}(\alpha_1+\alpha_3) \,,\\
    \mathfrak{T}^{(1)}_c &=\tfrac{1}{2}(\alpha_1+\alpha_0) \,,
\end{align}
while for the second $D_4$ singularity it is
\begin{align}
    \mathfrak{T}^{(2)}_{s} &=\tfrac{1}{2}(\alpha_5+\alpha_7) \,,\\
    \mathfrak{T}^{(2)}_c &=\tfrac{1}{2}(\alpha_5+\alpha_8) \,.
\end{align}
By the defect group calculation before, M2 branes are forbidden to wrap $\mathfrak{T}^{(i)}_{s}$
or $\mathfrak{T}^{(i)}_{c}$. We then obtain the evenness from the relation $\mathfrak{T}_{s}=\mathfrak{T}^{(1)}_{s}+\mathfrak{T}^{(2)}_{s}$ and $\mathfrak{T}_{c}=\mathfrak{T}^{(1)}_{s}+\mathfrak{T}^{(2)}_{c}$.

If one instead considers an M-theory compactification on a $I^*_{4+k}$ singularity where $k>0$, then we can consider a central fiber with $D^{(1/2)}_4$ and $D^{(1/2)}_{4+k}$ frozen singularities\footnote{If we more generally consider, a $D^{(1/2)}_{k_1}$ and $D^{(1/2)}_{k_2}$ singularity where $k_1+k_2=8+k$ then in F-theory language this involves turning on Wilson lines along the F-theory circle direction for the D7 brane stack probing the O$7^+$ \cite{Tachikawa:2015wka}.}. In F-theory this system consists of $k$ D7-branes probing the O$7^+$. Using the fact that the $D^{(1/2)}_{4+k}$ singularity has a non-trivial 1-form symmetry piece when $k>0$, we see that the evenness condition no longer holds as an M2 brane can wrap a $I^*_{4+k}$ relative 2-cycle that ends on the $D^{(1/2)}_{4+k}$ singularity. Concretely, let us take $k$ to be even, then the relative 2-cycles for the $D^{(1/2)}_{4+k}$ singularity are
\begin{align}
    \mathfrak{T}^{(2)}_{s} &=\tfrac{1}{2}(\alpha_5+\alpha_7+...+\alpha_{k+1}+\alpha_{k+3}) \,,\\
    \mathfrak{T}^{(2)}_c &=\tfrac{1}{2}(\alpha_5+\alpha_7+...+\alpha_{k+1}+\alpha_{k+4}) \,.
\end{align}
The $I^*_{4+k}$ relative 2-cycles $\mathfrak{T}_{s}$ and $\mathfrak{T}_{c}$ are given as in Table \ref{table:relative 2-cycles}, and we have the relation $\mathfrak{T}_{s}+\mathfrak{T}_{c}=\mathfrak{T}^{(2)}_{s}+\mathfrak{T}^{(2)}_{c}$. Therefore, M2 branes are allowed to wrap a $\mathbb{Z}_2$ subgroup of the possible $\mathbb{Z}^2_2$ group of $I^*_{4+k}$ relative 2-cycles. These are none other than the invariant combination of $(p,q)$ strings which end on this 7-brane system which appear as Wilson lines for the $\mathfrak{sp}(k)$ gauge theory with a non-trivial charge under $\mathbb{Z}^{(1)}_2$, which is consistent with \cite{Cvetic:2022uuu}.

In the M-theory frame the M5 branes are not restricted and can end on the frozen singularities. For F-theory this lifts to the fact that the $(p,q)$-5-branes do not have to satisfy an evenness condition, which was crucial for the coroot and coweight lattices discussed in \cite{Cvetic:2022uuu}. We therefore find a microscopic derivation of the evenness conditions from our results about defects groups on frozen M-theory singularities.

Finally, we comment that while we have understood how the defect groups of the individual frozen singularities ``glue together" to understand the defect group of a larger one, we can also apply the SymTree formalism of \cite{Baume:2023kkf} to glue together the SymTFT data of the individual singularities. This may be useful for understanding the topological sectors and SymTFT of F-theory 7-branes, but we leave this for future work.

\subsection{M-theory on a Compact K3 with Frozen Singularities}

We first cover the cases of M-theory compactified on a compact K3 manifold of the form $X=T^4/\Gamma$ with frozen singularities. Since we require $X$ to have $D$- and/or $E$-type singularities, we are restricted to the cases of $\Gamma$ being non-Abelian which are classified to be $D_4$, $D_5$, or $\Gamma_{E_6}$ (the binary tetrahedral group) \cite{19881}. There are, however, three different consistent choices for the action of $D_4$, only one for $D_5$, and two for $\Gamma_{E_6}$. The $ADE$ singularities present in each of these cases is tabulated in Table \ref{tab:ADET4}. Due to the requirement that the total flux vanishes \eqref{eq:fluxvanishes} and the lack of $A$-type frozen singularities, there are essentially only four possible configurations of frozen singularities which we also list in Table \ref{tab:ADET4}.

\begin{table}
\centering
\begin{tabular}{|c||c|c|}
\hline
 $X$ &  $ADE$ Singularities & Frozen Singularity Configuration(s) \\
 \hline \hline
$T^4/D_4$ &  $D^2_4\oplus A^3_3\oplus A^2_1$  &  $D^{(1/2)}_4\oplus D^{(1/2)}_4\oplus A^3_3\oplus A^2_1$\\
\hline
$T^4/D^{\prime}_4$ & $D^4_4\oplus  A^3_1$  & $D^{(1/2)}_4\oplus D^{(1/2)}_4\oplus D^2_4\oplus  A^3_1$ \\
&  & $\left(D^{(1/2)}_4\right)^4\oplus  A^3_1$ \\
\hline
$T^4/D^{\prime \prime}_4$ &  $A^6_3\oplus A_1$ & None  \\
\hline
$T^4/D_5$ & $D_5\oplus A^3_3\oplus A^2_2\oplus A_1$ & None  \\
\hline
$T^4/\Gamma_{E_6}$ & $\mathfrak{e}_6\oplus D_4\oplus A^4_2\oplus A_1$ &  $\mathfrak{e}_6^{(1/2)}\oplus D^{(1/2)}_4\oplus A^4_2\oplus A_1$ \\
\hline
$T^4/\Gamma_{E_6}^{\prime}$ &  $A_5\oplus A^2_3\oplus A^4_2$ & None \\ \hline
\end{tabular}
\caption{Collection of $ADE$ singularities for each possible $T^4/\Gamma$ such that $\Gamma$ is non-Abelian and their possible configuration of frozen singularities. The primes are used to differentiate the different groups actions on $T^4$.}
\label{tab:ADET4}
\end{table}

Firstly, note that M2 and M5 branes wrapped on $H_2(X)$ are stable electric/magnetic states under the full 7D gauge group $G_{\mathrm{full}}$. In our conventions we take $G_{\mathrm{full}}$ to include the $U(1)$ factors from graviphotons so it will be of rank 22 for M-theory on K3 with no frozen singularities, see \cite{Cvetic:2023pgm} for more details. The fact that these M2 brane states lead to the complete set of electrically charge particle states implies that $Z(G_{\mathrm{full}})^\vee=H_2(X)$, where the $\vee$ denotes Pontryagin dual. The analogous statement for M5 branes implies\footnote{Together this means that the 7D gauge group satisfies the particular relation $Z(G_{\mathrm{full}})^\vee=\pi_1(G_{\mathrm{full}})$ which is called a ``maximally mixed polarization" \cite{Cvetic:2023pgm, Gould:2023wgl}. We will see that in the frozen cases, this relation need not hold in general.} $\pi_1(G_{\mathrm{full}})=H_2(X)$ using the fact that $\pi_0(G_{\mathrm{full}})=1$.

While these relations give us some information on the global structure of the 7D gauge groups, for deriving the full data, we employ the methods of \cite{Cvetic:2023pgm} which addresses this problem for M-theory on K3 manifolds and spells out the details for $X=T^4/\Gamma$ in particular\footnote{See also \cite{Gould:2023wgl} for similar work in the context of IIB.}.
The general procedure of \cite{Cvetic:2023pgm} is to consider the decomposition of $X$ into open sets $X^\circ:=X\backslash \mathcal{S}$ and $U_{\mathcal{S}}$ where $\mathcal{S}$ is the collection of singular points of the K3 manifold and $U_{\mathcal{S}}$ is a disjoint union of open patches around each singularity. Topologically $\partial U_{\mathcal{S}}$ is equivalent to a disjoint union of spaces of the form $S^3/\Gamma_i$ where each $\Gamma_i$ characterizes an $ADE$ singularity in the K3. Next, consider the Mayer--Vietoris (MV) sequence with respect to this decomposition:
\begin{equation}\label{eq:mvseq}
    0 \; \rightarrow \; H_2(X^\circ)\; \xrightarrow[]{j_2} \;  H_2(X)\; \xrightarrow[]{\partial_2} \;  \oplus_i H_1(S^3/\Gamma_i )\; \xrightarrow[]{\iota_1}  \; H_1(X^\circ)\; \xrightarrow[]{j_1} \;  0 \,,
\end{equation}
where we have implicitly used the fact that $H_1(X)=0$ and $H_k(U_{\mathcal{S}})=0$ for $k>0$. The maps on the homology groups descend from the embeddings $j:X^\circ\hookrightarrow X$ and $\iota: \partial U_{\mathcal{S}} \hookrightarrow X^\circ$, as well as the boundary map on $k$-cycles, $\partial_k$. The main idea of \cite{Cvetic:2023pgm} was to extract $\mathcal{Z}$ and $\mathcal{Z}'$ of \eqref{eq:7Dgrp} from the various maps of this exact sequence. Namely, we have that
\begin{equation}\label{eq:zzprime}
    \mathcal{Z} \simeq \mathrm{coker}(\partial_2)=\mathrm{im}(\iota_1) \,, \quad  \mathcal{Z}'\simeq \mathrm{im}(\partial_2)|_{\mathrm{free}}=\mathrm{ker}(\iota_1) \,,
\end{equation}
where we have used the symbol $\simeq$ as those isomorphisms are not generally canonical. The map $\partial_2$ then contains the information of how M2/M5 branes wrapping classes in $H_2(X)$ are charged under $\mathfrak{g}_{ADE}$, and $|_{\mathrm{free}}$ denotes the restriction to the free part of $H_2(X)$. In particular, the image of $\partial_2$ on M2 brane states tells us what possible charges of the center $Z(G_{ADE}/\mathcal{Z})$ that the electric particles may take. Conversely, the image of $\partial_2$ on M5 brane states tell us the possible monopole charges which are valued in $\pi_1(G_{ADE}/\mathcal{Z})$. Together this is enough to fully fix $\mathcal{Z}$ by charge completeness\footnote{This is not a conjectural statement as one can derive this statement simply from the careful reduction of the $C_3$ field and analyzing the Mayer--Vietoris sequence for cohomology.} \cite{Polchinski:2003bq}, as well as the group $\mathcal{Z}'$ which tells us how $G_{ADE}/\mathcal{Z}$ mixes with the $U(1)$ factors.

For compact K3 manifolds with frozen singularities, all we have to do is essentially modify the defect groups of the $D$- and $E$-type frozen singularities according to Table \ref{table:frozendessert}. From the frozen configurations in Table \ref{tab:ADET4}, we see that all of these cases involve at least one $D^{(1/2)}_4$ singularity which forbids\footnote{Note that an $\mathfrak{e}^{(1/2)}_6$ singularity also appears but this does not cause any modification as the defect group is the same as the unfrozen $\mathfrak{e}_6$ singularity. } M2 branes from wrapping 2-cycles in $H_2(X)$ which, in a local neighborhood of $D^{(1/2)}_4$, appear as a relative 2-cycle whose boundary is valued in $H_1(S^3/D_4)=\mathbb{Z}^2_2$. In other words, by forbidding cycles in $H_1(S^3/D_4)\subset \oplus_i H_1(S^3/\Gamma_i )$ that an M2 branes can wrap, the representations of electric states of the 7D gauge group will be restricted modifying the center $Z(G_{\mathrm{full}})$, as well as $\mathcal{Z}$ and/or $\mathcal{Z}'$ in \eqref{eq:7Dgrp}. Since we are not restricting M5 branes in any way, we still see that
\begin{equation}\label{eq:pifrozengrp}
    \pi_1(G_{\mathrm{full}})=H_2(X) \,,
\end{equation}
whenever\footnote{One may have $\pi_0(G_{\mathrm{full}})\neq 1$ when $X$ contains $\mathfrak{e}^{(1/3)}_6$ or $\mathfrak{e}^{(1/4)}_7$ frozen singularities because their defect groups are non-trivial, are not modified by the frozen flux, and reduce a simple Lie algebra to zero rank which possibly leaves a discrete gauge group factor in $G_{\mathrm{full}}$ depending on the geometric details.}  $\pi_0(G_{\mathrm{full}})=1$, which holds for all of the $T^4/\Gamma$ cases. As we will see, the relation \eqref{eq:pifrozengrp} will be quite useful in lieu of full knowledge of the full presentation of the all the maps in the MV sequence.

For concreteness, let us work through the derivation of $G_{\mathrm{full}}$ in the case of $T^4/\Gamma_{E_6}$. The MV sequence \eqref{eq:mvseq} in this case is\footnote{For details on how to calculate each of the homology groups appearing in the MV sequence for each of the torus quotient K3 manifolds see \cite{Cvetic:2023pgm}.}
\begin{equation}\label{eq:mvseqe6}
    0 \; \rightarrow \; \mathbb{Z}^3\; \xrightarrow[]{j_2} \; \mathbb{Z}^3\oplus \mathbb{Z}_3\; \xrightarrow[]{\partial_2} \;  \mathbb{Z}_3\oplus \mathbb{Z}^2_2\oplus \mathbb{Z}^4_3\oplus \mathbb{Z}_2\simeq \Z^2_3\oplus \Z^3_6 \; \xrightarrow[]{\iota_1} \; \mathbb{Z}_3\; \xrightarrow[]{j_1} \;  0 \,,
\end{equation}
where we have decomposed the $H_1(\partial U_{\mathcal{S}})$ entry as $Z(E_6)\oplus Z(Spin(8))\oplus Z(SU(3))^4\oplus Z(SU(2))$. From Appendix C of \cite{Cvetic:2023pgm}, we have that in the unfrozen case the full gauge group is:
\begin{equation}
    \big(T^4/\Gamma_{E_6}\big)_{\text{unfrozen}} : G_{\mathrm{full}}=\frac{\left( E_6\times Spin(8) \times SU(3)^4\times SU(2)\right)/\mathbb{Z}_3\times U(1)^3}{\mathbb{Z}^3_6} \,.
\end{equation}
Which from \eqref{eq:zzprime} tells us that $\mathrm{coker}(\partial_2)=\Z_3$ and $\mathrm{Im}(\partial_2)|_{\mathrm{free}}=\Z^3_6$. The freezing of the $D_4$ informs us that M2 branes cannot wrap a cycle in a $\mathbb{Z}^2_2$ subgroup of $\Z^2_3\oplus \Z^3_6$. By exactness, the image of the map $j_2$ is given by $(6^3,0)$ in $\mathbb{Z}^3\oplus \mathbb{Z}_3$ where the superscript denotes multiplicity. Therefore the freezing must restrict M2 branes to wrap states in $\mathbb{Z}^2\subset H_2(X)$ with even charge. This implies that the frozen gauge group is given by
\begin{equation}\label{eq:notappear2}
    \big(T^4/\Gamma_{E_6}\big)_{\text{frozen}}: G_{\mathrm{full}}=\frac{\left( SU(3)\times SU(3)^4\times SU(2)\right)/\mathbb{Z}_3\times U(1)^3}{\mathbb{Z}^2_3\times \mathbb{Z}_6}
\end{equation}
which is consistent with \eqref{eq:pifrozengrp}.

Performing these steps for the other frozen singularity configurations of the various $T^4/\Gamma$ we arrive at
\begin{equation}\label{eq:nonmatch}
    T^4/D_4 \; (\textnormal{with 2$\times D^{(1/2)}_4$}): \; \; G_{\mathrm{full}}=\frac{\left( SU(4)^3\times SU(2)^2\right)/\mathbb{Z}^4_2\times U(1)^3}{\Z^4_2} \,,
\end{equation}
\begin{equation}\label{eq:match}
    T^4/D^\prime_4 \; (\textnormal{with 2$\times D^{(1/2)}_4$}): \; \; G_{\mathrm{full}}=\frac{\left( Spin(8)^2\times SU(4)^3\times SU(2)\right)/\mathbb{Z}^4_2\times U(1)^3}{\Z^3_2} \,,
\end{equation}
\begin{equation}\label{eq:notappear}
    T^4/D^\prime_4 \; (\textnormal{with 4$\times D^{(1/2)}_4$}): \; \; G_{\mathrm{full}}=\left( SU(2)^3/\mathbb{Z}^3_2\right)\times U(1)^3 \,.
\end{equation}
In the context of a heterotic dual of these 7D supergravity theories, several hundred examples of the torsion part of $\pi_1(G_{\mathrm{full}})$ were calculated in \cite{Fraiman:2021soq}. We see from entry 140 of their Table 5, our answer indeed matches $\mathrm{Tor}\; \big(\pi_1(G_{\mathrm{full}}) \big)=\Z^4_2$ for \eqref{eq:match}. However entry 1 of their Table 5 appears to disagree with \eqref{eq:nonmatch} as our answer is $\Z^4_2$ while theirs is $\Z^3_2$. These mismatch may have to do with mixing with the $U(1)$ factors which was not covered in \cite{Fraiman:2021soq}. The remaining cases \eqref{eq:notappear2} and \eqref{eq:notappear} do not appear in the list of possible Lie algebras in \cite{Fraiman:2021soq}, however we point out that this list is not completely exhaustive.

Finally, we mention M-theory on a K3 which is not a torus quotient but rather is an elliptic fibration over $\mathbb{P}^1$ with two type $II^*$ singularities with each host an $\mathfrak{e}_8$ gauge algebra. If we take projective coordinates $[u, v]$ on the $\mathbb{P}^1$ base, then the following Weierstrass model
\begin{equation}\label{eq:e8e8example}
    y^2=x^3+u^4v^4 x+u^5v^7+u^7v^5 \,,
\end{equation}
is an example of such a K3. In this case we can consider frozen configuration $\mathfrak{e}^{(1/4)}_8\oplus \mathfrak{e}^{(3/4)}_8$ which engineers a gauge algebra
\begin{equation}
    \mathfrak{g}_{\mathrm{full}}=\mathfrak{su}(2)\oplus \mathfrak{su}(2) \oplus \mathfrak{u}(1)^6 \,.
\end{equation}
From our remarks in Section \ref{ssec:e148} we see that the gauge group of the non-Abelian factors can be presented as either $SO(3)$ or $SU(2)$ but with the additional counterterm of \eqref{eq:e8counter} which forbids electric/magnetically charged states in the fundamental of either $\mathfrak{su}(2)$ factor. Its heterotic dual is a $T^3$ compactification with a $\mathbb{Z}_4$ almost-commuting triple for which there \textit{do} exist non-BPS perturbative states transforming under the fundamental representation of an $SU(2)$\footnote{We thank Hector Parra de Freitas for pointing this out to us.}. One possibility is that masses of these non-BPS states diverge in the M-theory limit where the string coupling is taken to infinity. Such a scenario would match our M-theory prediction, but we leave such an understanding of this subtle aspect of heterotic/M-theory duality for future work.

\section{Conclusions}\label{sec:conc}

In this paper, we have studied several aspects of M-theory frozen $ADE$ singularities and have seen an interesting interplay between the geometry and the flux data. In contrast to previous works, we have argued how the flux directly reduces the rank of the gauge algebra without appealing to a long-chain of string dualities. We have also studied how the frozen flux can alter the simple dictionary relating the $ADE$ geometry to the symmetries of the 7D theory. In particular, a general feature we found is that while the frozen flux can reduce the amount of line operators with conserved 1-form symmetry charge, the magnetic dual defect operators charged under 4-form symmetries are preserved. We have confirmed such features in multiple dualities frames, and provided a natural description for them in terms of symmetry breaking boundary conditions of an 8D SymTFT.

We also applied the SymTFT framework to understanding subtle symmetry features of of the $\mathfrak{e}^{(1/2)}_6$ and $\mathfrak{e}^{(1/4)}_8$ singularities. In particular, the latter engineers a 7D $\mathfrak{su}(2)$ gauge theory with neither fundamental Wilson nor 't Hooft operators which is a new feature of a pure gauge theory in any dimensions. We made sense of such an exotic theory by adding a counterterm which explicitly broke the higher-form symmetries. Such a procedure, which can be done for any non-Abelian gauge theory of any dimension to reduce the number of naively expected higher-form symmetries, may have pure QFT applications. Additionally, for such theories coupled to gravity, it would be interesting to properly understand the connection between charge completeness and lack of global symmetries, generalizing the considerations in \cite{Heidenreich:2021xpr}.

We also considered geometries with multiple singularites and have applied this symmetry data above to understanding the curious ``evenness condition", a property of O$7^+$ planes, as well as to calculating the global gauge groups of 7D $\mathcal{N}=1$ supergravities for certain points in the rank-reduced moduli space of M-theory on K3 manifolds. In Appendix \ref{app:appA} we also clarified why IIA $ADE$ singularities with a boundary $\int C_1$ monodromy freeze, in contrast to the a conjectural Freed--Witten-like anomaly mentioned in \cite{deBoer:2001wca}, via a confinement mechanism.

There are numerous examples of singular string theory backgrounds which would presumably freeze via a similar mechanisms described in this paper, and considering that there are more $p$-form potentials in string theory, it is clear frozen singularities occupy an understudied yet ubiquitous corner of the string theory landscape. Such backgrounds are in some sense vast generalizations of discrete torsion orbifolds \cite{Vafa:1994rv}. A particular realm of applications that we hope to apply some of our insights is in better understanding the string landscape of asymptotically Minkowski vacua. In the context of backgrounds with sixteen supercharges, it is not fully known whether all $D\leq 7$ vacua have been discovered or whether there are connected components of moduli space yet to be discovered. In particular, it would be interesting to know if such frozen M-theory/IIA vacua can be realized as strong coupling limits of exotic string vacua such as asymmetric orbifolds (see for instance the recent studies \cite{Baykara:2023plc, Baykara:2024vss}) or string islands \cite{Dabholkar:1998kv}. Since frozen $ADE$ singularities can be embedded in higher dimensional special holonomy manifolds, we also expect applications to vacua with lower amounts of supercharges. One could also embed such frozen singularities in AdS flux vacua where the rank reduction caused by the flux can be useful in moduli stabilization.

Finally, we mention that while were able to extract some of the IIA quiver data of the frozen singularity, it would be interesting if there was a procedure for deriving the full frozen quiver which generalizes the Douglas--Moore construction \cite{Douglas:1996sw} to the presence of the $\int C_3$ monodromy.

\section*{Acknowledgments}

MC, LL, ET, and HYZ thank the Harvard Swampland Initiative and the 2024 Simons Physics Summer Workshop for hospitality during the completion of this work. MD and ET thank the ESI in Vienna, in particular the program ``The Landscape vs. the Swampland'', for their hospitality during part of the time in which this work was completed. We thank F. Baume, P. Cheng, J.J. Heckman, M. H\"ubner, C.~Lawrie, M. Montero, H. Parra De Freitas, and A. Tomasiello for helpful discussions. The work of ET is supported in part by the ERC Starting Grant QGuide-101042568 - StG 2021. MC is supported by the by DOE Award (HEP) DE-SC0013528, the Simons Foundation Collaboration grant $\#$724069, Slovenian Research Agency (ARRS No. P1-0306) and Fay R. and Eugene L. Langberg Endowed Chair funds.

\newpage

\appendix

\section{Freezing ADE Singularities with 2-Form RR Flux}\label{app:appA}

In this Appendix, we study another class of frozen $ADE$ singularities that exist in IIA where we take a non-zero monodromy for the $C_1$ RR potential:
\begin{equation}\label{eq:basic}
\mathrm{IIA}\left(\mathbb{R}^{1,5}\times \mathbb{C}^2/\Gamma \; , \; \; \int_{\gamma_1}C_1\neq 0, \;\textnormal{where }\gamma_1\in H_1(S^3/\Gamma)=\mathrm{Ab}(\Gamma)  \right).
\end{equation}
Such vacua were mentioned in Section 4 of \cite{deBoer:2001wca} and are related to the IIA frozen singularities with $\int C_3$ frozen flux if one embeds the $ADE$ singularity into an elliptically-fibered K3 and performs a double T-duality. According to \cite{deBoer:2001wca}, these singularities freeze or partially freeze because the D2 branes that wrap the blown-up exceptional cycles suffer from a dual version of a Freed--Witten anomaly due to a $G_2$-flux along the exceptional cycle. Such an anomaly is still only conjectured to exist as far as we are aware, so we rather appeal to a IIA 10D coupling to argue that the 6D SYM theory freezes to a lower rank gauge group due to confinement.

For concreteness, let us consider the following IIA background
\begin{equation}\label{eq:IIAwarmup}
\mathrm{IIA}\left(\mathbb{R}^{1,5}\times \mathbb{C}^2/\mathbb{Z}_{2} \; , \; \; \int_{\gamma_1}C_1=  1/2 \right) \,.
\end{equation}
If we instead chose $\int_{\gamma_1} C_1=0$ then this would simply engineer a 6D $\mathcal{N}=(1,1)$ $\mathfrak{su}(2)$ gauge theory. As noted in \cite{Witten:1997kz, deBoer:2001wca}, the background $C_1$ field implies a non-trivial $S^1$ fibration structure in the uplift to M-theory. In particular, the background \eqref{eq:IIAwarmup} is dual to
\begin{equation}\label{eq:Mthwarmup}
    \textnormal{M-theory}\left(\mathbb{R}^{1,5}\times (\mathbb{C}^2\times S^1)/\mathbb{Z}_2\right) \,.
\end{equation}
We now see that low-energy 6D theory we obtain in the presence of the background $C_1$ flux is a trivial theory as it is the dimensional reduction of a trivial 7D. To get a more hands-on perspective of this freezing, consider the blow-up $\widetilde{X}\rightarrow \mathbb{C}^2/\mathbb{Z}_2$. $\widetilde{X}$ is a smooth hyper-K\"ahler space (also known as the Eguchi-Hanson space) with a generating exceptional 2-cycle which we denote by $E\in H_2(\widetilde{X})$. Geometrically, the long exact sequence of relative homology contains the short exact sequence\footnote{The homology coefficients are taken to be $\mathbb{Z}$ implicitly.}
\begin{align}
    & 0\rightarrow H_2(\widetilde{X})\rightarrow H_2(\widetilde{X},\partial \widetilde{X})\xrightarrow{\partial}H_1(\partial \widetilde{X})\rightarrow 0 \,, \\
    & 0\rightarrow \; \; \;  \mathbb{Z} \quad  \;  \rightarrow \quad \; \; \;  \mathbb{Z} \quad \quad \; \; \xrightarrow{\partial} \quad  \mathbb{Z}_2 \; \; \; \;\;   \rightarrow 0 \,,
\end{align}
which motivates denoting the generator of $H_2(\widetilde{X},\partial \widetilde{X})$ by $\frac{1}{2}E$. Concretely, the class $\frac{1}{2}E$ can be represented by a non-compact 2-cycle $\Gamma_2$ such that $\partial \Gamma_2=\gamma_1$. Let $G_2:=dC_1$, then from $\int_{\gamma_1} C_1=1/2 \; \mathrm{mod}\; 1$, we see that $\int_{\Gamma_2} G_2=1/2+M$ for some $M\in \mathbb{Z}$ which then implies that there is non-zero RR flux along the exceptional cycle:
\begin{equation}\label{eq:fluxoncycle}
    \int_{E}G_2=1+2M\neq 0 \,.
\end{equation}

We now explain why the flux \eqref{eq:fluxoncycle} leads to the freezing of the $\mathfrak{su}(2)$ gauge algebra to a trivial gauge algebra. Recall that in the absence of the boundary monodromy, a generic point on the Coulomb branch of the 6D $\mathfrak{su}(2)$ gauge theory corresponds to blowing up the singularity, i.e. taking $\mathrm{Vol}(E)\neq 0$. Since the adjoint scalar has a vacuum expectation value, the gauge algebra is Higgsed as $\mathfrak{su}(2)\rightarrow \mathfrak{u}(1)$ and the off-diagonal vector modes are now massive W-bosons. The IIA interpretation of such massive modes are D2 branes that wrap $E$. Recall also that such D2 branes contain a topological worldvolume term
\begin{equation}\label{eq:wzterm}
    S_{D2}\supset \int_{E\times L}a_1\wedge G_2\; \; \rightarrow \; \;  \int_{L}a_1 \,,
\end{equation}
where $a_1$ is the $U(1)$ gauge field on the D2 brane, which means that the flux \eqref{eq:fluxoncycle} generates a Wilson line defect along some line $L\in \mathbb{R}^{1,5}$. Such a defect corresponds to an F1 string attached to the D2 with a worldvolume $\Sigma_2$ such that $\partial \Sigma_2=L$. This is due to the fact that the combination $f_2-B_2$ is gauge invariant. Now such an F1 string can have a second endpoint if we wrap a D2 on $-E$, or equivalently an anti-D2 on $E$, which means that W-bosons of opposite charge under $\mathfrak{u}(1)$ are attached to each other by a string whose expectation value obeys an area law at leading order. In equations, this W-boson is a charge-2 Wilson line of the 6D $\mathfrak{u}(1)$ theory with a 1D worldvolume $L$ has expectation value
\begin{equation}
    \langle W_{+2}(L)\rangle = \langle W_{+2}(L)\rangle_{\mathfrak{u}(1)} \cdot \exp(-\mathrm{Area} (\Sigma_2)/\ell^2_s) \,,
\end{equation}
where $\langle W_{+2}\rangle_{\mathfrak{u}(1)}$ denotes the usual evaluation of the Wilson line in a 6D $\mathcal{N}=(1,1)$ $\mathfrak{u}(1)$ gauge theory and the dominating exponential term comes from evaluating the F1 string action, and $\ell_s$ is the string length scale. This means that the W-bosons are confined which is true even as we take $\mathrm{Area}(\Sigma_2)\rightarrow 0$ because the term \eqref{eq:wzterm} is topological. Having shown the off-diagonal components of the original $\mathfrak{su}(2)$ gauge algebra confine, what about the $\mathfrak{u}(1)$ factor? We can argue for this by reducing the 10D topological term
\begin{equation}
    \int_{10D}G_2\wedge C_5\wedge H_3 \,,
\end{equation}
to 6D under the decomposition $G_2=\omega_2$ and ($\omega$ the dual to $E$) $C_5=\widetilde{A}_3\wedge \omega_2$ which produces a term
\begin{equation}\label{eq:magstuck}
    \int_{\rm 6D} \widetilde{A}_3\wedge H_3 \,.
\end{equation}
where $\widetilde{A}_3$ is just the electromagnetic dual to the $\mathfrak{u}(1)$ gauge potential. Since \eqref{eq:magstuck} is a St\"uckelberg term for $\widetilde{A}_3$ we see that the dual $U(1)$ potential is Higgsed or, equivalently, the original gauge potential is confined. We have now showed that the full $\mathfrak{su}(2)$ vector multiplet is confined due to the $C_1$ boundary monodromy.

We close by mentioning a more non-trivial example. consider the following duality\footnote{Recall that we are assuming that $\gamma_1$ is a generating element of $H_1(S^3/\mathbb{Z}_4)$.}:
\begin{equation}
    \textnormal{M-theory}\left(\mathbb{R}^{1,5}\times (\mathbb{C}^2/\mathbb{Z}_{2}\times S^1)/\mathbb{Z}_2\right)\simeq \mathrm{IIA}\left(\mathbb{R}^{1,5}\times \mathbb{C}^2/\mathbb{Z}_{4} \; , \; \; \int_{\gamma_1}C_1=  1/2 \right).
\end{equation}
We expect then that the RR monodromy at the boundary partially freezes the gauge algebra as $\mathfrak{su}(4)\rightarrow \mathfrak{su}(2)$. To see how this works, consider first the full blow-up of $\widetilde{X}\rightarrow \mathbb{C}^2/\mathbb{Z}_4$ where $\widetilde{X}$ has three exceptional 2-cycles $E_{i=1,2,3}$. The short exact sequence of (relative) homology groups is
\begin{align}
    & 0\rightarrow H_2(\widetilde{X})\rightarrow H_2(\widetilde{X},\partial \widetilde{X})\xrightarrow{\partial}H_1(\partial \widetilde{X})\rightarrow 0 \,, \\
    & 0\rightarrow \; \; \;  \mathbb{Z}^3 \quad  \;  \rightarrow \quad \; \; \;  \mathbb{Z}^3 \quad \quad \; \; \xrightarrow{\partial} \quad  \mathbb{Z}_4 \; \; \; \;\;   \rightarrow 0 \,,
\end{align}
where the pullback of the generator of $\mathbb{Z}_4$ can be given as
\begin{equation}
    [\Gamma_2]=\tfrac{1}{4}(E_1+2E_2+3E_3) \,,
\end{equation}
i.e. $\partial\Gamma_2=\gamma_1$. The boundary RR-monodromy then tells us that
\begin{equation}
    \int_{E_1+2E_2+3E_3}G_2=2+4M, \quad \textnormal{for some $M\in \mathbb{Z}$} \,,
\end{equation}
which can be solved, for example, by taking $\int_{E_1}G_2=1$, $\int_{E_2}G_2=0$, $\int_{E_3}G_2=-1$, and $M=-1$. Focusing on this solution for concreteness, although each solution will give the same physics, we see from the previous example that D2 branes wrapping the $E_1$ and $E_3$ cycles are confined. We thus keep $E_1$ and $E_3$ blown down and do not associated gauge algebras with them, and see that (when $\mathrm{Vol}(E_2)\neq 0$) that we have a pair of frozen $A_1$ singularities each with $\int_{\alpha^{(i)}_1} C_1=1/2$ where $\alpha^{(1)}$ and $\alpha^{(3)}$ denote 1-cycles in the boundary of a local neighborhood $(S^3/\mathbb{Z}_2)^{(i)}$ surrounding the $A_1$ singularity gotten by blowing-down $E_1$ and $E_3$ respectively. This agrees with our RR flux values because $\frac{1}{2}E_i$ restricts to $\alpha^{(i)}_1$ along $(S^3/\mathbb{Z}_2)^{(i)}$. Finally, notice now that since $E_1\cdot E_2=E_3\cdot E_2=1$, the $G_2$ flux vanishes automatically on $E_2$ thus allowing it to not be frozen and leaving an $\mathfrak{su}(2)$ gauge algebra.

One can repeat these arguments for more general $ADE$ singularities by using the generalizations of the expressions of the relative 2-cycles in terms of rational linear combinations of compact exceptional cycles which can be found in \cite{Hubner:2022kxr}.

\section{Counterterm for Frozen Theories}\label{app:fullyfrozen}

This appendix is dedicated to deriving the claim of equation \eqref{eq:7Dcounterterm2} that for the frozen 7D theories with $\mathbb{D}(\mathfrak{h}_{\Gamma,d}) \neq \mathbb{D}(\mathfrak{g}^{(1/d)})$, there exists a counterterm of the form
\begin{equation}
    S^{\mathrm{c.t.}}_{\rm 7D}=2\pi (L_{\Gamma})_{ij}\int_{\rm 7D} b^i_2 \cup b^j_5 \,,
\end{equation}
localized on the $\mathcal{T}^{(\mathrm{M})}[\mathfrak{g}^{(n/d)}]$ boundary of the 8D SymTFT. Recall that the fully frozen singularities are $D^{(1/2)}_{4}$, $\mathfrak{e}^{(1/3)}_6$, $\mathfrak{e}^{(1/4)}_7$, $\mathfrak{e}^{(1/5)}_8$, and $\mathfrak{e}^{(1/6)}_8$, where the 2-form and 5-form potentials are respectively valued in $\mathbb{Z}^2_2$, $\mathbb{Z}_3$, $\mathbb{Z}_2$, $0$, and $0$. On its own, the above action would not be invariant under gauge transformations of $b^i_2$ and $b^i_5$ without in-flowing from the 8D SymTFT action
\begin{equation}\label{eq:symtftactionappendix}
    2\pi (L_{\Gamma})_{ij}\int_{\rm 8D} b^i_2 \cup \delta b^j_5 \,.
\end{equation}
On one level, it may be obvious that because the 7D theory localized on $ADE$ singularity has trivial gauge dynamics, which normally have a mixed 1-form/4-form anomaly one must include such a 7D counterterm to correctly produce to the 8D anomaly theory. Clear as this may be to some readers, we find it enlightening to understand how the counterterm appears from the geometric reduction of the 11D supergravity action.

 Our derivation will closely follow Appendix B of \cite{Baume:2023kkf} where the action \eqref{eq:symtftactionappendix} was derived for the unfrozen $ADE$ singularities by reducing the kinetic term of the 11D supergravity action, see also \cite{GarciaEtxebarria:2024fuk}. Intuitively, this SymTFT action carries the topological data of the linking of M2 and M5 branes wrapped on the torsion 1-cycles of $S^3/\Gamma$ despite originating from a non-topological kinetic term. A key difference with our approach is that we will be working in a dual frame where the frozen singularity is realizes as an F-theory compactification on $Z=(I_0\times S^1)/\mathbb{Z}_d$ (the action of $\mathbb{Z}_d$ on the Kodaira surface $I_0$ can be found in the main text). Or more specifically, we will work in a further circle compactification of this where we discuss M-theory on $Z$. We will obtain a 7D SymTFT action which is a circle reduction of \eqref{eq:symtftactionappendix} by reducing the M-theory kinetic term on $\partial Z$, and obtain the counterterm by reducing on $Z$. The latter is a little subtle because $Z$ has an asymptotic boundary, which requires properly understanding the integration of forms on $Z$ which we will untangle.

We first reduce M-theory on $\partial Z$, recall that its (integer) homology groups in this case are (recall that in these sets we are listing the entries as $*=0,...,4$, see Section \ref{ssec:outerauto})
\begin{equation}
    H_*(\partial Z)=\{ \mathbb{Z}, \mathbb{Z}^2\oplus \text{Ab}[\Gamma_X], \mathbb{Z}^2\oplus \mathbb{Z}_2, \mathbb{Z}^2, \mathbb{Z} \} \,,
\end{equation}
which, from Poincar\'e duality, means that the integer cohomology is
\begin{equation}
    H^*(\partial Z)=\{ \mathbb{Z}, \mathbb{Z}^2, \mathbb{Z}^2\oplus \text{Ab}[\Gamma_X], \mathbb{Z}^2\oplus \mathbb{Z}_2, \mathbb{Z} \} \,,
\end{equation}
where Ab$[\Gamma_X] = Z(G_\Gamma)$ is the center of the unfrozen gauge group, equipped with the linking pairing $L_\Gamma$.

For ease of notation, we will focus on the $\mathfrak{e}^{(1/4)}_7$ case, where $L_\Gamma = 1/2$ and the potentials are $\mathbb{Z}_2$-valued.
We can regard the torsional generators of $\mathbb{Z}_2=\mathrm{Tor} \big( H^k(\partial Z) \big)$ ($k=2,3$) as pairs of differential forms $(\alpha_k, \beta_{k-1})$ which satisfy\footnote{Such an approach to KK-reducing string theory $p$-form potentials was notably explored in \cite{Camara:2011jg}. See also \cite{Apruzzi:2021nmk} for a relatively recent introduction to using differential cohomology which is an alternative approach to dealing with torsional cocycles.} ($d^{\dagger} = * d *$)
\begin{equation}
    2\alpha_k= d\beta_{k-1}, \quad \quad d^\dagger \beta_{k-1}=0 \,.
\end{equation}
We can expand the $G_4$ and $G_7$ M-theory fluxes on these cocycles as:
\begin{align}
    G_4\supset (dA_1+2B_2)\wedge \alpha_2 +dB_2\wedge \beta_1+(dA_0+2B_1)\wedge \alpha_3 +dB_1\wedge \beta_2 \,, \\
    G_7\supset (dA_4+2B_5)\wedge \alpha_2 +dB_5\wedge \beta_1+(dA_3+2B_4)\wedge \alpha_3 +dB_4\wedge \beta_2  \,.
\end{align}
Expanding the kinetic terms and keeping relevant terms:
\begin{align}
    -2\pi S_{11D}&=\frac{1}{2}\int_{11D} G_4\wedge G_7 \\
     &\supset 2\left( \int_{\partial Z} \alpha_2 \wedge \beta_2 \right) \int_{\rm 7D} B_2 \wedge dB_4 + 2\left( \int_{\partial Z} \alpha_3 \wedge \beta_1 \right) \int_{\rm 7D} B_1 \wedge dB_5 \,.\label{eq:expansionappendix}
\end{align}
Notice that each of these terms are topological as opposed to, say, terms proportional to $dA_1\wedge dB_4$ which depend explicitly on the 7D metric because $*_{\rm 7D}dA_1=dB_4$. Also the integrals over $\partial Z$ are in fact equal to $1 \; \mathrm{mod} \; 2$ (in this normalization) as they are both valid integral representations of the torsional pairing \cite{Camara:2011jg}:
\begin{equation}
 L_{\Gamma}: \mathrm{Tor} H^2(\partial Z)\times \mathrm{Tor}H^3(\partial Z)\rightarrow \mathbb{Q}/\mathbb{Z} \,.
\end{equation}
In all we have the 7D action
\begin{equation}
    \frac{1}{2\pi}\times (2)\int_{\rm 7D}\left( B_2\wedge dB_4 +B_1\wedge dB_5 \right) \,,
\end{equation}
which can of course be presented as the circle reduction of
\begin{equation}
    \frac{1}{2\pi}\times (2)\int_{\rm 8D}\left( B_2\wedge dB_5\right) \,.
\end{equation}
As is standard for a BF-theory, because the observables of this 8D SymTFT are $\mathbb{Z}_2$-valued, we can trade off the action for $\mathbb{Z}_2$ valued potentials via $\frac{2}{2\pi}B_i=b_i$ to arrive at the action \eqref{eq:symtftactionappendix} for the $\mathfrak{e}^{(1/4)}_7$ case.

To derive the 7D counterterm, we now perform a similar reduction of the 11D kinetic term but over $Z$. We first mention that, because $Z$ has non-zero boundary, the correct pairing to integrate forms uses Poincar\'e--Lefschetz duality $H^p(Z, \partial Z)\simeq H_{5-p}(Z)$ which results from the cap product with the fundamental class $[Z]\in H_5(Z,\partial Z)$:
\begin{equation}
  \cap: \; H_5(Z,\partial Z)\times H^p(Z, \partial Z)\rightarrow H_{5-p}(Z) \,.
\end{equation}
Additionally, we have $\mathrm{Tor}\big(H_p(Z,\partial Z)\big)\simeq \mathrm{Tor}\big(H^{p+1}(Z,\partial Z)\big)$ from the Universal Coefficient Theorem, which follows from a similar pairing
\begin{equation}
    \mathrm{Tor}\big(H_p(Z,\partial Z)\big)\times \mathrm{Tor}\big(H^{p+1}(Z,\partial Z)\big)\rightarrow H_0(Z,\mathbb{Q}/\mathbb{Z})\simeq \mathbb{Q}/\mathbb{Z} \,,
\end{equation}
from integrating torsion cocycles over their dual cycles\footnote{The appearance of the $\mathbb{Q}/\mathbb{Z}$ coefficients can be derived more carefully using the long exact sequence with respect to the short exact sequence on the coefficients $0\rightarrow \mathbb{Z}\rightarrow \mathbb{Q}\rightarrow \mathbb{Q}/\mathbb{Z}\rightarrow 0$. See for instance \cite{CueMaudeZee} for more details. }. Putting these together, we have a perfect pairing
\begin{equation}\label{eq:relativenonrelative}
    \mathrm{Tor}\big(H^{5-p}(Z)\big)\times \mathrm{Tor}\big(H^{p+1}(Z,\partial Z)\big)\rightarrow \mathbb{Q}/\mathbb{Z} \,.
\end{equation}
What \eqref{eq:relativenonrelative} implies is that if we wedge a torsional $(5-p)$-form of $Z$ with a relative torsional $(p+1)$-form, we can integrate over $Z$. Relevant to our case is $p=3$ where $\mathrm{Tor}\big(H^2(Z)\big)\simeq \mathrm{Tor}\big(H^4(Z,\partial Z)\big) \simeq \mathrm{Tor}\big( H_1(Z) \big)$ are the only non-trivial torsion pieces in $H^*(Z)$ and $H^*(Z,\partial Z)$.
This is precisely the case whenever $\mathbb{D}( \mathfrak{h}_{\Gamma,d} ) \neq \mathbb{D}( \mathfrak{g}^{(1/d)})$, and the arguments spelled out for $\mathfrak{e}_7^{(1/4)}$ below apply mutatis mutandis to all of them.

Before expanding $G_4$ and $G_7$ on these cocycles, we note the relation of these forms to forms in $H^*(\partial Z)$. In particular, we have the \emph{isomorphisms} which follow from the long exact sequence of relative cohomology:
\begin{align}
    \mathrm{Tor}\big(H^2(Z)\big)\xrightarrow[]{\mathrm{restriction}} \mathrm{Tor}\big(H^2(\partial Z)\big) \,, \\
    \mathrm{Tor}\big(H^3(\partial Z)\big)\xrightarrow[]{\delta} \mathrm{Tor}\big(H^4(Z,\partial Z)\big) \,.
\end{align}
The top line means that the class $(\alpha_2, \beta_1)$ generating $\mathbb{Z}_2=\mathrm{Tor}\big(H^2(\partial Z)\big)$ can be extended to the $Z$ bulk as $(\widetilde{\alpha}_2, \widetilde{\beta}_1)$ which are related simply by $\alpha_2=\widetilde{\alpha}_2|_{\partial Z}$ and $\beta_1=\widetilde{\beta}_1|_{\partial Z}$. Meanwhile the bottom line means that $(\alpha_3, \beta_2)$ generating $\mathbb{Z}_2=\mathrm{Tor}\big(H^3(\partial Z)\big)$ when extended to the $Z$ bulk as $(\widetilde{\alpha}_3, \widetilde{\beta}_2)$ are related to the torsion generator of $H^4(Z,\partial Z)$, $(\widetilde{\alpha}_4, \widetilde{\beta}_3)$, as
\begin{equation}\label{eq:that}
    d\widetilde{\alpha}_3=\widetilde{\alpha}_4|_{\partial Z} \,, \quad \quad d\widetilde{\beta}_2=\widetilde{\beta}_3|_{\partial Z} \,.
\end{equation}
This means we can write the torsion pairing as an integral
\begin{equation}\label{eq:this}
    \int_{Z} \widetilde{\alpha}_4\wedge \widetilde{\beta}_1=\int_{Z} \widetilde{\alpha}_2 \wedge \widetilde{\beta}_3 \,.
\end{equation}
From the expansion \eqref{eq:expansionappendix} we see that we can rewrite, say the first term, as
\begin{equation}
    2\left( \int_{\partial Z} \alpha_2 \wedge \beta_2 \right) \int_{\rm 7D} B_2 \wedge dB_4=2\left( \int_{Z} \widetilde{\alpha}_2 \wedge \widetilde{\beta}_3 \right) \int_{\rm 6D} B_2 \wedge B_4 \,,
\end{equation}
by using the identities \eqref{eq:that} and the fact that the 7D spacetime in the SymTFT action in \eqref{eq:expansionappendix} contains the radial direction of the base of $Z$. Essentially we are just moving the exterior derivative in the 7D action to the link pairing piece in a sensible fashion. Doing this with the second term in \eqref{eq:expansionappendix} gives in total
\begin{equation}
    \frac{1}{2\pi}\times (2)\int_{\rm 6D}\left( B_2\wedge B_4 +B_1\wedge B_5 \right) \,,
\end{equation}
which can be taken as the circle reduction of the 7D counterterm
\begin{equation}
     \frac{1}{2\pi}\times (2)\int_{\rm 7D}\left( B_2\wedge B_5\right) \,.
\end{equation}

\section{Consistency with Resolving Singularities}\label{app:resolving}

In Section \ref{sec:defectgrps}, we calculated the defect group of F-theory on a twisted circle compactification $Z\equiv (Y\times S_1)/\mathbb{Z}_d$ where $Y$ is a Kodaira singularity. We did this by calculating the groups $ H_k(Z,\partial Z)/H_k(Z)$ which by exactness is equivalent to $\mathrm{ker}(H_{k-1}(\partial Z)\rightarrow H_{k-1}(Z))$ where the map is induced from the inclusion $\partial Z\hookrightarrow Z$. These groups specify $k$-cycles in $Z$ which restrict in $\partial Z$ to a non-trivial $(k-1)$-cycles in $\partial Z$. For $k=2$, this tells us physically distinct string junctions charges in 7D since if we consider F-theory on $Z\times S^1$, and these string junction equivalence classes are equivalently phrased in terms of M-theory on $Z$ as M2 branes wrapped on classes of $H_2(Z,\partial Z)/H_2(Z)$ which wrap a 1-cycle in the elliptic fiber.

In this Appendix, we complement these results by computing how the homology groups of $Z$ change when resolving the Kodaira singularity. Namely, resolving the fiber introduces extra $\mathbb{P}^1$ 2-cycles in $Y$ which thus adds 2- and 3-cycles to $Z$. We denote the fully resolved manifolds as $\widetilde{Y}$ and $\widetilde{Z}$. By exactness and the fact that $\partial Z=\partial \widetilde{Z}$, we see that $H_2(\widetilde{Z},\partial \widetilde{ Z})$ will gain numerous free factors from resolving, while the defect group does not change:
\begin{equation}
   \mathbb{D}^{(1)}(Z)= H_2(\widetilde{Z},\partial \widetilde{Z})/H_2(\widetilde{Z})=H_2(Z,\partial Z)/H_2(Z) \,.
\end{equation}
We will find in general that $H_2(\widetilde{Z})$ is a free lattice whose rank is reduced from that of $H_2(\widetilde{Y})$, provided we are considering a case where $\rho$ acts non-trivially on the elliptic fiber. Conceptually, this is just due to the fact that one can only switch on hyperk\"ahler parameters of $\widetilde{Y}$ that are invariant under $\rho$.

Similar to Section \ref{sec:defectgrps}, we can calculate $H_2(\widetilde{Z})$ and $H_3(\widetilde{Z})$ by knowing the monodromy action on the $k$-cycles $\rho_k:H_k(\widetilde{Y})\rightarrow H_k(\widetilde{Y})$ and using the following short exact sequences
\begin{align}\label{eq:h2tildeZ}
    0\rightarrow \mathrm{coker}(\rho_2-1)\rightarrow H_2(\widetilde{Z})\rightarrow 0\rightarrow 0 \,, \\
   0\rightarrow 0 \rightarrow H_3(\widetilde{Z})\rightarrow \mathrm{ker}(\rho_2-1)\rightarrow 0 \,.\label{eq:h3tildeZ}
\end{align}
In \eqref{eq:h3tildeZ}, we have used the fact that $H_3(\widetilde{Y})=0$ because $\widetilde{Y}$ is contractible to its central elliptic fiber at $z=0$, and in \eqref{eq:h2tildeZ} we used $\mathrm{ker}(\rho_1-1)=0$ following explicit calculations in Section \ref{sec:defectgrps}. In previous sections, $\rho_2$ was simply the identity which is now no longer the case in general. Since the data of $\rho_1$ is unchanged after the resolution, the key quantities we compute in this section are $\mathrm{coker}(\rho_2-1)$ (from which $H_2(\widetilde{Z})$ follows from the calculation of $\mathrm{ker}(\rho_2-1)$ in Section \ref{sec:defectgrps}) and $\mathrm{ker}(\rho_2-1)=H_3(\widetilde{Z})$.

\begin{figure}[t!]
\centering
\includegraphics[scale=0.30, trim = {0cm 0cm 0cm 0cm}]{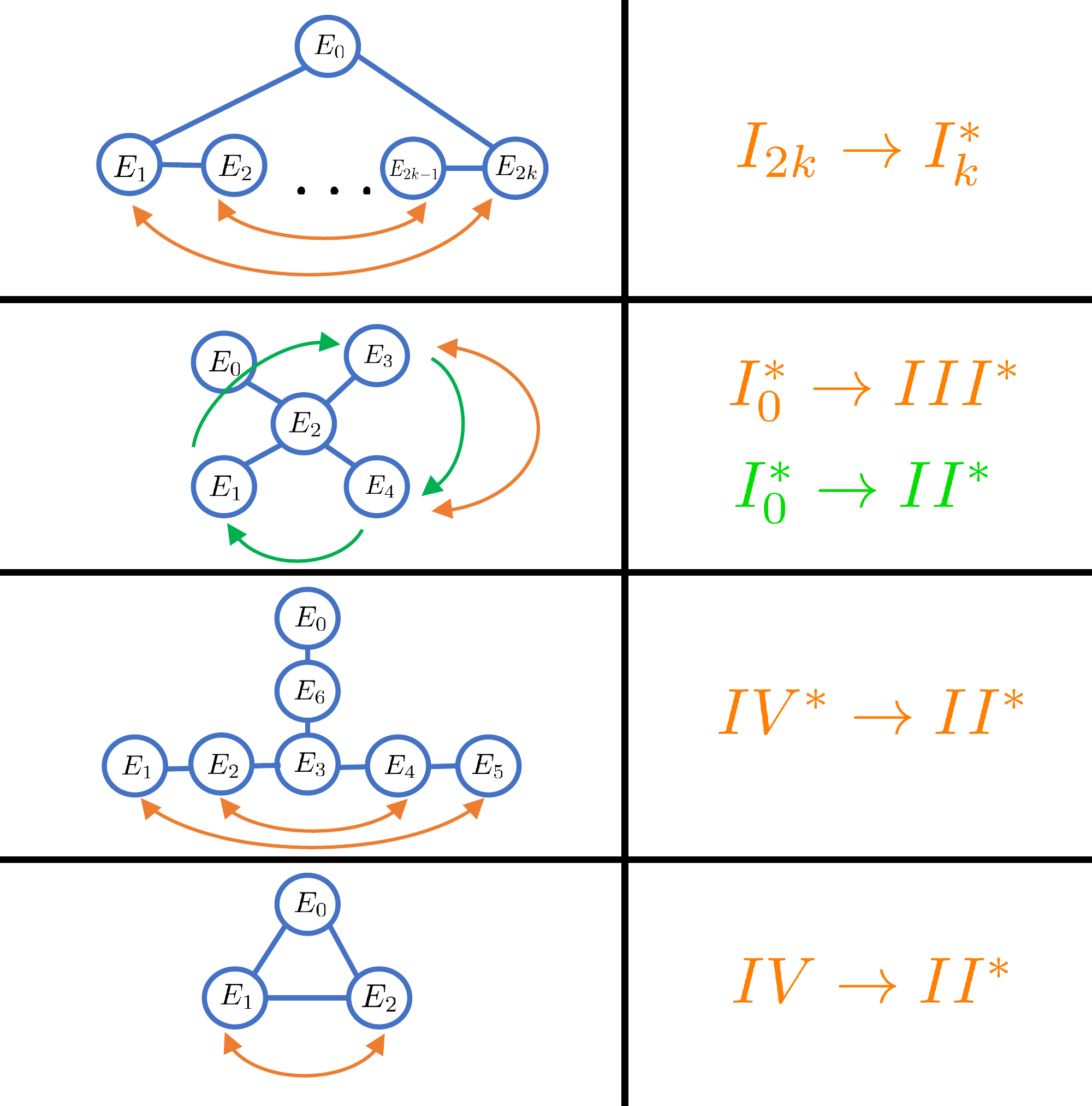}
\caption{$\mathbb{Z}_2$ (orange) and $\mathbb{Z}_3$ (green) actions on the resolved central fiber of a Kodaira surface $Y$. Physically these actions become outer automorphisms of the gauge algebra in the singular limit. On the right column, we denote the Kodaira classification of $Y$ and what it would be if we were to quotient it by $\rho$. In terms of equation \eqref{eq:fmduality}, this just means the Kodaira type of the surface $X$.}
\label{fig:outerauto}
\end{figure}

To understand the action of $\rho_2$ on $H_2(\widetilde{Y})$, we first recall the fact that $H_2(\widetilde{Y})$ is generated by cycles $E_i$, $i=0,..., \mathrm{rank}(\mathfrak{g})$ where $\mathfrak{g}$ is the Lie algebra associated with the Kodaira singularity. The elliptic fiber 2-cycle homology class $[\mathbb{E}_z]$ away from $z=0$ decomposes into these $\mathbb{P}^1$ cycles as
\begin{equation}
    [\mathbb{E}_z]=\sum_{i=0}^{\mathrm{rank}(\mathfrak{g})}E_i \,.
\end{equation}
The action of $\rho$ on the elliptic fiber $\mathbb{E}_z$ away from $z=0$ is specified by an element $\rho_1\in SL(2,\mathbb{Z})$ acting on its 1-cycles, which is given in all cases of interest in Section \ref{ssec:outerauto}. Note that a common phenomenon we found across section \ref{ssec:outerauto} was that the order of $\rho_1$ as an element of $SL(2,\mathbb{Z})$ may not match the order of $\rho_1$ regarded as an automorphism $\rho_1:\mathbb{D}^{(1)}(\widetilde{Y})\rightarrow \mathbb{D}^{(1)}(\widetilde{Y})$ which is given in the ``Outer" column of Table \ref{table:frozendinner}. These $\mathbb{Z}_2$ or $\mathbb{Z}_3$ actions on the resolved central fiber of $Y$ is then equivalent to the usual involution of an affine Dynkin diagram associated to an outer-automorphism of a Lie algebra. These are summarized in Figure \ref{fig:outerauto}.

As a sanity check, we can see that these actions are consistent with the actions on $\mathbb{D}^{(1)}(\widetilde{Y})=H_2(\widetilde{Y},\partial \widetilde{Y})/H_2(\widetilde{Y})$ as these can be regarded as fractional linear combinations of 2-cycles in $H_2(\widetilde{Y})$ modulo integer 2-cycles. For example, in the case of $Y$ being an $IV$ singularity, the generator of $\mathbb{D}^{(1)}(\widetilde{Y})=\mathbb{Z}_3$ can be presented as (for more general Kodaira singularities see Section 3 of \cite{Hubner:2022kxr})
\begin{equation}
    \textnormal{Generator of $\mathbb{D}^{(1)}(\widetilde{Y})$:
      } \quad \frac{1}{3}\left( E_1+2E_2\right) \,,
\end{equation}
where $E_i$ denote the exceptional cycles as in Figure \ref{fig:outerauto}. We see that the automorphism exchanges $E_1$ and $E_2$ which sends the above torsional element to
\begin{equation}
    \rho:\; \; \frac{1}{3}\left( E_1+2E_2\right)\quad \mapsto \quad \frac{1}{3}\left( E_2+2E_1\right)\simeq -\frac{1}{3}\left( E_1+2E_2\right)
\end{equation}
where the last equation follows from adding $-E_1-E_2$ to the LHS. Thus, we see that the automorphism on $H_2(\widetilde{Y})$ in Figure \ref{fig:outerauto} induces the $\mathbb{Z}_2$ automorphism of $\mathrm{Tor}\big(H_1(\partial \widetilde{Y})\big)=\mathbb{Z}_3$.

We are now in a position to calculate $H_2(\widetilde{Z})$ and $H_3(\widetilde{Z})$ using \eqref{eq:h2tildeZ} and \eqref{eq:h3tildeZ} which we handle case-by-case. In general we find that
\begin{align}\label{eq:h2h3ztilde}
    H_2(\widetilde{Z})=H_3(\widetilde{Z})= \mathbb{Z}^{R_f+1}.
\end{align}
where $R_f$ is the rank of the frozen gauge algebra listed in the ``$\mathfrak{h}_{\Gamma,d}$" column in Table \ref{table:frozendinner}. Physically, this is consistent with the duality \eqref{eq:fmduality} as the number of possible moduli match: from \eqref{eq:h2h3ztilde}, we have $3R_{f}+3$ moduli which in terms of the frozen elliptic fibration $X$ consists of $3R_{f}$ Coulomb branch parameters of the frozen gauge algebra as well as the hyper-K\"ahler parameters of the elliptic fiber.

\paragraph{D-Type Frozen Singularities}
Our task is to compute the kernel and cokernel of the permutation matrix associated with the top row in Figure \ref{fig:outerauto} minus the identity. For $2k=4$, this is (using the ordering $E_0$, $E_1$, ..., $E_{4}$)
\begin{equation}
   \rho_2- 1=\begin{pmatrix}
        0 & 0 & 0 & 0 & 0 \\
        0 & -1 & 0 & 0 & 1 \\
        0 & 0 & -1 & 1 & 0 \\
        0 & 0 & 1 & -1 & 0 \\
        0 & 1 & 0 & 0 & -1 \\
    \end{pmatrix} \,,
\end{equation}
which has a Smith normal form
\begin{equation}
   \mathrm{SNF}(\rho_2- 1)=\begin{pmatrix}
        1 & 0 & 0 & 0 & 0 \\
        0 & 1 & 0 & 0 & 0 \\
        0 & 0 & 0 & 0 & 0 \\
        0 & 0 & 0 & 0 & 0 \\
        0 & 0 & 0 & 0 & 0 \\
    \end{pmatrix} \,.
\end{equation}
For general $2k$, the Smith normal form is
\begin{equation}
    \mathrm{SNF}(\rho_2- 1)=\mathrm{diag}(1^{k},0^{k+1}) \,,
\end{equation}
where superscript denotes multiplicity. This means that $\mathrm{coker}(\rho_2 -1)=\mathrm{ker}(\rho_2 - 1)=\mathbb{Z}^{k+1}$ so we have
\begin{align}
    H_2(\widetilde{Z})=H_3(\widetilde{Z})= \mathbb{Z}^{k+1} \,.
\end{align}
\paragraph{$\mathbf{\mathfrak{e}^{(1/2)}_7}$ Frozen Singularity}
From the second row in Figure \ref{fig:outerauto} (in orange) we have that
\begin{align}
   \rho_2-1=\begin{pmatrix}
        0 & 0 & 0 & 0 & 0 \\
        0 & 0 & 0 & 0 & 0 \\
        0 & 0 & 0 & 0 & 0 \\
        0 & 0 & 0 & -1 & 1 \\
        0 & 0 & 0 & 1 & -1 \\
    \end{pmatrix} \,,\\
    \mathrm{SNF}(\rho_2-1)=\begin{pmatrix}
        1 & 0 & 0 & 0 & 0 \\
        0 & 0 & 0 & 0 & 0 \\
        0 & 0 & 0 & 0 & 0 \\
        0 & 0 & 0 & 0 & 0 \\
        0 & 0 & 0 & 0 & 0 \\
    \end{pmatrix} \,,
\end{align}
which gives
\begin{align}
    H_2(\widetilde{Z})=H_3(\widetilde{Z})= \mathbb{Z}^{4} \,.
\end{align}
\paragraph{$\mathbf{\mathfrak{e}^{(n/d)}_8}$ Frozen Singularities}
Beginning with $\widetilde{Z}$ dual to the $\mathfrak{e}^{(1/2)}_8$ singularity, the relevant matrices follow from the third row of Figure \ref{fig:outerauto} to be
\begin{align}
   \rho_2- 1=\begin{pmatrix}
        0 & 0 & 0 & 0 & 0 & 0 & 0\\
        0 & -1 & 0 & 0 & 0 & 1 & 0\\
        0 & 0 & -1 & 0 & 1 & 0 & 0\\
        0 & 0 & 0 & 0 & 0 & 0 & 0\\
        0 & 0 & 1 & 0 & -1 & 0 & 0\\
        0 & 1 & 0 & 0 & 0 & -1 & 0\\
       0 & 0 & 0 & 0 & 0 & 0 & 0
    \end{pmatrix} \,, \\
    \mathrm{SNF}(\rho_2- 1)=
      \begin{pmatrix}
        1 & 0 & 0 & 0 & 0 & 0 & 0\\
        0 & 1 & 0 & 0 & 0 & 0 & 0\\
        0 & 0 & 0 & 0 & 0 & 0 & 0\\
        0 & 0 & 0 & 0 & 0 & 0 & 0\\
        0 & 0 & 0 & 0 & 0 & 0 & 0\\
        0 & 0 & 0 & 0 & 0 & 0 & 0\\
        0 & 0 & 0 & 0 & 0 & 0 & 0
    \end{pmatrix} \,,
\end{align}
from which it follows that
\begin{align}
    H_2(\widetilde{Z})=H_3(\widetilde{Z})= \mathbb{Z}^{5} \,.
\end{align}

As for $\mathfrak{e}^{(1/3)}_8$ we have (from the second row of Figure \ref{fig:outerauto} in green)
\begin{align}
   \rho_2-1=\begin{pmatrix}
        0 & 0 & 0 & 0 & 0 \\
        0 & -1 & 0 & 1 & 0 \\
        0 & 0 & 0 & 0 & 0 \\
        0 & 0 & 0 & -1 & 1 \\
        0 & 1 & 0 & 0 & -1 \\
    \end{pmatrix} \,,\\
    \mathrm{SNF}(\rho_2-1)=
    \begin{pmatrix}
        1 & 0 & 0 & 0 & 0 \\
        0 & 1 & 0 & 0 & 0 \\
        0 & 0 & 0 & 0 & 0 \\
        0 & 0 & 0 & 0 & 0 \\
        0 & 0 & 0 & 0 & 0 \\
    \end{pmatrix} \,,
\end{align}
which yields
\begin{align}
    H_2(\widetilde{Z})=H_3(\widetilde{Z})= \mathbb{Z}^{3} \,.
\end{align}

Finally, the calculation of $H_2(\widetilde{Z})$ and $H_3(\widetilde{Z})$ in the $\mathfrak{e}^{(1/4)}_8$ case follows from the $2k=2$ $D$-type example. Therefore
\begin{align}
    H_2(\widetilde{Z})=H_3(\widetilde{Z})= \mathbb{Z}^{2} \,.
\end{align}

\newpage

\bibliographystyle{utphys}
\bibliography{frozen}

\providecommand{\href}[2]{#2}\begingroup\raggedright\begin{thebibliography}{100}

\bibitem{Witten:1997bs}
E.~Witten, ``{Toroidal compactification without vector structure},''
  \href{http://dx.doi.org/10.1088/1126-6708/1998/02/006}{{\em JHEP} {\bfseries
  02} (1998) 006},
\href{http://arxiv.org/abs/hep-th/9712028}{{\ttfamily arXiv:hep-th/9712028
  [hep-th]}}.
%%CITATION = HEP-TH/9712028;%%.

\bibitem{deBoer:2001wca}
J.~de~Boer, R.~Dijkgraaf, K.~Hori, A.~Keurentjes, J.~Morgan, D.~R. Morrison,
  and S.~Sethi, ``{Triples, fluxes, and strings},''
  \href{http://dx.doi.org/10.4310/ATMP.2000.v4.n5.a1}{{\em Adv. Theor. Math.
  Phys.} {\bfseries 4} (2002) 995--1186},
  \href{http://arxiv.org/abs/hep-th/0103170}{{\ttfamily arXiv:hep-th/0103170}}.

\bibitem{Atiyah:2001qf}
M.~Atiyah and E.~Witten, ``{M theory dynamics on a manifold of G(2)
  holonomy},'' \href{http://dx.doi.org/10.4310/ATMP.2002.v6.n1.a1}{{\em Adv.
  Theor. Math. Phys.} {\bfseries 6} (2003) 1--106},
  \href{http://arxiv.org/abs/hep-th/0107177}{{\ttfamily arXiv:hep-th/0107177}}.

\bibitem{Tachikawa:2015wka}
Y.~Tachikawa, ``{Frozen singularities in M and F theory},''
  \href{http://dx.doi.org/10.1007/JHEP06(2016)128}{{\em JHEP} {\bfseries 06}
  (2016) 128}, \href{http://arxiv.org/abs/1508.06679}{{\ttfamily
  arXiv:1508.06679 [hep-th]}}.

\bibitem{Bhardwaj:2018jgp}
L.~Bhardwaj, D.~R. Morrison, Y.~Tachikawa, and A.~Tomasiello, ``{The frozen
  phase of F-theory},'' \href{http://dx.doi.org/10.1007/JHEP08(2018)138}{{\em
  JHEP} {\bfseries 08} (2018) 138},
  \href{http://arxiv.org/abs/1805.09070}{{\ttfamily arXiv:1805.09070
  [hep-th]}}.

\bibitem{Fraiman:2021hma}
B.~Fraiman and H.~P. de~Freitas, ``{Freezing of gauge symmetries in the
  heterotic string on T$^{4}$},''
  \href{http://dx.doi.org/10.1007/JHEP04(2022)007}{{\em JHEP} {\bfseries 04}
  (2022) 007}, \href{http://arxiv.org/abs/2111.09966}{{\ttfamily
  arXiv:2111.09966 [hep-th]}}.

\bibitem{ParraDeFreitas:2022wnz}
H.~Parra De~Freitas, ``{New supersymmetric string moduli spaces from frozen
  singularities},'' \href{http://dx.doi.org/10.1007/JHEP01(2023)170}{{\em JHEP}
  {\bfseries 01} (2023) 170}, \href{http://arxiv.org/abs/2209.03451}{{\ttfamily
  arXiv:2209.03451 [hep-th]}}.

\bibitem{Cecotti:2023mlc}
S.~Cecotti, ``{Hwang-Oguiso invariants and frozen singularities in special
  geometry},'' \href{http://dx.doi.org/10.1007/JHEP04(2024)012}{{\em JHEP}
  {\bfseries 04} (2024) 012}, \href{http://arxiv.org/abs/2304.04481}{{\ttfamily
  arXiv:2304.04481 [hep-th]}}.

\bibitem{Morrison:2023hqx}
D.~R. Morrison and B.~Sung, ``{On the frozen F-theory landscape},''
  \href{http://dx.doi.org/10.1007/JHEP05(2024)126}{{\em JHEP} {\bfseries 05}
  (2024) 126}, \href{http://arxiv.org/abs/2310.11432}{{\ttfamily
  arXiv:2310.11432 [hep-th]}}.

\bibitem{Donagi:2023sbk}
R.~Donagi and M.~Wijnholt, ``{The $M$-Theory Three-Form and Singular
  Geometries},'' \href{http://arxiv.org/abs/2310.05838}{{\ttfamily
  arXiv:2310.05838 [hep-th]}}.

\bibitem{Oehlmann:2024cyn}
P.-K. Oehlmann, F.~Ruehle, and B.~Sung, ``{The frozen phase of heterotic
  F-theory duality},'' \href{http://dx.doi.org/10.1007/JHEP07(2024)295}{{\em
  JHEP} {\bfseries 07} (2024) 295},
  \href{http://arxiv.org/abs/2404.02191}{{\ttfamily arXiv:2404.02191
  [hep-th]}}.

\bibitem{Aharony:2013hda}
O.~Aharony, N.~Seiberg, and Y.~Tachikawa, ``{Reading between the lines of
  four-dimensional gauge theories},''
  \href{http://dx.doi.org/10.1007/JHEP08(2013)115}{{\em JHEP} {\bfseries 08}
  (2013) 115}, \href{http://arxiv.org/abs/1305.0318}{{\ttfamily arXiv:1305.0318
  [hep-th]}}.

\bibitem{Gaiotto:2014kfa}
D.~Gaiotto, A.~Kapustin, N.~Seiberg, and B.~Willett, ``{Generalized Global
  Symmetries},'' \href{http://dx.doi.org/10.1007/JHEP02(2015)172}{{\em JHEP}
  {\bfseries 02} (2015) 172}, \href{http://arxiv.org/abs/1412.5148}{{\ttfamily
  arXiv:1412.5148 [hep-th]}}.

\bibitem{Cordova:2022ruw}
C.~Cordova, T.~T. Dumitrescu, K.~Intriligator, and S.-H. Shao, ``{Snowmass
  White Paper: Generalized Symmetries in Quantum Field Theory and Beyond},'' in
  {\em {Snowmass 2021}}.
\newblock 5, 2022.
\newblock \href{http://arxiv.org/abs/2205.09545}{{\ttfamily arXiv:2205.09545
  [hep-th]}}.

\bibitem{Schafer-Nameki:2023jdn}
S.~Schafer-Nameki, ``{ICTP lectures on (non-)invertible generalized
  symmetries},'' \href{http://dx.doi.org/10.1016/j.physrep.2024.01.007}{{\em
  Phys. Rept.} {\bfseries 1063} (2024) 1--55},
  \href{http://arxiv.org/abs/2305.18296}{{\ttfamily arXiv:2305.18296
  [hep-th]}}.

\bibitem{Bhardwaj:2023kri}
L.~Bhardwaj, L.~E. Bottini, L.~Fraser-Taliente, L.~Gladden, D.~S.~W. Gould,
  A.~Platschorre, and H.~Tillim, ``{Lectures on generalized symmetries},''
  \href{http://dx.doi.org/10.1016/j.physrep.2023.11.002}{{\em Phys. Rept.}
  {\bfseries 1051} (2024) 1--87},
  \href{http://arxiv.org/abs/2307.07547}{{\ttfamily arXiv:2307.07547
  [hep-th]}}.

\bibitem{Luo:2023ive}
R.~Luo, Q.-R. Wang, and Y.-N. Wang, ``{Lecture notes on generalized symmetries
  and applications},''
  \href{http://dx.doi.org/10.1016/j.physrep.2024.02.002}{{\em Phys. Rept.}
  {\bfseries 1065} (2024) 1--43},
  \href{http://arxiv.org/abs/2307.09215}{{\ttfamily arXiv:2307.09215
  [hep-th]}}.

\bibitem{Shao:2023gho}
S.-H. Shao, ``{What's Done Cannot Be Undone: TASI Lectures on Non-Invertible
  Symmetry},'' \href{http://arxiv.org/abs/2308.00747}{{\ttfamily
  arXiv:2308.00747 [hep-th]}}.

\bibitem{DelZotto:2015isa}
M.~Del~Zotto, J.~J. Heckman, D.~S. Park, and T.~Rudelius, ``{On the Defect
  Group of a 6D SCFT},''
  \href{http://dx.doi.org/10.1007/s11005-016-0839-5}{{\em Lett. Math. Phys.}
  {\bfseries 106} no.~6, (2016) 765--786},
  \href{http://arxiv.org/abs/1503.04806}{{\ttfamily arXiv:1503.04806
  [hep-th]}}.

\bibitem{Garcia-Etxebarria:2019cnb}
I.~Garc{\' i}a~Etxebarria, B.~Heidenreich, and D.~Regalado, ``{IIB flux
  non-commutativity and the global structure of field theories},''
  \href{http://dx.doi.org/10.1007/JHEP10(2019)169}{{\em JHEP} {\bfseries 10}
  (2019) 169}, \href{http://arxiv.org/abs/1908.08027}{{\ttfamily
  arXiv:1908.08027 [hep-th]}}.

\bibitem{Morrison:2020ool}
D.~R. Morrison, S.~Schafer-Nameki, and B.~Willett, ``{Higher-Form Symmetries in
  5d},'' \href{http://dx.doi.org/10.1007/JHEP09(2020)024}{{\em JHEP} {\bfseries
  09} (2020) 024}, \href{http://arxiv.org/abs/2005.12296}{{\ttfamily
  arXiv:2005.12296 [hep-th]}}.

\bibitem{Albertini:2020mdx}
F.~Albertini, M.~Del~Zotto, I.~Garc\'\i{}a~Etxebarria, and S.~S. Hosseini,
  ``{Higher Form Symmetries and M-theory},''
  \href{http://dx.doi.org/10.1007/JHEP12(2020)203}{{\em JHEP} {\bfseries 12}
  (2020) 203}, \href{http://arxiv.org/abs/2005.12831}{{\ttfamily
  arXiv:2005.12831 [hep-th]}}.

\bibitem{Bah:2020uev}
I.~Bah, F.~Bonetti, and R.~Minasian, ``{Discrete and higher-form symmetries in
  SCFTs from wrapped M5-branes},''
  \href{http://dx.doi.org/10.1007/JHEP03(2021)196}{{\em JHEP} {\bfseries 03}
  (2021) 196}, \href{http://arxiv.org/abs/2007.15003}{{\ttfamily
  arXiv:2007.15003 [hep-th]}}.

\bibitem{DelZotto:2020esg}
M.~Del~Zotto, I.~Garcia~Etxebarria, and S.~S. Hosseini, ``{Higher form
  symmetries of Argyres-Douglas theories},''
  \href{http://dx.doi.org/10.1007/JHEP10(2020)056}{{\em JHEP} {\bfseries 10}
  (2020) 056}, \href{http://arxiv.org/abs/2007.15603}{{\ttfamily
  arXiv:2007.15603 [hep-th]}}.

\bibitem{Apruzzi:2020zot}
F.~Apruzzi, M.~Dierigl, and L.~Lin, ``{The Fate of Discrete 1-Form Symmetries
  in 6d},'' \href{http://dx.doi.org/10.21468/SciPostPhys.12.2.047}{{\em SciPost
  Phys.} {\bfseries 12} (2022) 047},
  \href{http://arxiv.org/abs/2008.09117}{{\ttfamily arXiv:2008.09117
  [hep-th]}}.

\bibitem{Bhardwaj:2020phs}
L.~Bhardwaj and S.~Sch\"afer-Nameki, ``{Higher-form symmetries of 6d and 5d
  theories},'' \href{http://dx.doi.org/10.1007/JHEP02(2021)159}{{\em JHEP}
  {\bfseries 02} (2021) 159}, \href{http://arxiv.org/abs/2008.09600}{{\ttfamily
  arXiv:2008.09600 [hep-th]}}.

\bibitem{Cvetic:2020kuw}
M.~Cveti\v{c}, M.~Dierigl, L.~Lin, and H.~Y. Zhang, ``{String Universality and
  Non-Simply-Connected Gauge Groups in 8d},''
  \href{http://dx.doi.org/10.1103/PhysRevLett.125.211602}{{\em Phys. Rev.
  Lett.} {\bfseries 125} no.~21, (2020) 211602},
  \href{http://arxiv.org/abs/2008.10605}{{\ttfamily arXiv:2008.10605
  [hep-th]}}.

\bibitem{DelZotto:2020sop}
M.~Del~Zotto and K.~Ohmori, ``{2-Group Symmetries of 6D Little String Theories
  and T-Duality},'' \href{http://dx.doi.org/10.1007/s00023-021-01018-3}{{\em
  Annales Henri Poincare} {\bfseries 22} no.~7, (2021) 2451--2474},
  \href{http://arxiv.org/abs/2009.03489}{{\ttfamily arXiv:2009.03489
  [hep-th]}}.

\bibitem{Bhardwaj:2021pfz}
L.~Bhardwaj, M.~Hubner, and S.~Schafer-Nameki, ``{1-form Symmetries of 4d N=2
  Class S Theories},'' \href{http://arxiv.org/abs/2102.01693}{{\ttfamily
  arXiv:2102.01693 [hep-th]}}.

\bibitem{Apruzzi:2021vcu}
F.~Apruzzi, L.~Bhardwaj, J.~Oh, and S.~Schafer-Nameki, ``{The Global Form of
  Flavor Symmetries and 2-Group Symmetries in 5d SCFTs},''
  \href{http://arxiv.org/abs/2105.08724}{{\ttfamily arXiv:2105.08724
  [hep-th]}}.

\bibitem{Cvetic:2021sxm}
M.~Cveti{\v c}, M.~Dierigl, L.~Lin, and H.~Y. Zhang, ``{Higher-form symmetries
  and their anomalies in M-/F-theory duality},''
  \href{http://dx.doi.org/10.1103/PhysRevD.104.126019}{{\em Phys. Rev. D}
  {\bfseries 104} no.~12, (2021) 126019},
  \href{http://arxiv.org/abs/2106.07654}{{\ttfamily arXiv:2106.07654
  [hep-th]}}.

\bibitem{Braun:2021sex}
A.~P. Braun, M.~Larfors, and P.-K. Oehlmann, ``{Gauged 2-form symmetries in 6D
  SCFTs coupled to gravity},''
  \href{http://dx.doi.org/10.1007/JHEP12(2021)132}{{\em JHEP} {\bfseries 12}
  (2021) 132}, \href{http://arxiv.org/abs/2106.13198}{{\ttfamily
  arXiv:2106.13198 [hep-th]}}.

\bibitem{Bhardwaj:2021wif}
L.~Bhardwaj, ``{2-Group symmetries in class S},''
  \href{http://dx.doi.org/10.21468/SciPostPhys.12.5.152}{{\em SciPost Phys.}
  {\bfseries 12} no.~5, (2022) 152},
  \href{http://arxiv.org/abs/2107.06816}{{\ttfamily arXiv:2107.06816
  [hep-th]}}.

\bibitem{Tian:2021cif}
J.~Tian and Y.-N. Wang, ``{5D and 6D SCFTs from $\mathbb{C}^3$ orbifolds},''
  \href{http://dx.doi.org/10.21468/SciPostPhys.12.4.127}{{\em SciPost Phys.}
  {\bfseries 12} no.~4, (2022) 127},
  \href{http://arxiv.org/abs/2110.15129}{{\ttfamily arXiv:2110.15129
  [hep-th]}}.

\bibitem{Bhardwaj:2021mzl}
L.~Bhardwaj, S.~Giacomelli, M.~H\"ubner, and S.~Sch\"afer-Nameki, ``{Relative
  Defects in Relative Theories: Trapped Higher-Form Symmetries and Irregular
  Punctures in Class S},'' \href{http://arxiv.org/abs/2201.00018}{{\ttfamily
  arXiv:2201.00018 [hep-th]}}.

\bibitem{DelZotto:2022fnw}
M.~Del~Zotto, J.~J. Heckman, S.~N. Meynet, R.~Moscrop, and H.~Y. Zhang,
  ``{Higher symmetries of 5D orbifold SCFTs},''
  \href{http://dx.doi.org/10.1103/PhysRevD.106.046010}{{\em Phys. Rev. D}
  {\bfseries 106} no.~4, (2022) 046010},
  \href{http://arxiv.org/abs/2201.08372}{{\ttfamily arXiv:2201.08372
  [hep-th]}}.

\bibitem{Cvetic:2022imb}
M.~Cveti\v{c}, J.~J. Heckman, M.~H\"ubner, and E.~Torres, ``{0-form, 1-form,
  and 2-group symmetries via cutting and gluing of orbifolds},''
  \href{http://dx.doi.org/10.1103/PhysRevD.106.106003}{{\em Phys. Rev. D}
  {\bfseries 106} no.~10, (2022) 106003},
  \href{http://arxiv.org/abs/2203.10102}{{\ttfamily arXiv:2203.10102
  [hep-th]}}.

\bibitem{DelZotto:2022joo}
M.~Del~Zotto, I.~Garc\'\i{}a~Etxebarria, and S.~Schafer-Nameki, ``{2-Group
  Symmetries and M-Theory},''
  \href{http://dx.doi.org/10.21468/SciPostPhys.13.5.105}{{\em SciPost Phys.}
  {\bfseries 13} (2022) 105}, \href{http://arxiv.org/abs/2203.10097}{{\ttfamily
  arXiv:2203.10097 [hep-th]}}.

\bibitem{Hubner:2022kxr}
M.~Hubner, D.~R. Morrison, S.~Schafer-Nameki, and Y.-N. Wang, ``{Generalized
  Symmetries in F-theory and the Topology of Elliptic Fibrations},''
  \href{http://dx.doi.org/10.21468/SciPostPhys.13.2.030}{{\em SciPost Phys.}
  {\bfseries 13} no.~2, (2022) 030},
  \href{http://arxiv.org/abs/2203.10022}{{\ttfamily arXiv:2203.10022
  [hep-th]}}.

\bibitem{Bashmakov:2022jtl}
V.~Bashmakov, M.~Del~Zotto, and A.~Hasan, ``{On the 6d origin of non-invertible
  symmetries in 4d},'' \href{http://dx.doi.org/10.1007/JHEP09(2023)161}{{\em
  JHEP} {\bfseries 09} (2023) 161},
  \href{http://arxiv.org/abs/2206.07073}{{\ttfamily arXiv:2206.07073
  [hep-th]}}.

\bibitem{vanBeest:2022fss}
M.~van Beest, D.~S.~W. Gould, S.~Schafer-Nameki, and Y.-N. Wang, ``{Symmetry
  TFTs for 3d QFTs from M-theory},''
  \href{http://dx.doi.org/10.1007/JHEP02(2023)226}{{\em JHEP} {\bfseries 02}
  (2023) 226}, \href{http://arxiv.org/abs/2210.03703}{{\ttfamily
  arXiv:2210.03703 [hep-th]}}.

\bibitem{Bashmakov:2022uek}
V.~Bashmakov, M.~Del~Zotto, A.~Hasan, and J.~Kaidi, ``{Non-invertible
  symmetries of class S theories},''
  \href{http://dx.doi.org/10.1007/JHEP05(2023)225}{{\em JHEP} {\bfseries 05}
  (2023) 225}, \href{http://arxiv.org/abs/2211.05138}{{\ttfamily
  arXiv:2211.05138 [hep-th]}}.

\bibitem{Acharya:2023bth}
B.~S. Acharya, M.~Del~Zotto, J.~J. Heckman, M.~Hubner, and E.~Torres,
  ``{Junctions, Edge Modes, and $G_2$-Holonomy Orbifolds},''
  \href{http://arxiv.org/abs/2304.03300}{{\ttfamily arXiv:2304.03300
  [hep-th]}}.

\bibitem{Dierigl:2023jdp}
M.~Dierigl, J.~J. Heckman, M.~Montero, and E.~Torres, ``{R7-branes as charge
  conjugation operators},''
  \href{http://dx.doi.org/10.1103/PhysRevD.109.046004}{{\em Phys. Rev. D}
  {\bfseries 109} no.~4, (2024) 046004},
  \href{http://arxiv.org/abs/2305.05689}{{\ttfamily arXiv:2305.05689
  [hep-th]}}.

\bibitem{Cvetic:2023plv}
M.~Cveti\v{c}, J.~J. Heckman, M.~H\"ubner, and E.~Torres, ``{Fluxbranes,
  generalized symmetries, and Verlinde\textquoteright{}s metastable
  monopole},'' \href{http://dx.doi.org/10.1103/PhysRevD.109.046007}{{\em Phys.
  Rev. D} {\bfseries 109} no.~4, (2024) 046007},
  \href{http://arxiv.org/abs/2305.09665}{{\ttfamily arXiv:2305.09665
  [hep-th]}}.

\bibitem{Bashmakov:2023kwo}
V.~Bashmakov, M.~Del~Zotto, and A.~Hasan, ``{Four-manifolds and Symmetry
  Categories of 2d CFTs},'' \href{http://arxiv.org/abs/2305.10422}{{\ttfamily
  arXiv:2305.10422 [hep-th]}}.

\bibitem{Apruzzi:2023uma}
F.~Apruzzi, F.~Bonetti, D.~S.~W. Gould, and S.~Schafer-Nameki, ``{Aspects of
  categorical symmetries from branes: SymTFTs and generalized charges},''
  \href{http://dx.doi.org/10.21468/SciPostPhys.17.1.025}{{\em SciPost Phys.}
  {\bfseries 17} no.~1, (2024) 025},
  \href{http://arxiv.org/abs/2306.16405}{{\ttfamily arXiv:2306.16405
  [hep-th]}}.

\bibitem{Closset:2023pmc}
C.~Closset and H.~Magureanu, ``{Reading between the rational sections: Global
  structures of 4d $\mathcal{N}=2$ KK theories},''
  \href{http://dx.doi.org/10.21468/SciPostPhys.16.5.137}{{\em SciPost Phys.}
  {\bfseries 16} no.~5, (2024) 137},
  \href{http://arxiv.org/abs/2308.10225}{{\ttfamily arXiv:2308.10225
  [hep-th]}}.

\bibitem{Heckman:2022xgu}
J.~J. Heckman, M.~Hubner, E.~Torres, X.~Yu, and H.~Y. Zhang, ``{Top down
  approach to topological duality defects},''
  \href{http://dx.doi.org/10.1103/PhysRevD.108.046015}{{\em Phys. Rev. D}
  {\bfseries 108} no.~4, (2023) 046015},
  \href{http://arxiv.org/abs/2212.09743}{{\ttfamily arXiv:2212.09743
  [hep-th]}}.

\bibitem{Zhang:2024oas}
H.~Y. Zhang, ``{K-theoretic Global Symmetry in String-constructed QFT and
  T-duality},'' \href{http://arxiv.org/abs/2404.16097}{{\ttfamily
  arXiv:2404.16097 [hep-th]}}.

\bibitem{Torres:2024sbl}
E.~Torres, ``{Giving a $KO$ to 8D Gauge Anomalies},''
  \href{http://arxiv.org/abs/2405.08809}{{\ttfamily arXiv:2405.08809
  [hep-th]}}.

\bibitem{Apruzzi:2021nmk}
F.~Apruzzi, F.~Bonetti, I.~Garc\'\i{}a~Etxebarria, S.~S. Hosseini, and
  S.~Schafer-Nameki, ``{Symmetry TFTs from String Theory},''
  \href{http://dx.doi.org/10.1007/s00220-023-04737-2}{{\em Commun. Math. Phys.}
  {\bfseries 402} no.~1, (2023) 895--949},
  \href{http://arxiv.org/abs/2112.02092}{{\ttfamily arXiv:2112.02092
  [hep-th]}}.

\bibitem{Apruzzi:2022rei}
F.~Apruzzi, I.~Bah, F.~Bonetti, and S.~Schafer-Nameki, ``{Noninvertible
  Symmetries from Holography and Branes},''
  \href{http://dx.doi.org/10.1103/PhysRevLett.130.121601}{{\em Phys. Rev.
  Lett.} {\bfseries 130} no.~12, (2023) 121601},
  \href{http://arxiv.org/abs/2208.07373}{{\ttfamily arXiv:2208.07373
  [hep-th]}}.

\bibitem{GarciaEtxebarria:2022vzq}
I.~Garc\'\i{}a~Etxebarria, ``{Branes and Non-Invertible Symmetries},''
  \href{http://dx.doi.org/10.1002/prop.202200154}{{\em Fortsch. Phys.}
  {\bfseries 70} no.~11, (2022) 2200154},
  \href{http://arxiv.org/abs/2208.07508}{{\ttfamily arXiv:2208.07508
  [hep-th]}}.

\bibitem{Heckman:2022muc}
J.~J. Heckman, M.~H\"ubner, E.~Torres, and H.~Y. Zhang, ``{The Branes Behind
  Generalized Symmetry Operators},''
  \href{http://dx.doi.org/10.1002/prop.202200180}{{\em Fortsch. Phys.}
  {\bfseries 71} no.~1, (2023) 2200180},
  \href{http://arxiv.org/abs/2209.03343}{{\ttfamily arXiv:2209.03343
  [hep-th]}}.

\bibitem{Freed:2012bs}
D.~S. Freed and C.~Teleman, ``{Relative quantum field theory},''
  \href{http://dx.doi.org/10.1007/s00220-013-1880-1}{{\em Commun. Math. Phys.}
  {\bfseries 326} (2014) 459--476},
  \href{http://arxiv.org/abs/1212.1692}{{\ttfamily arXiv:1212.1692 [hep-th]}}.

\bibitem{Freed:2022qnc}
D.~S. Freed, G.~W. Moore, and C.~Teleman, ``{Topological symmetry in quantum
  field theory},'' \href{http://arxiv.org/abs/2209.07471}{{\ttfamily
  arXiv:2209.07471 [hep-th]}}.

\bibitem{Witten:1998wy}
E.~Witten, ``{AdS / CFT correspondence and topological field theory},''
  \href{http://dx.doi.org/10.1088/1126-6708/1998/12/012}{{\em JHEP} {\bfseries
  12} (1998) 012}, \href{http://arxiv.org/abs/hep-th/9812012}{{\ttfamily
  arXiv:hep-th/9812012}}.

\bibitem{Belov:2006xj}
D.~M. Belov and G.~W. Moore, ``{Type II Actions from 11-Dimensional
  Chern-Simons Theories},''
  \href{http://arxiv.org/abs/hep-th/0611020}{{\ttfamily arXiv:hep-th/0611020}}.

\bibitem{Gukov:2020btk}
S.~Gukov, P.-S. Hsin, and D.~Pei, ``{Generalized Global Symmetries of $T[M]$
  Theories. I},'' \href{http://arxiv.org/abs/2010.15890}{{\ttfamily
  arXiv:2010.15890 [hep-th]}}.

\bibitem{Apruzzi:2022dlm}
F.~Apruzzi, ``{Higher Form Symmetries TFT in 6d},''
  \href{http://arxiv.org/abs/2203.10063}{{\ttfamily arXiv:2203.10063
  [hep-th]}}.

\bibitem{Bergman:2022otk}
O.~Bergman and S.~Hirano, ``{The holography of duality in $ \mathcal{N} $ = 4
  Super-Yang-Mills theory},''
  \href{http://dx.doi.org/10.1007/JHEP11(2022)069}{{\em JHEP} {\bfseries 11}
  (2022) 069}, \href{http://arxiv.org/abs/2208.09396}{{\ttfamily
  arXiv:2208.09396 [hep-th]}}.

\bibitem{Lawrie:2023tdz}
C.~Lawrie, X.~Yu, and H.~Y. Zhang, ``{Intermediate defect groups, polarization
  pairs, and noninvertible duality defects},''
  \href{http://dx.doi.org/10.1103/PhysRevD.109.026005}{{\em Phys. Rev. D}
  {\bfseries 109} no.~2, (2024) 026005},
  \href{http://arxiv.org/abs/2306.11783}{{\ttfamily arXiv:2306.11783
  [hep-th]}}.

\bibitem{Cvetic:2023pgm}
M.~Cveti\v{c}, J.~J. Heckman, M.~H\"ubner, and E.~Torres, ``{Generalized
  symmetries, gravity, and the swampland},''
  \href{http://dx.doi.org/10.1103/PhysRevD.109.026012}{{\em Phys. Rev. D}
  {\bfseries 109} no.~2, (2024) 026012},
  \href{http://arxiv.org/abs/2307.13027}{{\ttfamily arXiv:2307.13027
  [hep-th]}}.

\bibitem{Baume:2023kkf}
F.~Baume, J.~J. Heckman, M.~H\"ubner, E.~Torres, A.~P. Turner, and X.~Yu,
  ``{SymTrees and Multi-Sector QFTs},''
  \href{http://dx.doi.org/10.1103/PhysRevD.109.106013}{{\em Phys. Rev. D}
  {\bfseries 109} no.~10, (2024) 106013},
  \href{http://arxiv.org/abs/2310.12980}{{\ttfamily arXiv:2310.12980
  [hep-th]}}.

\bibitem{Yu:2023nyn}
X.~Yu, ``{Noninvertible symmetries in 2D from type IIB string theory},''
  \href{http://dx.doi.org/10.1103/PhysRevD.110.065008}{{\em Phys. Rev. D}
  {\bfseries 110} no.~6, (2024) 065008},
  \href{http://arxiv.org/abs/2310.15339}{{\ttfamily arXiv:2310.15339
  [hep-th]}}.

\bibitem{Gould:2023wgl}
D.~S.~W. Gould, L.~Lin, and E.~Sabag, ``{Swampland constraints on the SymTFT of
  supergravity},'' \href{http://dx.doi.org/10.1103/PhysRevD.109.126005}{{\em
  Phys. Rev. D} {\bfseries 109} no.~12, (2024) 126005},
  \href{http://arxiv.org/abs/2312.02131}{{\ttfamily arXiv:2312.02131
  [hep-th]}}.

\bibitem{Heckman:2024oot}
J.~J. Heckman, M.~H\"ubner, and C.~Murdia, ``{On the holographic dual of a
  topological symmetry operator},''
  \href{http://dx.doi.org/10.1103/PhysRevD.110.046007}{{\em Phys. Rev. D}
  {\bfseries 110} no.~4, (2024) 046007},
  \href{http://arxiv.org/abs/2401.09538}{{\ttfamily arXiv:2401.09538
  [hep-th]}}.

\bibitem{Braeger:2024jcj}
N.~Braeger, V.~Chakrabhavi, J.~J. Heckman, and M.~Hubner, ``{Generalized
  Symmetries of Non-Supersymmetric Orbifolds},''
  \href{http://arxiv.org/abs/2404.17639}{{\ttfamily arXiv:2404.17639
  [hep-th]}}.

\bibitem{Cvetic:2024dzu}
M.~Cveti\v{c}, R.~Donagi, J.~J. Heckman, M.~H\"ubner, and E.~Torres,
  ``{Cornering Relative Symmetry Theories},''
  \href{http://arxiv.org/abs/2408.12600}{{\ttfamily arXiv:2408.12600
  [hep-th]}}.

\bibitem{Apruzzi:2024htg}
F.~Apruzzi, F.~Bedogna, and N.~Dondi, ``{SymTh for non-finite symmetries},''
  \href{http://arxiv.org/abs/2402.14813}{{\ttfamily arXiv:2402.14813
  [hep-th]}}.

\bibitem{GarciaEtxebarria:2024fuk}
I.~Garc\'\i{}a~Etxebarria and S.~S. Hosseini, ``{Some aspects of symmetry
  descent},'' \href{http://arxiv.org/abs/2404.16028}{{\ttfamily
  arXiv:2404.16028 [hep-th]}}.

\bibitem{Heckman:2024zdo}
J.~J. Heckman and M.~H\"ubner, ``{Celestial Topology, Symmetry Theories, and
  Evidence for a Non-SUSY D3-Brane CFT},''
  \href{http://arxiv.org/abs/2406.08485}{{\ttfamily arXiv:2406.08485
  [hep-th]}}.

\bibitem{Cvetic:2022uuu}
M.~Cveti\v{c}, M.~Dierigl, L.~Lin, and H.~Y. Zhang, ``{All eight- and
  nine-dimensional string vacua from junctions},''
  \href{http://dx.doi.org/10.1103/PhysRevD.106.026007}{{\em Phys. Rev. D}
  {\bfseries 106} no.~2, (2022) 026007},
  \href{http://arxiv.org/abs/2203.03644}{{\ttfamily arXiv:2203.03644
  [hep-th]}}.

\bibitem{Font:2021uyw}
A.~Font, B.~Fraiman, M.~Gra{\~ n}a, C.~A. N\'u{\~ n}ez, and H.~Parra
  De~Freitas, ``{Exploring the landscape of CHL strings on T$^{d}$},''
  \href{http://dx.doi.org/10.1007/JHEP08(2021)095}{{\em JHEP} {\bfseries 08}
  (2021) 095}, \href{http://arxiv.org/abs/2104.07131}{{\ttfamily
  arXiv:2104.07131 [hep-th]}}.

\bibitem{Cvetic:2021sjm}
M.~Cveti{\v c}, M.~Dierigl, L.~Lin, and H.~Y. Zhang, ``{Gauge group topology of
  8D Chaudhuri-Hockney-Lykken vacua},''
  \href{http://dx.doi.org/10.1103/PhysRevD.104.086018}{{\em Phys. Rev. D}
  {\bfseries 104} no.~8, (2021) 086018},
  \href{http://arxiv.org/abs/2107.04031}{{\ttfamily arXiv:2107.04031
  [hep-th]}}.

\bibitem{Fraiman:2021soq}
B.~Fraiman and H.~P. De~Freitas, ``{Symmetry enhancements in 7d heterotic
  strings},'' \href{http://dx.doi.org/10.1007/JHEP10(2021)002}{{\em JHEP}
  {\bfseries 10} (2021) 002}, \href{http://arxiv.org/abs/2106.08189}{{\ttfamily
  arXiv:2106.08189 [hep-th]}}.

\bibitem{Fraiman:2022aik}
B.~Fraiman and H.~Parra De~Freitas, ``{Unifying the 6D $ \mathcal{N} $ = (1, 1)
  string landscape},'' \href{http://dx.doi.org/10.1007/JHEP02(2023)204}{{\em
  JHEP} {\bfseries 02} (2023) 204},
  \href{http://arxiv.org/abs/2209.06214}{{\ttfamily arXiv:2209.06214
  [hep-th]}}.

\bibitem{Tachikawa:Slides}
Y.~Tachikawa, ``{On fractional M5 branes and frozen singularities},'' 2015.
\newblock \url{https://member.ipmu.jp/yuji.tachikawa/transp/kiastalk.pdf}.

\bibitem{Witten:1996md}
E.~Witten, ``{On flux quantization in M theory and the effective action},''
  \href{http://dx.doi.org/10.1016/S0393-0440(96)00042-3}{{\em J. Geom. Phys.}
  {\bfseries 22} (1997) 1--13},
  \href{http://arxiv.org/abs/hep-th/9609122}{{\ttfamily arXiv:hep-th/9609122}}.

\bibitem{Ohmori:2014kda}
K.~Ohmori, H.~Shimizu, Y.~Tachikawa, and K.~Yonekura, ``{Anomaly polynomial of
  general 6d SCFTs},'' \href{http://dx.doi.org/10.1093/ptep/ptu140}{{\em PTEP}
  {\bfseries 2014} no.~10, (2014) 103B07},
  \href{http://arxiv.org/abs/1408.5572}{{\ttfamily arXiv:1408.5572 [hep-th]}}.

\bibitem{Fuchs:1997jv}
J.~Fuchs and C.~Schweigert, {\em {Symmetries, Lie algebras and representations:
  A graduate course for physicists}}.
\newblock Cambridge University Press, 10, 2003.

\bibitem{Borel:1999bx}
A.~Borel, R.~Friedman, and J.~W. Morgan, ``{Almost commuting elements in
  compact Lie groups},'' \href{http://arxiv.org/abs/math/9907007}{{\ttfamily
  arXiv:math/9907007}}.

\bibitem{Meyers:1979pd}
C.~Meyers, M.~De~Roo, and P.~Sorba, ``{GROUP THEORETICAL ASPECTS OF
  INSTANTONS},'' \href{http://dx.doi.org/10.1007/BF02770858}{{\em Nuovo Cim. A}
  {\bfseries 52} (1979) 519--530}.

\bibitem{DelZotto:2014hpa}
M.~Del~Zotto, J.~J. Heckman, A.~Tomasiello, and C.~Vafa, ``{6d Conformal
  Matter},'' \href{http://dx.doi.org/10.1007/JHEP02(2015)054}{{\em JHEP}
  {\bfseries 02} (2015) 054}, \href{http://arxiv.org/abs/1407.6359}{{\ttfamily
  arXiv:1407.6359 [hep-th]}}.

\bibitem{Hanany:1996ie}
A.~Hanany and E.~Witten, ``{Type IIB superstrings, BPS monopoles, and
  three-dimensional gauge dynamics},''
  \href{http://dx.doi.org/10.1016/S0550-3213(97)00157-0}{{\em Nucl. Phys. B}
  {\bfseries 492} (1997) 152--190},
  \href{http://arxiv.org/abs/hep-th/9611230}{{\ttfamily arXiv:hep-th/9611230}}.

\bibitem{Esole:2020tby}
M.~Esole and M.~J. Kang, ``{Matter representations from geometry: under the
  spell of Dynkin},'' \href{http://arxiv.org/abs/2012.13401}{{\ttfamily
  arXiv:2012.13401 [hep-th]}}.

\bibitem{Yamatsu:2015npn}
N.~Yamatsu, ``{Finite-Dimensional Lie Algebras and Their Representations for
  Unified Model Building},'' \href{http://arxiv.org/abs/1511.08771}{{\ttfamily
  arXiv:1511.08771 [hep-ph]}}.

\bibitem{Douglas:1996sw}
M.~R. Douglas and G.~W. Moore, ``{D-branes, quivers, and ALE instantons},''
  \href{http://arxiv.org/abs/hep-th/9603167}{{\ttfamily arXiv:hep-th/9603167}}.

\bibitem{Johnson:1996py}
C.~V. Johnson and R.~C. Myers, ``{Aspects of type IIB theory on ALE spaces},''
  \href{http://dx.doi.org/10.1103/PhysRevD.55.6382}{{\em Phys. Rev. D}
  {\bfseries 55} (1997) 6382--6393},
  \href{http://arxiv.org/abs/hep-th/9610140}{{\ttfamily arXiv:hep-th/9610140}}.

\bibitem{Gross:2000wc}
D.~J. Gross and N.~A. Nekrasov, ``{Monopoles and strings in noncommutative
  gauge theory},'' \href{http://dx.doi.org/10.1088/1126-6708/2000/07/034}{{\em
  JHEP} {\bfseries 07} (2000) 034},
  \href{http://arxiv.org/abs/hep-th/0005204}{{\ttfamily arXiv:hep-th/0005204}}.

\bibitem{Douglas:1995bn}
M.~R. Douglas, ``{Branes within branes},'' {\em NATO Sci. Ser. C} {\bfseries
  520} (1999) 267--275, \href{http://arxiv.org/abs/hep-th/9512077}{{\ttfamily
  arXiv:hep-th/9512077}}.

\bibitem{Aspinwall:2000xs}
P.~S. Aspinwall and M.~R. Plesser, ``{D-branes, discrete torsion and the McKay
  correspondence},''
  \href{http://dx.doi.org/10.1088/1126-6708/2001/02/009}{{\em JHEP} {\bfseries
  02} (2001) 009}, \href{http://arxiv.org/abs/hep-th/0009042}{{\ttfamily
  arXiv:hep-th/0009042}}.

\bibitem{Weigand:2018rez}
T.~Weigand, ``{F-theory},'' {\em PoS} {\bfseries TASI2017} (2018) 016,
  \href{http://arxiv.org/abs/1806.01854}{{\ttfamily arXiv:1806.01854
  [hep-th]}}.

\bibitem{Cvetic:2018bni}
M.~Cveti\v{c} and L.~Lin, ``{TASI Lectures on Abelian and Discrete Symmetries
  in F-theory},'' \href{http://dx.doi.org/10.22323/1.305.0020}{{\em PoS}
  {\bfseries TASI2017} (2018) 020},
  \href{http://arxiv.org/abs/1809.00012}{{\ttfamily arXiv:1809.00012
  [hep-th]}}.

\bibitem{Nahm:2001kh}
W.~Nahm and K.~Wendland, ``{Mirror symmetry on Kummer type K3 surfaces},''
  \href{http://dx.doi.org/10.1007/s00220-003-0985-3}{{\em Commun. Math. Phys.}
  {\bfseries 243} (2003) 557--582},
  \href{http://arxiv.org/abs/hep-th/0106104}{{\ttfamily arXiv:hep-th/0106104}}.

\bibitem{Cvetic:2021maf}
M.~Cveti\v{c}, J.~J. Heckman, E.~Torres, and G.~Zoccarato, ``{Reflections on
  the matter of 3D N=1 vacua and local Spin(7) compactifications},''
  \href{http://dx.doi.org/10.1103/PhysRevD.105.026008}{{\em Phys. Rev. D}
  {\bfseries 105} no.~2, (2022) 026008},
  \href{http://arxiv.org/abs/2107.00025}{{\ttfamily arXiv:2107.00025
  [hep-th]}}.

\bibitem{Heckman:2017uxe}
J.~J. Heckman and L.~Tizzano, ``{6D Fractional Quantum Hall Effect},''
  \href{http://dx.doi.org/10.1007/JHEP05(2018)120}{{\em JHEP} {\bfseries 05}
  (2018) 120}, \href{http://arxiv.org/abs/1708.02250}{{\ttfamily
  arXiv:1708.02250 [hep-th]}}.

\bibitem{Kapustin:2014gua}
A.~Kapustin and N.~Seiberg, ``{Coupling a QFT to a TQFT and Duality},''
  \href{http://dx.doi.org/10.1007/JHEP04(2014)001}{{\em JHEP} {\bfseries 04}
  (2014) 001}, \href{http://arxiv.org/abs/1401.0740}{{\ttfamily arXiv:1401.0740
  [hep-th]}}.

\bibitem{19881}
A.~Fujiki, ``Finite automorphism groups of complex tori of dimension two,''
  \href{http://dx.doi.org/10.2977/prims/1195175326}{{\em Publications of the
  Research Institute for Mathematical Sciences} {\bfseries 24} no.~1, (1988)
  1--97}.

\bibitem{Polchinski:2003bq}
J.~Polchinski, ``{Monopoles, duality, and string theory},''
  \href{http://dx.doi.org/10.1142/S0217751X0401866X}{{\em Int. J. Mod. Phys. A}
  {\bfseries 19S1} (2004) 145--156},
  \href{http://arxiv.org/abs/hep-th/0304042}{{\ttfamily arXiv:hep-th/0304042}}.

\bibitem{Heidenreich:2021xpr}
B.~Heidenreich, J.~McNamara, M.~Montero, M.~Reece, T.~Rudelius, and
  I.~Valenzuela, ``{Non-invertible global symmetries and completeness of the
  spectrum},'' \href{http://dx.doi.org/10.1007/JHEP09(2021)203}{{\em JHEP}
  {\bfseries 09} (2021) 203}, \href{http://arxiv.org/abs/2104.07036}{{\ttfamily
  arXiv:2104.07036 [hep-th]}}.

\bibitem{Vafa:1994rv}
C.~Vafa and E.~Witten, ``{On orbifolds with discrete torsion},''
  \href{http://dx.doi.org/10.1016/0393-0440(94)00048-9}{{\em J. Geom. Phys.}
  {\bfseries 15} (1995) 189--214},
  \href{http://arxiv.org/abs/hep-th/9409188}{{\ttfamily arXiv:hep-th/9409188}}.

\bibitem{Baykara:2023plc}
Z.~K. Baykara, Y.~Hamada, H.-C. Tarazi, and C.~Vafa, ``{On the string landscape
  without hypermultiplets},''
  \href{http://dx.doi.org/10.1007/JHEP04(2024)121}{{\em JHEP} {\bfseries 04}
  (2024) 121}, \href{http://arxiv.org/abs/2309.15152}{{\ttfamily
  arXiv:2309.15152 [hep-th]}}.

\bibitem{Baykara:2024vss}
Z.~K. Baykara, H.-C. Tarazi, and C.~Vafa, ``{The Quasicrystalline String
  Landscape},'' \href{http://arxiv.org/abs/2406.00129}{{\ttfamily
  arXiv:2406.00129 [hep-th]}}.

\bibitem{Dabholkar:1998kv}
A.~Dabholkar and J.~A. Harvey, ``{String islands},''
  \href{http://dx.doi.org/10.1088/1126-6708/1999/02/006}{{\em JHEP} {\bfseries
  02} (1999) 006}, \href{http://arxiv.org/abs/hep-th/9809122}{{\ttfamily
  arXiv:hep-th/9809122}}.

\bibitem{Witten:1997kz}
E.~Witten, ``{New 'gauge' theories in six-dimensions},''
  \href{http://dx.doi.org/10.1088/1126-6708/1998/01/001}{{\em JHEP} {\bfseries
  01} (1998) 001}, \href{http://arxiv.org/abs/hep-th/9710065}{{\ttfamily
  arXiv:hep-th/9710065}}.

\bibitem{Camara:2011jg}
P.~G. Camara, L.~E. Ibanez, and F.~Marchesano, ``{RR photons},''
  \href{http://dx.doi.org/10.1007/JHEP09(2011)110}{{\em JHEP} {\bfseries 09}
  (2011) 110}, \href{http://arxiv.org/abs/1106.0060}{{\ttfamily arXiv:1106.0060
  [hep-th]}}.

\bibitem{CueMaudeZee}
T.~M.~A. Project.
\newblock \url{http://www.map.mpim-bonn.mpg.de/Linking_form}.

\end{thebibliography}\endgroup

\end{document}